\documentclass[acmsmall,natbib=false,nonacm]{acmart}

\usepackage{xspace}
\usepackage{csquotes}
\usepackage{tikz}
\usetikzlibrary{math,calc,shapes,arrows,positioning,fit}

\usepackage{longtable}
\usepackage{array}
\newcolumntype{x}[1]{>{\centering\arraybackslash\hspace{0pt}}p{#1}}
\usepackage{enumitem}
\usepackage{booktabs}
\usepackage{adjustbox}
\usepackage{pdflscape}

\usepackage{lineno}
\usepackage{algorithm}
\usepackage{algpseudocode}
\usepackage{listings}

\newcommand{\Iltis}{\emph{Iltis}\xspace}
\newcommand{\eps}{\varepsilon}
\renewcommand{\epsilon}{\varepsilon}
\newcommand{\curl}{\ensuremath{\rightsquigarrow}}
\newcommand{\rulesep}{;\allowbreak\hspace{2mm}}
\newcommand{\NP}{\textsc{NP}\xspace}

\newcommand{\N}{\ensuremath{\mathbb{N}}}

\newcommand{\bigO}{\ensuremath{\mathcal{O}}}

\newcommand{\df}{\ensuremath{\mathrel{\smash{\stackrel{\scriptscriptstyle{
    \text{def}}}{=}}}} \;}

\providecommand {\calC}      {{\mathcal C}\xspace}

\providecommand {\calG}      {{\mathcal G}\xspace}

\providecommand {\calI}      {{\mathcal I}\xspace}

\providecommand {\calN}      {{\mathcal N}\xspace}

\providecommand {\calV}      {{\mathcal V}\xspace}

\RequirePackage[
  datamodel=acmdatamodel,
  style=acmnumeric,
  maxnames=99
  ]{biblatex}

\setcopyright{cc}
\setcctype{by}

\begin{document}

\title{Detecting and Explaining (In-)equivalence of Context-Free Grammars}

\author{Marko Schmellenkamp}
\orcid{0000-0003-3966-6590}
\affiliation{\institution{Ruhr University Bochum}
  \city{Bochum}
  \country{Germany}
}
\email{marko.schmellenkamp@rub.de}

\author{Thomas Zeume}
\orcid{0000-0002-5186-7507}
\affiliation{\institution{Ruhr University Bochum}
  \city{Bochum}
  \country{Germany}
}
\email{thomas.zeume@rub.de}

\author{Sven Argo}
\orcid{0009-0009-5544-6417}
\affiliation{\institution{Ruhr University Bochum}
  \city{Bochum}
  \country{Germany}
}
\email{sven.argo@rub.de}

\author{Sandra Kiefer}
\orcid{0000-0003-4614-9444}
\affiliation{\institution{University of Oxford}
  \city{Oxford}
  \country{United Kingdom}
}
\email{sandra.kiefer@cs.ox.ac.uk}

\author{Cedric Siems}
\orcid{0009-0007-5671-2209}
\affiliation{\institution{Ruhr University Bochum}
  \city{Bochum}
  \country{Germany}
}
\email{cedric.siems@tu-dortmund.de}

\author{Fynn Stebel}
\orcid{0009-0009-1352-7892}
\affiliation{\institution{Ruhr University Bochum}
  \city{Bochum}
  \country{Germany}
}
\email{fynn.stebel@rub.de}

\begin{abstract}

We propose a scalable framework for deciding, proving, and explaining (in-)equivalence of context-free grammars. We present an implementation of the framework and evaluate it on large data sets collected within educational support systems. Even though the equivalence problem for context-free languages is undecidable in general, the framework is able to handle a large portion of these datasets. It introduces and combines techniques from several areas, such as an abstract grammar transformation language to identify equivalent grammars as well as sufficiently similar inequivalent grammars, theory-based comparison algorithms for a large class of context-free languages, and a graph-theory-inspired grammar canonization that allows to efficiently identify isomorphic grammars.

 \end{abstract}



\maketitle

\section{Motivation and Introduction}\label{sec:motivation}

The syntax, specification, and parsing of programming languages is an integral part of most undergraduate computer science curricula (see, e.g., \cite{CS2023,GI2016,Dougherty24}). Courses on this topic, such as introductory courses on formal languages, typically cover modelling with representations for regular and context-free languages, as well as properties and algorithms for these classes of languages. 

The objective of this paper is to explore the algorithmic foundations of educational support systems for formal languages and, specifically, to develop a practically relevant framework for modelling with context-free representations. 
Ideally, such educational systems provide immediate feedback and witnessing explanations to students, such as those shown in this interaction:

\noindent
\textbf{Assignment:} \hspace{0.4cm}	\textit{Design a context-free grammar for the language} $L = \{a^nb^{n+2} \mid n \in \mathbb{N}\}$\\
\textbf{Student:} \hspace{1.2cm}$S  \rightarrow a S b \mid abb$\\
\textbf{Potential feedback (provided in customizable stages):}
		\begin{itemize}
		 \item[(1)]  Correctness: \textit{Your grammar is not correct.}
			\item[(2)] Pinpointing mistakes:
		\begin{itemize}
		 \item[(a)] Relating the solution to a non-equivalent language: \\
			\textit{Your grammar describes the language}  $L = \{a^nb^{n+1} \mid n \in \mathbb{N}\}$.

		 \item[(b)] Hinting at a wrong production: \\
		    \textit{It might help to have a look at the production $S \rightarrow abb$.}

		 \item[(c)] Providing a counterexample: \\
		    \textit{The language described by your grammar contains the word $abb$, which is not in $L$.}

		\end{itemize}
		\end{itemize}

Providing such feedback algorithmically is difficult. Already testing a solution attempt for correctness -- which is required for (1) -- is undecidable in general, since it requires testing the equivalence of two context-free grammars. It is even less clear how to algorithmically produce more meaningful feedback such as (2a) and (2b).

\begin{description}
 \item[Challenge:] Providing feedback for context-free modelling assignments requires handling inherently hard algorithmic problems.
\end{description}

And yet, despite the complexity-theoretic obstacles, human tutors manage to provide useful feedback, such as witnesses for mistakes.

In this paper, we provide the theoretical foundations and a prototypical implementation of a framework for testing and proving (in-)equivalence of grammars and producing advanced explanations. For its relevance in practice, we require our framework to be efficient and scalable.

Our motivation for this line of work is that providing human tutoring on this topic is very resource-intense, especially with the rapidly increasing number of students enrolled in computer science courses from diverse backgrounds \cite{CRA2017}. This can be addressed by \enquote{leveraging technologies to make the most effective use of students' time, shifting from information delivery to sense-making and practice in class} \cite{Singer2012,Beach2012} which has been outlined as a general strategy by the US National Research Council for improving learning outcomes across STEM disciplines. The combination of lectures and tutorials with dedicated, content-specific educational support systems has proven to be an effective and motivating way of enhancing learning outcomes. Immediate and personalised feedback helps students and can be  used flexibly to prepare them for tutorials as well as exams.

For formal languages, there is a plethora of educational support systems, which vary greatly in the number of supported exercise types and the quality of user interactions. Well-known systems include
\emph{AutomataTutor} \cite{AntoniKAGV2015,AntoniHKRW2020},
\emph{FLACI} \cite{HielscherW19},
\emph{Iltis} \cite{SchmellenkampVZ24},
\emph{JFLAP} \cite{GramondR1999,Rodger1999}, and
\emph{RASCO} \cite{CreusG14}.

While most of these systems support exercises in which users are asked to construct a context-free grammar for a given formal language, the  explanatory feedback by the system is very limited. Usually, the system tries to provide counterexamples as witnesses when the submitted grammar is incorrect. If the system fails to find a counterexample, it is often (and sometimes incorrectly!) concluded that the submitted solution is correct. These shortcomings with respect to feedback are not surprising, due to the algorithmic hardness of testing equivalence outlined above. However, particularly in an educational context, misclassifications of solutions should be avoided at all costs.

\subsection{Contributions} We propose a scalable framework for proving and explaining (in-)equivalence of context-free grammars. We present an implementation and evaluate it with data collected within the educational support systems \emph{AutomataTutor} and \emph{Iltis}. Our framework can be used to provide feedbacks (1), (2a), (2b), and (2c) as outlined in the interaction above. 

\subsubsection{Conceptual Insights} Our framework uses three conceptual steps for proving and explaining (in-)equivalence of context-free grammars: grammar canonization; a grammar-transformation framework; and algorithms for bounded context-free languages. Before discussing how these three elements combine to form a scalable framework, we sketch each of them.
\begin{itemize}
 \item The idea of \emph{grammar canonization} is an adaption of graph canonization, which is used in the context of graph isormorphism tests. Grammars are \emph{isomorphic} when they are the same up to renaming non-terminals and/or the order of productions. In that case, they are equivalent because they describe the same language. In grammar canonization, a canonical representation for each grammar is computed, which does not depend on the names of non-terminals and order of productions. Testing equivalence of isormorphic grammars then amounts to testing equality of these canonical representatives. 

\item There are also non-isomorphic grammars that are equivalent. Our \emph{grammar transformation framework} builds on the observation that grammars for the same context-free language are often based on the same idea and therefore look very similar. The idea of our grammar transformation framework is to identify such similar grammars via transformations. The framework includes a rule-based grammar transformation language as well as an interface to include hard-coded grammar transformations.  The \emph{rule-based language for grammar transformations} allows to declaratively specify pattern-based transformations of grammars. The language is designed so that it allows to specify (i) rules for simple equivalence transformations of grammars, as well as (ii) rules for identifying ``buggy'' grammars that are inequivalent to some grammar due to a typical modelling mistake. The latter can be used to identify wrong productions in grammars and to provide feedback (2b) from above.

\item Many languages encountered in introductory courses on formal languages are \emph{bounded context-free languages}. Those are a subclass of context-free languages that is structurally and algorithmically well-behaved and well-understood, e.g., it is  known that testing equivalence of an arbitrary context-free language $L_1$ and a bounded context-free language $L_2$ (both provided as a context-free grammar) is decidable \cite{ginsburgs64}. Among others, we exploit the shape of bounded languages to compute set representations for context-free languages specified as grammars or pushdown automata and provide feedback of the form (2a) from above. The literature only implicitly provides algorithmic results for bounded languages in the shape of technical and partly unconstructive proofs; here we engineer explicit algorithms.
\end{itemize}

By itself, each of these ingredients can prove and/or explain (in-)equivalence of some grammars. 

We also combine these ingredients to achieve scalability for testing large sets $\calG$ of student submitted grammars against a solution grammar. As an example, our data suggest that identifying equivalent grammars is typically more difficult than identifying inequivalent grammars. To address this we, among other techniques, use equivalence transformations from our grammar transformation framework to cluster $\calG$ into disjoint clusters $\calG_1, \calG_2, \dots$ of grammars such that each cluster only contains equivalent grammars. Then, if equivalence of the solution grammar to some representative of a cluster is established by some technique (e.g. grammar canonization or algorithms for bounded languages), then all grammars from this cluster are classified equally.

\subsubsection{Implementation} We implemented this framework for proving and explaining (in-)equivalence of context-free grammars. Besides modules that implement the conceptual insights sketched above, our framework includes many other more na\"{\i}ve tests, among them tests for emptiness and finiteness, searching for counter examples, and comparison of symbol frequencies. The grammar transformation framework allows for easy specification of additional rules and combination of rules into transformation pipelines.  The high-level structure of our framework, in particular how the different modules interact, is outlined in Figure \ref{fig:flowchart_all}.

\subsubsection{Evaluation} We evaluate our framework on datasets from  the educational support systems \emph{AutomataTutor} and \emph{Iltis} with in total 55\,667 attempts for 67 exercises. Our evaluation addresses several research questions along several dimensions. We outline the research questions and present a high-level summary of our findings. The detailed evaluation can be found in Section \ref{section:evaluation}.

One of the main goals of our framework is that correctness of student attempts can be decided for as many attempts as possible. Also, instructors should be required to manually check correctness for as few input grammars as possible and these manual checks should ideally classify as many other student attempts as possible (e.g., via the clustering approach described above). We address the effectiveness of our framework with the following research questions:

    \begin{enumerate}[leftmargin=30pt]
        \item[RQ1:] For how many student attempts can our framework decide the equivalence to the solution?
        \item[RQ2:] How many grammars have to be evaluated manually to classify all student attempts?
    \end{enumerate}

From the 45\,458 incorrect attempts, our framework can prove the inequivalence to the respective solution for all but 7 attempts. From the 10\,209 correct attempts in our data sets, our framework can automatically prove 9113 as equivalent to the respective solution.  All in all, only 260 grammars have to be manually evaluated to have all 55\,667 attempts classified. Only 7 exercises require manually evaluating more than 10 grammars.

To support students in identifying their mistakes, incorrectness of their attempts should be explained on a high level for many attempts. In our evaluation we focus on quantitative aspects, as our framework addresses inherent algorithmic questions; we leave the effectiveness of explanations provided to follow-up CS education research studies. 

    \begin{enumerate}[leftmargin=30pt]
        \item[RQ3:] For how many incorrect attempts can our framework compute high-level explanations?
    \end{enumerate}

We use different types of explanations, supplementing each other. Overall, for more than 34\,000 out of the 45\,458 incorrect attempts, we can provide at least one of the high-level explanations.

Efficiency is not the most relevant factor for our framework, in particular in the presence of advanced caching (described in Section \ref{section:framework}). However, a good understanding of the efficiency of individual methods and the effectiveness of caching across all inputs is helpful.

    \begin{enumerate}[leftmargin=30pt]
        \item[RQ4:] How many student attempts can be classified and explained by caching?
        \item[RQ5:] How efficient is testing student attempts for equivalence to the solution grammar and how efficient is computing high-level explanations?
    \end{enumerate}

While many methods for proving (in-)equivalence are very performant, some methods like testing bounded languages for equivalence or applying a pipeline of transformations may take several seconds for some attempts. However, we show that equivalent attempts on the same exercise in different data sets often match and therefore caching can be used to avoid many computations.

Throughout this paper we assume familiarity with basics of formal languages, in particular with context-free grammars (CFGs), pushdown automata (PDAs), deterministic finite state automata (DFAs) etc. (see, e.g., the textbook \cite{HopcroftU79}).

\subsection{Related Work}
We discuss how our work relates to research on theoretical foundations for educational support systems for topics in formal foundations of computer science, and to other frameworks.

\subsubsection{Theoretical Foundations for Educational Support Systems} Most educational support systems for formal foundations of computer science provide only elementary feedback to students. The reason is that already providing elementary feedback usually requires solving algorithmically hard problems,  and more meaningful feedback often requires a combination of theoretical insights and creativity. Recently, theoretical foundations have been established and implemented for several topics in formal foundations of computer science: 
\begin{itemize}
 \item For regular languages, an enriched specification language based on monadic second-order logic, \emph{MOSEL}, was designed; studied from an algorithmic perspective; integrated into the educational support system \emph{AutomataTutor}; and used in large courses and as basis for CS education research in formal language instruction \cite{AlurDGKV13,AntoniKAGV2015,AntoniHKRW2020}.
 \item For propositional logic, rule-based languages have been designed to describe  (1)  ``buggy'' equivalence transformation; and (2) typical modeling mistakes. Both formalisms have been integrated into educational support systems -- the former in \emph{LogEx} \cite{LodderH11, LodderHJ15}, the latter in \emph{Iltis} \cite{GeckLPSVZ18,SchmellenkampVZ24} -- and used in large courses and as basis for CS education research on formal foundations.
 \item For computational reductions, two specification languages have been proposed recently  for educational contexts. \emph{Karp} is a logic-inspired specification language for defining reductions between \NP problems which has been implemented and evaluated with respect to efficiency in \cite{ZhangHD22}. The framework of \emph{Cookbook reductions} provides a graphical specification language for reductions based on quantifier-free reductions. Its algorithmic properties have been explored and it has been implemented within the educational support system \emph{Iltis} \cite{KneiselRSVZ25, GrangeVVZ24}.
\end{itemize}

\subsubsection{Relation to Other Grammar Frameworks}
Madhavan et al. \cite{Madhavan15} proposed a framework with a similar goal of proving the (in-)equivalence of context-free grammars. Their framework includes a derivation tree enumerator for efficiently testing for counterexamples and a partial decision procedure for grammar equivalence which is, however, only complete for LL grammars. Our framework differs from their framework in two aspects that are, in particular, important in educational contexts. First, for bounded languages, which are widely used in assignments in introductory courses, our framework provides a decision procedure against arbitrary context-free grammars. In particular, no constraints are posed on student attempts within our framework. Second, while both frameworks employ grammar transformations, our transformation framework  is designed with the  focus to allow easy specification of both equivalence and ``buggy'' transformations. The latter enables our framework to provide high-level explanations designed for educational contexts.

Creus and Godoy \cite{CreusG14} describe and benchmark a framework that consists of four heuristics to prove inequivalence of context-free grammars. However -- unlike in our framework -- none of their methods allows for proving equivalence (for a subclass of context-free languages) and no high-level explanations for the inequivalence of grammars other than counterexamples are given.

Grammar transformations are also performed in various other contexts with different requirements on the formalism. In  \emph{grammar convergence} (and the closely related fields grammar comparison and grammar transformation), e.g. \cite{laemmelz11}, the goal is to transform grammars into a \enquote{simpler} form while preserving the semantics. While the approach is similar to ours from a high-level perspective, the focus is different in that grammar convergence is seen as a tool for software language engineers who manually decide on how to apply transformations. 
For parsing,
grammar transformations are used, for instance, to remove left-recursion. In this context, transformations are hard-coded by hand. Other grammar transformations are based on rewriting systems and equational theories \cite{mcallester92,michaelis01}.

\subsection{Structure of This Paper} We start with a conceptual perspective on  our framework and a sketch of its main ingredients in Section \ref{section:framework}. Then, in Sections \ref{section:transformations} and \ref{section:bounded-languages} we provide technical details on our abstract grammar transformation framework and the framework for handling bounded context-free languages. We then report on the evaluation of the implementation of the framework in Section \ref{section:evaluation}.

 \section{Conceptual Framework and Key Ingredients}
\label{section:framework}

The goal of our framework is to decide equivalence for a sequence of grammars with respect to a fixed target grammar and to provide proofs and explanations for the inequivalence decisions. More precisely, we aim to solve the following algorithmic problem:

\begin{itemize}\setlength{\itemindent}{-6mm}
	\item[] \textbf{Algorithmic problem:} Context-free grammar (in-)equivalence with explanations
	\begin{itemize}\setlength{\itemindent}{-4mm}
		\item[] \textbf{A priori input:} A context-free grammar $H$
		\item[]    \textbf{Input:} A sequence $G_1, G_2, \ldots$ of context-free grammars\item[] \textbf{Output:} Decide, for each $i$, whether $L(G_i)=L(H)$ and provide explanations if not
	\end{itemize}
\end{itemize}

As outlined in the introduction, the scenario that we have in mind is an educational support system where a teacher has provided a context-free grammar $H$ for a context-free target language $L(H)$ and student  attempts $G_1, G_2, \dots$ for modelling $L(H)$ come in over time. 

From a high-level perspective, our framework works as follows. Suppose $H$ is a context-free grammar given a priori. We store a set $\calG$ of grammars for which we already know the decisions and explanations. The database $\calG$ contains the already seen grammars $G_1, G_2, \dots, G_k$  as well as (possibly) other grammars which are derived from the $G_i$'s. When the next grammar $G \df G_{k+1}$ arrives, it is compared with the grammars in $\calG$. If the grammar (or an isomorphic copy) is contained in $\calG$, the grammar $G$ is classified accordingly; otherwise $G$ is analyzed in detail in order to classify it correctly and, if it is not equivalent to $H$, to prove or explain the inequivalence.

To make this approach work, we need to be able to decide the equivalence of an arriving grammar $G$ to the grammar $H$ (i.e. to decide whether $L(G) = L(H)$) and to provide explanations for the decision.  We next outline our approaches for making decisions and finding explanations. A large fraction of grammars $H$ inequivalent to $G$ can be identified by (1) searching for counterexamples, and (2) testing whether $H$ describes the empty language or a finite language. Both approaches also provide basic explanations, yet can only prove inequivalence. To prove equivalence as well as to provide advanced explanations, we use three more approaches, namely (3) canonizations, (4) grammar transformations, and (5) algorithms for bounded context-free languages. We will outline these approaches and how they are used in our framework below. The technical details of (4) and (5) will be explained later (in Sections \ref{section:transformations} and  \ref{section:bounded-languages}).

Figure \ref{fig:flowchart_all} outlines how the approaches (1) -- (5) are combined into a coherent framework for solving context-free grammar (in-)equivalence for sequences of grammars.

\definecolor{iltisBeige1}{HTML}{fef6ee}
\definecolor{iltisBeige2}{HTML}{e3d5c8}
\definecolor{iltisBeige3}{HTML}{cfbeb0}
\definecolor{iltisBeige4}{HTML}{bfac9b}

\definecolor{iltisGrey1}{HTML}{edebe8}
\definecolor{iltisGrey2}{HTML}{d9d6d2}
\definecolor{iltisGrey3}{HTML}{c2bfbc}
\definecolor{iltisGrey4}{HTML}{a6a4a1}

\definecolor{iltisLightGreen1}{HTML}{f4ffe9}
\definecolor{iltisLightGreen2}{HTML}{d4f7b2}
\definecolor{iltisLightGreen3}{HTML}{bde697}
\definecolor{iltisLightGreen4}{HTML}{a3d177}

\definecolor{iltisGreen1}{HTML}{cfffe1}
\definecolor{iltisGreen2}{HTML}{a2e8bd}
\definecolor{iltisGreen3}{HTML}{86d1a2}
\definecolor{iltisGreen4}{HTML}{6fbf8d}

\definecolor{iltisYellow1}{HTML}{fef2d0}
\definecolor{iltisYellow2}{HTML}{ffe3a2}
\definecolor{iltisYellow3}{HTML}{ffd77d}
\definecolor{iltisYellow4}{HTML}{f2c55e}

\definecolor{iltisRed1}{HTML}{ffe9e6}
\definecolor{iltisRed2}{HTML}{eda498}
\definecolor{iltisRed3}{HTML}{e08475}
\definecolor{iltisRed4}{HTML}{c26e60}

\definecolor{iltisOrange1}{HTML}{ffdfb3}
\definecolor{iltisOrange2}{HTML}{ffc97d}
\definecolor{iltisOrange3}{HTML}{ebb467}
\definecolor{iltisOrange4}{HTML}{e0a34c}

\definecolor{iltisCyan1}{HTML}{e0fffe}
\definecolor{iltisCyan2}{HTML}{b4e0df}
\definecolor{iltisCyan3}{HTML}{95c7c5}
\definecolor{iltisCyan4}{HTML}{81b3b0}

\definecolor{iltisBlue1}{HTML}{cce8ff}
\definecolor{iltisBlue2}{HTML}{8db8d9}
\definecolor{iltisBlue3}{HTML}{6f9abd}
\definecolor{iltisBlue4}{HTML}{5c86a8}
\definecolor{iltisBlue5}{HTML}{2e5b80}

\definecolor{iltisViolet1}{HTML}{ede6ff}
\definecolor{iltisViolet2}{HTML}{b1a5cc}
\definecolor{iltisViolet3}{HTML}{9b8eba}
\definecolor{iltisViolet4}{HTML}{8578a6}

\begin{figure}[t]
  \centering

  \pgfdeclarelayer{backgroundA}
  \pgfdeclarelayer{backgroundB}
  \pgfsetlayers{backgroundB,backgroundA,main}
  \newcommand{\leftBorder}{0cm}
  \newcommand{\rightBorder}{1.4\linewidth}
  \newcommand{\topBorder}{100mm}
  \newcommand{\bottomBorder}{7mm}
  \newcommand{\verticalZero}{14.5mm}
  \newcommand{\verticalThird}{35mm}
  \newcommand{\verticalMid}{52.8mm}
  \newcommand{\verticalTwoThird}{70mm}
  \newcommand{\phaseZeroSkip}{5mm}
  \newcommand{\phaseOneSkip}{17mm}
  \newcommand{\phaseTwoSkip}{15mm}
  \newcommand{\phaseThreeSkip}{10mm}
  \newcommand{\phaseFourSkip}{8mm}
  \newcommand{\phaseFiveSkip}{5mm}
  \newcommand{\groupVerticalSkip}{4mm}
  \newcommand{\groupHeadlineVerticalSkip}{1mm}
  \newcommand{\stepVerticalSkip}{1mm}
  \newcommand{\headlineVerticalSkip}{2mm}
  \scalebox{.755}{
  \begin{tikzpicture}[
    any-node/.style={align=center,font={\scriptsize}},
    any-block/.style={any-node,draw,text badly centered,rounded corners},
    input-node/.style={any-block,fill=iltisGrey1},
    output-node/.style={any-block,fill=iltisGreen1},
    heading-node/.style={any-node,font={\scriptsize\bfseries},inner xsep=0},
    step-node/.style={any-block,fill=iltisBlue1},
    group-node/.style={any-node,font={\scriptsize\bfseries}},
    collection-node/.style={any-block,fill=white},
    column 0 node/.style={minimum width=11mm},
    column 1 node/.style={minimum width=1.5cm},
    column 2 node/.style={minimum width=2.4cm},
    column 3 node/.style={minimum width=1.6cm},
    column 4 node/.style={minimum width=2.6cm},
    column 5 node/.style={minimum width=1.3cm},
    column 6 node/.style={minimum width=17mm},
    any-path/.style={draw=iltisBlue5,very thick},
    result-path/.style={any-path,-stealth',shorten >= 1pt},
    result-inter-path/.style={any-path},
    result-path-nodec/.style={result-path,dashed},
    result-inter-path-nodec/.style={result-inter-path,dashed},
    label-node/.style={any-node,text=iltisBlue5,font={\scriptsize},
        inner ysep=2pt,inner xsep=2pt,outer ysep=1.5mm,outer ysep=1.5mm},
    label-node with background/.style={label-node,fill=white,fill opacity=.5,text opacity=1,rounded corners,},
    groupbox/.style={draw,rounded corners,fill=iltisBlue2},
    phasebox/.style={inner xsep=2mm,inner ysep=1mm,fill=iltisBeige1,draw=iltisBeige3,rounded corners},
    white-path/.style={draw,line width=1.5mm,iltisBeige1},
    wide-white-path/.style={white-path,line width=3mm},
    crossing/.style={circle,fill=iltisBlue5,draw=iltisBlue5,inner sep=.5mm},
    ]
    
    \coordinate (top left) at (\leftBorder,\topBorder);
    \coordinate (top right) at (\rightBorder,\topBorder);
    \coordinate (bottom left) at (\leftBorder,\bottomBorder);
    \coordinate (zero) at (\leftBorder,\verticalZero);
    \coordinate (third) at (\leftBorder,\verticalThird);
    \coordinate (mid) at (\leftBorder,\verticalMid);
    \coordinate (twothird) at (\leftBorder,\verticalTwoThird);

\node[column 0 node,heading-node,anchor=north west] (input-head) at (top left) {};
    \node[column 1 node,heading-node,anchor=north west,xshift=\phaseZeroSkip] (precom-head) at (input-head.north east) {Caching};
    \node[column 2 node,heading-node,anchor=north west,xshift=\phaseOneSkip] (com-head) at (precom-head.north east) {\strut\\\strut};
    \node[column 3 node,heading-node,anchor=north west,xshift=\phaseTwoSkip] (collection-head) at (com-head.north east) {\strut\\\strut};
    \node[column 4 node,heading-node,anchor=north west,xshift=\phaseThreeSkip] (postcom1-head) at (collection-head.north east) {Computation of \\further explanations\\after inequivalence \\has been established};
    \node[column 4 node,heading-node,anchor=north west,xshift=\phaseThreeSkip,yshift=-4mm] (postcom2-head) at (collection-head.north east |- mid) {Deferred computations\\for filling the cache};
    \node[column 5 node,heading-node,anchor=north west,xshift=\phaseFourSkip] (cache-head) at (postcom1-head.north east) {Caching};
    \node[column 6 node,heading-node,anchor=north west,xshift=\phaseFiveSkip] (output-head) at (cache-head.north east) {};
    
    \node[heading-node,anchor=north] (com+result-head) at ($(com-head.north west)!0.5!(collection-head.north east)$) {Equivalence tests \& \\computation of explanations};

\node[input-node,column 0 node] (input) at (input-head |- mid) {input\\grammar};

\node[step-node,column 1 node] (canon) at (precom-head |- mid) {canonization\\and \\cache lookup};

\node[group-node,column 2 node,below=\headlineVerticalSkip of com-head] (generallang) {all\\target languages};
    \node[step-node,column 2 node,below=\groupHeadlineVerticalSkip of generallang] (counterexample) {counterexample test};
    \node[step-node,column 2 node,below=\stepVerticalSkip of counterexample] (empty) {emptiness test};
    \node[step-node,column 2 node,below=\stepVerticalSkip of empty] (finite) {finiteness test};
    \node[step-node,column 2 node,below=\stepVerticalSkip of finite] (symbolfreq) {symbol frequency test};
    \node[step-node,column 2 node,below=\stepVerticalSkip of symbolfreq] (normalization) {normalization\\\& cache lookup};

    \node[group-node,column 2 node,below=\groupVerticalSkip of normalization] (boundedlang) {bounded\\target languages};
    \node[step-node,column 2 node,below=\groupHeadlineVerticalSkip of boundedlang] (bounded-test) {test for boundedness};
    \node[step-node,column 2 node,below=8mm of bounded-test] (bounded-equiv-test) {equivalence test};

    \begin{pgfonlayer}{backgroundA}
      \node (generallang-group) [groupbox,fit={(generallang) (counterexample) (empty) (finite) (symbolfreq) (normalization)}] {};
      \node (boundedlang-group) [groupbox,fit={(boundedlang) (bounded-test) (bounded-equiv-test)}] {};
    \end{pgfonlayer}

\node[collection-node,column 3 node] (inequiv) at (collection-head |- twothird) {inequivalence\\proven};
    \node[collection-node,column 3 node] (equiv) at (collection-head |- third) {equivalence\\proven};
    \node[collection-node,column 3 node] (nodec) at (collection-head |- zero) {no decision};

\node[step-node,column 4 node] (bugfixing) at (postcom1-head |- twothird) {search for\\ bug-fixing\\ transformations};
    \node[step-node,column 4 node] (variation) at (postcom1-head |- third) {construct\\ equivalent variations\\ via transformations};
    \node[step-node,column 4 node] (manual) at (postcom1-head |- zero) {manual evaluation\\ by instructor};

\node[step-node,column 5 node] (output-cache) at (cache-head |- third) {add \\decision \& \\explanation \\to cache};
    \node[output-node,column 6 node] (output) at (output-head |- mid) {output:\\(in-)equivalent\\\& explanation};

\coordinate (inequiv-steps-col) at ($(counterexample.east)+(3mm,0)$);
    \path[result-inter-path] (counterexample) -- (inequiv-steps-col);
    \path[result-inter-path] (empty) -| (inequiv-steps-col);
    \path[result-inter-path] (finite) -| (inequiv-steps-col);
    \path[result-inter-path] (symbolfreq) -| (inequiv-steps-col);
    \path[result-inter-path] (normalization.10) -| (inequiv-steps-col);
    \path[result-path] (inequiv-steps-col) -| node[label-node,pos=0.2,above] {at least one test\\ is successful} (inequiv);
\path[white-path] (normalization.350 -| equiv) -| (equiv);
    \path[result-inter-path] (normalization.350) -| (normalization.350 -| equiv);
    \path[result-path] (normalization.350 -| equiv) -| (equiv);

    \coordinate (nodec-steps-col) at ($(normalization.west)+(-3mm,0)$);
    \path[result-inter-path-nodec] (counterexample) -| (nodec-steps-col);
    \path[result-inter-path-nodec] (empty) -- (nodec-steps-col |- empty);
    \path[result-inter-path-nodec] (finite) -| (nodec-steps-col |- finite);
    \path[result-inter-path-nodec] (symbolfreq) -| (nodec-steps-col |- symbolfreq);
    \path[result-inter-path-nodec] (normalization) -| (nodec-steps-col |- normalization);
    \path[result-path-nodec] (nodec-steps-col) |- node[label-node,pos=0.7,above] {no test is successful} (nodec);

\coordinate (bounded-equiv-test-col) at at ($(bounded-equiv-test.5)+(11mm,0)$);
    \path[white-path] (bounded-equiv-test-col) |- (inequiv);
    \path[result-inter-path] (bounded-test) -- node[label-node,pos=.55,above] {input \\is not\\bounded} (bounded-test -| bounded-equiv-test-col);
    \path[result-path] (bounded-test) -- node[label-node with background,pos=.5,left,xshift=-1mm] {input is\\bounded} (bounded-equiv-test);
    \path[result-inter-path] (bounded-equiv-test.5) -- (bounded-equiv-test-col);
    \path[result-path] (bounded-equiv-test-col) |- (inequiv);
    \path[result-path] (bounded-equiv-test.355) -| (equiv);

\node[crossing] (inequiv-split) at ($(inequiv.east)+(5mm,0)$) {};
    \node[crossing] (equiv-split) at (inequiv-split |- equiv) {};
    \node[crossing] (equiv-join) at (equiv-split |- mid) {};
    \node[crossing] (bugfixing-join) at (output-cache |- output) {};
    
    \path[result-path] (inequiv) -- (inequiv-split);
    \path[result-path] (inequiv-split) -- (bugfixing);
    \path[result-path] (inequiv-split) -- (equiv-join);
    \path[result-path] (equiv) -- (equiv-split);
    \path[result-path] (equiv-split) -- (variation);
    \path[result-path] (equiv-split) -- (equiv-join);

    \path[result-path] (equiv-join) -- (bugfixing-join);
    \path[result-path] (bugfixing) -| (bugfixing-join);
    \path[result-path] (bugfixing-join) -- (output);
    \path[result-path] (bugfixing-join) -- (output-cache);

    \path[result-path] (variation) -- (output-cache);
    \path[result-path] (nodec) -- (manual);
    \path[result-path] (manual) -| (output-cache);

\path[result-path] (input) -- (canon);
\node[crossing] (canon-join) at ($(canon.east)+(11mm,0)$) {};
    \path[result-path] (canon) -- node[label-node,pos=.5,above] {no\\cache\\hit} (canon-join);
    \path[result-path] (canon-join) |- ($(generallang.west)+(-1.5mm,0)$);
    \path[wide-white-path] (canon-join) |- ($(boundedlang.west)+(-1.5mm,0)$);
    \path[result-path] (canon-join) |- ($(boundedlang.west)+(-1.5mm,0)$);

    \path[result-path] (canon) -- node[label-node,left,pos=.15] {cache\\hit} ($(canon|-zero)+(0,-6mm)$) -| (output);

\begin{pgfonlayer}{backgroundB}
      \node (precom-phase) [phasebox,fit={(precom-head) (precom-head |- bottom left)}] {};
      \node (com-phase) [phasebox,fit={(com-head) (collection-head) (com-head |- bottom left) (canon-join)}] {};
      \node (postcom1-phase) [phasebox,fit={(postcom1-head) ($(postcom1-head |- mid)+(0,4mm)$)}] {};
      \node (postcom2-phase) [phasebox,fit={(postcom2-head) (postcom2-head |- bottom left)}] {};
      \node (cache-phase) [phasebox,fit={(cache-head) (cache-head |- bottom left)}] {};
    \end{pgfonlayer}

  \end{tikzpicture}
  }
  \caption{An illustration of the information flow through the single components of our framework to test equivalence of an input grammar to a target language and provide a high-level explanation in case of inequality. An arriving input grammar is first canonized to enable an efficient lookup in a cache. If no information are stored for the input grammar, several equivalence tests are applied.
  The result and the explanation derived by them is output and stored in the cache. In case inequivalence has been determined, bug-fixing transformations can be used to generate an additional high-level explanation. In case of equivalence, the chance of future cache hits for similar gramars can be increased by storing equivalent grammar variations. If no decision can be determined, the grammar is manually evaluated and the result stored in the cache.
  }
  \label{fig:flowchart_all}
  \Description{Given an input grammar, the process of determining (in-)equivalence to a target language and generating an explanation is structured in five stages: 1. Canonizing and cache lookup; 2. Testing for equivalence and computing explanations by various methods; 3. If inequivalence has been established, computations for additional explanations, namely the use of bug-fixing transformations; 4. If equivalence has been established, deferred computations for filling the cache for future computations, namely computing equivalent variations of the input grammar, and if equivalence is still unknown, using manual evaluation by the teacher; and finally 5. Storing the (in-)equivalence result an the computed explanation in the cache.}
\end{figure}

\subsection{Testing Equivalence via Canonization} 
\label{section:framework:canonization}
One way that $G$ and $H$ can differ, but still describe the same language, is that $G$ is obtained from $H$ by renaming variables and/or reordering the order of rules (i.e., the grammars are isomorphic). 
While there are 55\,662 solution attempts in our data, only 30\,302 are (pair-wise) non-isomorphic.

To handle isomorphic grammars, our framework stores, in $\calG$, a \emph{canonical representative} $\text{canon}(G)$ for each grammar $G$ seen so far. The canon of a grammar has the property that whenever $G$ and $H$ are isomorphic grammars (i.e. one is obtained by renaming variables in the other), then $\text{canon}(G) = \text{canon}(H)$. We obtain such a canon for a grammar $G$ by translating $G$ into a graph and computing a canonization of the graph using the state-of-the-art isomorphism tester \emph{bliss} \cite{JunttilaK07,JunttilaK11}.  Storing canonical representatives has the additional advantage that large sets $\calG$ can be handled efficiently. The set $\calG$ can be large (think of: ${}>1.000.000$), also because many variations of grammars are obtained via grammar transformation and added to $\calG$. When using canonizations, all canons can be inserted into a hashmap, which then allows for efficient lookups.

\subsection{Testing (In-)equivalence and Generating Explanations via Grammar Transformations}\label{sec:trafo-overview}

In sequences $G_1, G_2, \dots$ generated in the context of educational support systems, typically many grammars are based on the same idea and therefore look very similar, though are not identical. We introduce a grammar transformation framework that allows to identify similar grammars. The framework incorporates (1) a rule-based grammar transformation language, and (2) an interface to include to hard-coded grammar transformations. Both kind of transformations can be combined into pipelines that allow to transform a grammar into a set of grammars, using concatenation, iteration, and branching of transformations. Roughly speaking, we use the framework to (a) identify grammars that are equivalent to a solution grammar, and to (b) identify ``buggy'' grammars that are inequivalent to a solution grammar but ``almost'' correct.

Our grammar transformation language is based on the observation that grammars can be similar because they only differ ``locally''. For example, in grammars for describing the language $\{a^mb^nc^n \mid n, m \in \N\}$ one will often find one of $\{A\to aA\mid\eps\}$, $\{A\to Aa\mid\eps\}$, or $\{A\to AA\mid a\mid \eps\}$ as a subset of the productions (possibly with a different name for the non-terminal~$A$). Each of these sets of productions describes strings of the form $a^*$, so these sets of productions are equivalent. 

The idea is to specify such local replacements in grammars. Informally, a \emph{pattern-based transformation} $\rho :  \pi_\text{source} \rightsquigarrow \pi_\text{target}$ consists of a \emph{source pattern} $\pi_\text{source}$ and a \emph{target pattern} $\pi_\text{target}$. Both patterns are ``templates'' that can match parts of a grammar and can use variables to capture parts of a grammar. If $\pi_\text{source}$ matches parts of a grammar $G$, then applying $\rho$ to that part yields a grammar $G^\diamond$ obtained by locally ``rearranging'' the matched parts of $G$ according to $\pi_\text{target}$. For a discussion of how this approach is used, we can safely ignore the technical details, which we defer to Section~\ref{section:transformations}. 

The language allows, for example, to specify (i) typical equivalence transformations between grammars, and (ii) bug-fixing transformations that correct small ``bugs'' in grammars. For instance, the equivalence transformation that transforms $\{A\to aA\mid\eps\}$ into $\{A\to Aa\mid\eps\}$ can be specified. Also a generic transformation that ``corrects'' faulty recursion ends can be specified, e.g. for correcting $\{S \to aaSb \mid abb\}$ to  $\{S \to aSb \mid abbb\}$ for the language $L = \{a^nb^{n+2} \mid n \in \N\}$. Using an interface for including hard-coded grammar transformations, more ``global'' transformations such as removing unreachable or unproductive non-terminals can be easily included.

The whole grammar transformation framework is applied to incoming grammars $G$ to try to (a) prove equivalence to some grammar in $\calG$, i.e. to a grammar that we have already classified previously, to (b) provide explanations for the inequivalence to a previously classified grammar $H$, and to (c) sample further grammars for preprocessing. We sketch these applications briefly:
\begin{enumerate}
	\item[(a)] To prove equivalence to a previously seen grammar, we use \emph{grammar normalization}. To this end, a set $\Gamma_\text{normal}$ of transformations is used whose repeated application to a grammar $G$ results in a set $N(G)$ of normalized grammars which are equivalent to $G$. The idea is that the grammars in $N(G)$ are \enquote{simplified} and \enquote{harmonized} versions of $G$ in which clutter such as syntactically redundant productions has been removed and productions modeling common sub-languages such as the different sets of productions for modeling $a^*$ (mentioned above) have been normalized into a standard form.  Whenever a grammar is characterised by our framework, then its set of normalised grammars is permanently stored in $\calG$. Now, when a new grammar $G \df G_i$ arrives, we compute the set $N(G)$ of grammars and test whether one of them is already stored in $\calG$ (using grammar canonization). 
	\item[(b)] To explain the inequivalence of a grammar $G \df G_i$ classified as inequivalent, we try to transform $G$ into a grammar equivalent to the target grammar $H$ by applying a set of \emph{bug-fixing transformations}. If found, such a bug-fixing transformation provides an explanation.
	\item[(c)] Once a grammar $G \df G_i$ has been classified, equivalent variations of $G$ can be generated and also permanently stored in $\calG$. This kind of preprocessing works best for a well-chosen set of typical equivalence transformations, as it reduces later effort to prove equivalence.
\end{enumerate}

We emphasize that the sets of transformations used in (a) -- (c) need not be fixed in advance, so that new insights gained from data can be reflected. We have engineered a set of equivalence transformations that works well for our data from educational support systems. We also devised some prototypical bug-fixing transformations that we evaluate in Section \ref{sec:rq3}.

\subsection{Testing (In-)equivalence and Generating Explanations via Bounded Context-Free Languages}
\label{section:framework:bounded}
More than half of all languages in our data sets (see Section \ref{section:evaluation}) are bounded. As bounded context-free languages have good algorithmic properties, our framework has a dedicated component for them. 

Recall that a language $L$ is \emph{bounded} if there are words $w_1, \ldots, w_\ell$ such that $L \subseteq L(w_1^* \ldots w_\ell^*)$, where $L(w_1^* \ldots w_\ell^*)$ is the language described by the regular expression $w_1^* \ldots w_\ell^*$.  The words $w_1,\ldots,w_\ell\in\Sigma^*$ are called \emph{boundedness witnesses} for $L$. Many typical context-free languages considered in introductory formal language courses are bounded languages, e.g. $\{a^nb^m \mid n < m\}$ and $\{a^n b^m c^{n+m} \mid n, m \in \N\}$ are bounded via boundedness witnesses $a$,$b$ and $a$,$b$,$c$, respectively. Bounded context-free languages have the additional nice property, that they can be represented as $\{w_1^{x_1} \ldots w_\ell^{x_\ell} \mid \varphi(x_1, \ldots, x_\ell)\}$ where $\varphi$ is an (existential) formula of Presburger Arithmetic (i.e., a first-order formula over the structure $(\mathbb{Z},0,1,+,<)$) and $w_1, \ldots, w_\ell$ are a boundedness witness. 

In their seminal work,  Ginsburg and Spanier showed that one can algorithmically decide whether two context-free grammars $G$ and $H$ are equivalent, if one of them describes a bounded language \cite{ginsburgs64}. Unfortunately, even though Ginsburg's and Spanier's result is of algorithmic nature, the algorithms are hidden implicitly within their proofs. By close examination and algorithm engineering, we extract algorithms from their proofs as well as from a proof by Verma, Seidl, and Schwentick \cite{VermaSS05}.

The algorithmic idea for testing whether a context-free grammar $G$ for a bounded language $L(G)$ and an arbitrary context-free grammar $H$ are equivalent is to first test whether $H$ describes a bounded language as well: if it does not, then the grammars are clearly not equivalent. Otherwise, it remains to test equivalence of two bounded languages. Here, the main insight is that equivalence of two bounded languages reduces to testing equivalence of two derived Presburger formulas.

We provide, in Section \ref{section:bounded-languages}, algorithms for (a) testing whether a context-free grammar $H$ describes a bounded language $L(H)$ via witness $w_1, \ldots, w_\ell$ and, if so, (b) constructing a Presburger formula $\varphi_H$ such that $L(H) = \{w_1^{x_1} \ldots w_\ell^{x_\ell} \mid \varphi_H(x_1, \ldots, x_\ell)\}$. Combining (a) and (b) with a Presburger satisfiability algorithm (to test whether $\varphi_H$ and Presburger formula $\varphi_G$ for $G$ are equivalent) yields an algorithm for (c) testing equivalence of context-free grammars $G$ and $H$, if $L(G)$ is bounded. Detailed descriptions for these algorithms are provided in Section \ref{section:bounded-languages}.

Within our framework, these algorithms are applied to incoming grammars $G$ if the target grammar $H$ describes a bounded language. In the case that $G$ and $H$ are not equivalent, the objects computed by the algorithms can be used to derive explanations:

\begin{itemize}
 \item if algorithm (a) fails for $G$ and boundedness witness $w_1, \ldots, w_\ell$ for $H$, then a \emph{structural counterexample} $u \notin L(w_1^*\dots w^*_\ell)$ can be provided (indicating that some words in $G$ do not have the structure required for words in $L(H)$); and
 \item if algorithm (c) fails for $G$ and $H$, then a \emph{a natural description in set notation} can be provided for $G$ and $H$ (highlighting the difference between the languages $L(G)$ and $L(H)$).
\end{itemize}

For providing explanations of the second kind, we provide an algorithm for simplifying the Presburger formulas $\varphi_G$ and $\varphi_H$ (see Section  \ref{section:bounded-languages}).   

As an example for these kinds of explanations, consider the grammar $H: S \rightarrow aSb \mid abbb$ with $L(H) = \{a^nb^{n+2} \mid n \in \N\}$ and boundedness witness $w_1 = a, w_2 = b$. For a grammar $G_1: S \rightarrow aS \mid T \mid \epsilon;\;\; T \rightarrow bT \mid S \mid \epsilon$, algorithm (a) fails and an explanatory counterexample $aba$ not of the shape $a^*b^*$ can be provided. For a grammar $G_2: S \rightarrow aSb \mid abb$, algorithm (c) fails and an explanatory description $\{a^nb^{n+1} \mid n \in \N\}$ of $L(G)$ in set notation can be provided.

\section{A Pattern-Based Grammar Transformation Specification Language}
\label{section:transformations}

In this section, we outline our pattern-based specification language for grammar transformations. Recall that, informally, a \emph{pattern-based transformation} $\rho:  \pi_\text{source} \rightsquigarrow \pi_\text{target}$ consists of a \emph{source pattern} $\pi_\text{source}$ and a \emph{target pattern} $\pi_\text{target}$. Both patterns are essentially templates that can match parts of a grammar and can use variables to capture sentential forms of a grammar. If $\pi_\text{source}$ matches parts of a grammar $G$, then applying $\rho$ to that part yields a grammar $G'$ obtained by locally \enquote{rearranging} the matched parts of $G$ according to $\pi_\text{target}$.

We will first discuss examples in order to outline some key aspects of the language, before formally defining patterns and pattern-based transformations and discussing their implementation.

\begin{example}[A simple equivalence transformation] \label{example:simplifyingAStar}
	One use case of transformations is to normalize grammars in order to identify equivalent grammars. Two simple equivalence transformations that can be used to normalize productions that generate strings of the form $a^*$ are 
    \begin{align*}
        \rho_1 &:\quad X \rightarrow \sigma X\mid \epsilon \quad \rightsquigarrow \quad X \rightarrow XX\mid \sigma\mid \varepsilon \\
        \rho_2 &:\quad X \rightarrow X\sigma\mid \epsilon  \quad \rightsquigarrow \quad X \rightarrow XX\mid \sigma\mid \varepsilon
    \end{align*}
		
	Here, $X$ is a pattern variable for non-terminals and $\sigma$ is a pattern variable for a single alphabet symbol.  Such a transformation is applied to a grammar by matching the source pattern to the productions of the grammar such that pattern variables are mapped to components of appropriate type; and then rearranging the grammar according to the target pattern. As an example, applying $\rho_1$ to the grammar with productions $\{S  \rightarrow bA\mid aB\rulesep  A \rightarrow aA\mid \varepsilon\rulesep  B \rightarrow bB\mid \varepsilon\}$ can yield either of the grammars 
	$\{S  \rightarrow bA\mid aB\rulesep  A \rightarrow AA\mid a\mid \varepsilon\rulesep  B \rightarrow bB\mid \varepsilon\}$ or $\{S  \rightarrow bA\mid aB\rulesep A \rightarrow aA\mid \varepsilon\rulesep B \rightarrow BB\mid b\mid \varepsilon\}$, by matching $X$ to $A$ or $B$ and $\sigma$ to $a$ or $b$, respectively.
	
	An important aspect is that productions with the same non-terminal left-hand side have to be matched exhaustively by a pattern. For example, $\rho_1$ cannot be applied to the grammar $\{A \rightarrow aA\mid\varepsilon\mid b\}$ as the sentential form $b$ is not matched by the source pattern. This is sensible, as $\{A \rightarrow aA\mid\varepsilon\mid b\}$ is not equivalent to $\{A \rightarrow AA\mid a\mid\varepsilon\mid b\}$. \qed
\end{example}

\begin{example}[Another equivalence transformation with index variables] 
  The transformation $\rho_1$ from the last example cannot be applied to normalize $\{A \rightarrow aA \mid bA \mid \varepsilon\}$ as the source pattern of $\rho_1$ only matches non-terminals with exactly two right-hand sides. This is remedied by the following pattern-based transformation:
  \[\rho_3 :\quad X \rightarrow \sigma_i X \mid \varepsilon \quad \rightsquigarrow \quad  X \rightarrow XX \mid \sigma_i \mid \varepsilon\]
  Here, $\sigma_i$ is an indexed variable which allows that $\sigma_i X$ is matched to several right-hand sides with differing assignments for $\sigma_i$ but the same assignment for $X$. For instance, $\sigma_i X$  can match both $aA$ and $bA$ in $\{A \rightarrow aA \mid bA \mid \varepsilon\}$ such that applying $\rho_3$ yields $\{A \rightarrow AA \mid a \mid b \mid \varepsilon\}$. \qed
\end{example}

\begin{example}[Another equivalence transformation with several rules] \label{example:ElRecLevl}
    Patterns can also span over rules with different left-hand sides. For example, the transformation 
    \[\rho_4 :\quad\left\{\;
    \begin{aligned}
        \strut X &\rightarrow \sigma_i Y \tau_i \mid \alpha_i  \\
        \strut Y &\rightarrow \sigma_i Y \tau_i \mid \alpha_i \mid \beta_i
    \end{aligned}\;\right\}
    \quad \rightsquigarrow \quad 
        X \rightarrow \sigma_i X \tau_i \mid \alpha_i \mid \sigma_i\beta_j\tau_i
    \]
    describes the elimination of a redundant variable as part of a recursion. For instance, it can be used to simplify the grammar $\{S \rightarrow aAb \mid bAa\rulesep A \rightarrow aAb \mid bAa \mid a \mid b \}$ with a single application into the equivalent grammar $\{S \rightarrow aSb \mid bSa \mid aab \mid baa \mid abb \mid bba \}$. In this application of the transformation, $\alpha_i$ is matched by the empty set of sentential forms. The right-hand side $\sigma_i\beta_j\tau_i$ features two different index symbols. This expression describes all combinations of values matched for $(\sigma,\tau)$ (indexed by $i$) and values matched for $\beta$ (indexed by $j$). In our example, $(\sigma,\tau)$ matches $[(a,b),(b,a)]$ and $\beta$ matches $[a,b]$, resulting in the four right-hand sides $aab$, $baa$, $abb$, and $bba$. 

    To ensure that this transformation has the intended meaning, one has to constrain several of the variables of the source pattern. For example, $X$ and $Y$ have to be matched to different non-terminals and the non-terminal matched by $Y$ must not appear in any other production than the ones matched by the pattern. Our pattern language allows to specify such constraints.  
    
    We emphasize that the scope of index variables is local to single rules, so, for instance, the rules  $\{Y \rightarrow \sigma_i Y \tau_i \mid \alpha_i \mid \beta_i\}$ could also be written as $\{Y  \rightarrow \sigma_i Y \tau_i \mid \alpha_j \mid \beta_k\}$.
    \qed
\end{example}

\begin{example}[A bug-fixing transformation]\label{example:replaceEpsByCanonical}
 Another application of transformations is to identify bug-fixes for grammars that are almost equivalent. The following pattern-based transformation fixes buggy recursions that end in $\varepsilon$ (though they should not):
  \[\rho_5 :\quad X \rightarrow \sigma_i X \tau_i \mid  \varepsilon \mid  \gamma_i \quad \rightsquigarrow \quad  X \rightarrow \sigma_i X \tau_i \mid  \sigma_i\tau_i \mid  \gamma_i\]
 This transformation can, for instance, be used to repair the grammar $\{S \rightarrow aSb \mid  \varepsilon\}$ intended to describe the language $\{a^nb^n \mid n \geq 1\}$ into $\{S \rightarrow aSb \mid  ab\}$. It can also repair  the grammar $\{S \rightarrow aSb \mid \varepsilon \mid  C\rulesep C \rightarrow cC \mid \varepsilon\}$ intended to describe $\{a^nc^mb^n \mid n \geq 1 \text{ and } m\geq 0\}$ into $\{S \rightarrow aSb \mid  ab \mid  C\rulesep C \rightarrow cC \mid  \varepsilon\}$. \qed
\end{example}

\subsection{Formalizing Pattern-Based Grammar Transformations} Suppose $\calN$ is a set of variable names. Patterns may use \emph{(elementary) variables} $x \in \calN$ and \emph{indexed variables} of the form $x_i$ with $x \in \calN$ and $i \in \calI$ for some set $\calI$ of index names. We make the general assumption that in each pattern, a variable name can only be used either as name of an elementary variable or in indexed variables. The set of all variables, elementary or indexed, is denoted by $\calV$. 

A \emph{(grammar) pattern} $\pi=(V_\pi, R_\pi, C_\pi)$ consists of a set $V_\pi \subseteq \calV$ of \emph{(pattern) variables}, a set $R_\pi$ of \emph{(pattern) rules}, and a set $C_\pi$ of \emph{constraints}. Rules are of the form $x\to\beta$ for an elementary variable $x \in V_\pi$ and $\beta\in V_\pi^*$. Constraints in $C_\pi$ specify (global) properties of the variables of $V_\pi$:
\begin{itemize}
 \item it can be specified whether a variable $x$ can only match non-terminals (typically indicated by naming the variables $X, Y, Z$), terminals ($\sigma$, $\tau$), or arbitrary sentential forms ($\alpha$, $\beta$, $\gamma$);
 \item if a variable $x$ matches non-terminals, it can additionally be specified whether the non-terminal matched by $x$ must (not) match the start symbol of a grammar or whether it must (not) occur in any non-matched production of the grammar;
 \item for two strings of variables, it can be specified whether the sentential forms matched by them must (not) be equal or whether one must (not) be a prefix/infix/suffix of the other;
 \item for an indexed variable, it can be specified whether it has to match any production.
\end{itemize}

We now define what constitutes a \emph{match} of a pattern $\pi=(V_\pi, R_\pi, C_\pi)$ into a (context-free) grammar $G=(N_G, \Sigma_G, P_G, S_G)$ with non-terminals $N_G$, alphabet $\Sigma_G$, productions $P_G$, and start symbol $S_G\in N_G$.  The idea is that a match assigns sentential forms of $G$ to pattern variables such that each elementary variable gets assigned exactly one sentential form and indexed variables get assigned a set of sentential forms, one for each ``concrete'' index. If a pattern rule has several variables with the same index, then this shall be reflected in the match. For instance, matching the rule $X \rightarrow \sigma_i Y \tau_i$ to $S \rightarrow aAb \mid bAa$ shall result in assigning $[(a, b), (b,a)]$ to $(\sigma_i, \tau_i)$.  To formalize this, in a match, each indexed variable is instantiated by choosing concrete indices over the natural numbers. Then, each instantiation of an indexed variable is assigned  exactly one sentential form.

Formally, an \emph{instantiation} of the indexed variables in $V_\pi$ assigns a finite set $\text{inst}(x) \subset \N$ 
of concrete indices to each indexed variable $x \in V_\pi$, with the constraint that $\text{inst}(x) = \text{inst}(y)$ for two indexed variables $x$ and $y$ if (1) $x, y$ share the same name from $\calN$ or if (2) $x, y$ occur within the same right-hand side of a rule of $\pi$ with the same index from $\calI$. Denote by $V^\text{inst}_\pi$ the set of all elementary  and all instantiated indexed variables, i.e.
\[V^\text{inst}_\pi \df \{x \mid \text{$x \in V_\pi$ is elementary}\} \cup \{x_\ell \mid \text{$x_i \in V_\pi$ is an index variable and $\ell \in \text{inst}(x)$}\}\]
Let $\nu: V^\text{inst}_\pi \rightarrow (N_G \cup \Sigma_G)$ be a mapping that assigns sentential forms $N_G \cup \Sigma_G$ to each variable from $V^\text{inst}_\pi$. Such a mapping can be lifted to rules $\tau : x \rightarrow y_1 \ldots  y_k$ of pattern $\pi$ via $\nu(\tau) \df \nu(x) \rightarrow \nu(y_1)\ldots \nu(y_k)$. We say that $\mu$ \emph{matches} a pattern rule $\tau$ of $\pi$ to a production $\tau^\diamond$ of $G$ if $\nu(\tau) = \tau^\diamond$.

A \emph{match} $\mu = (\text{inst}, \nu)$ of pattern $\pi$ into the grammar $G$ consists of an instantiation $\text{inst}$ of the indexed variables of $\pi$ and a mapping $\nu: V^\text{inst}_\pi \rightarrow (N_G \cup \Sigma_G)$ such that
\begin{enumerate}
 \item[(1)] $\nu$ matches all pattern rules $\tau \in R_\pi$ to productions of $G$; and 
 \item[(2)] if some production of $G$ with left-hand side $X$ is matched by some pattern rule via $\nu$, then so are all other productions of $G$ with left-hand side $X$.
\end{enumerate}

A \emph{pattern-based transformation} $\rho:\pi \curl\pi^\diamond$ consists of a source pattern $\pi$ and a target pattern $\pi^\diamond$ over the same set $V$ of variables. A source grammar $G$ can be \emph{transformed} into a target grammar $G^\diamond$ via $\rho$, if there are matches $\mu = (\text{inst}, \nu)$ of $\pi$ into $G$ and $\mu^\diamond = (\text{inst}^\diamond, \nu^\diamond)$ of $\pi^\diamond$ into $G^\diamond$ such that 
\begin{itemize}
 \item[(1)] indexed variables occurring in both $\pi$ and $\pi^\diamond$ are instantiated in the same way in $\mu$ and $\mu^\diamond$, \item[(2)] $\nu$ and $\nu^\diamond$ assign the same values to variables occurring in both $V_\pi^\text{inst}$ and $V_{\pi^\diamond}^{\text{inst}^\diamond}$.
\end{itemize}

We write $\rho(G)$ for the set of grammars that can be obtained by transforming $G$ via $\rho$. Note that this set can contain multiple grammars, one for each match.

\subsection{Implementing Pattern-Based Transformations}
Our implementation of the grammar transformation framework allows for easy specification of additional rules and combining them into transformation pipelines. 

The application of a pattern-based transformation $\rho:\pi\curl\pi^\diamond$ on a source grammar $G$ is deferred to an SMT solver. This is sensible, because matching a pattern into a grammar is easily seen to be \NP-hard (e.g. by a reduction from the \textsc{Clique} problem, see Appendix~\ref{sec:np-hardness}). We encode the problem of matching the source pattern $\pi$ into $G$ as a propositional formula $\varphi_{\pi, G}$ such that any satisfying assignment of the variables in $\varphi_{\pi,G}$ represents a match of $\pi$ into $G$. To enumerate all substantially different matches, we gradually extend $\varphi_{\pi, G}$ by a formula describing that further matches have to be substantially different from the previous ones. We use the SMT solver \emph{Z3} \cite{Leonardo08} to solve the formulas. To generate a target grammar $G^\diamond$, we then apply the match $\mu$ on the target pattern $\pi^\diamond$ by replacing all the pattern variables by the respective sentential forms defined in $\mu$. Note that depending on the target pattern, different matches do not necessarily lead to different target grammars.

In our implementation, we extended pattern-based transformations by a variable substitution that is applied after a match has been applied to the target pattern. This helps in transformations that merge two non-terminals in which therefore all references to one of the non-terminals in remaining productions should be redirected to the other non-terminal.

 \section{An Algorithmic Framework for Bounded Context-Free Languages}
\label{section:bounded-languages}

Here, we outline an algorithm for deciding whether a context-free grammar $G$ for a bounded language and an arbitrary context-free grammar $H$ are equivalent. If they are inequivalent, the algorithm also provides an explanation. The algorithm is engineered from proofs in \cite{ginsburgs64, VermaSS05} and by extending algorithms from \cite{gawrychowskikrs10}. Its implementation is part of our framework (see Section \ref{section:evaluation} for an evaluation).

The algorithm for testing equivalence against a bounded language uses sub-algorithms for 
(a) computing a boundedness witness $w_1, \ldots, w_\ell$ for a grammar $G$ which describes a bounded language $L(G)$ (Section \ref{section:determining-witnesses}, Algorithm \ref{algo:compute-witness});
(b) testing whether a context-free grammar $H$ describes a bounded language $L(H)$ wrt. some boundedness witness $w_1, \ldots, w_\ell$ (Section \ref{section:testing-boundedness}, Algorithm \ref{algo:bounded-by-witness}); and, if so, 
(c) constructing a Presburger formula $\varphi_H$ such that $L(H) = \{w_1^{x_1} \ldots w_\ell^{x_\ell} \mid \varphi_H(x_1, \ldots, x_\ell)\}$ (Section \ref{section:constructing-presburger}, Algorithm \ref{algo:adapted-parikh-image}). These algorithms can be combined to an algorithm for deciding whether a context-free grammar $G$ for a bounded language $L(G)$ and an arbitrary context-free grammar $H$ are equivalent (Section \ref{section:bounded-equivalence-test}, Algorithm \ref{algo:bounded-equivalence}). 

At the end of this section (Section \ref{section:bounded-set-notation}), we will outline a heuristic for simplifying Presburger formulas $\varphi_H$ used in language descriptions such as $L(H) = \{w_1^{x_1} \ldots w_\ell^{x_\ell} \mid \varphi_H(x_1, \ldots, x_\ell)\}$.

\subsection{Testing Boundedness of a Language and Computing a Boundedness Witness}\label{section:determining-witnesses}

To test whether a language $L(G)$ over $\Sigma$ is bounded we rely on a polynomial time algorithm from \cite{gawrychowskikrs10} which refines the decidability result from use theoretical results from \cite{ginsburgs64}. We extend \cite{gawrychowskikrs10} by also computing boundedness witnesses (Algorithm \ref{algo:compute-witness}).  

For testing boundedness, the algorithm form \cite{gawrychowskikrs10} computes, for all non-terminals $X$, the sets 
\[\text{$\text{left}(X) \df \{u\in\Sigma^*\mid X\Rightarrow^*uXv,v\in\Sigma^*\}$ \hspace{2mm}and\hspace{2mm} $\text{right}(X) \df \{v\in\Sigma^*\mid X\Rightarrow^*uXv,v\in\Sigma^*\}$}\]
and tests whether  $\text{left}(X)\subseteq \{\ell_X\}^*$ and $\text{right}(X)\subseteq \{r_X\}^*$ for some words $\ell_X,r_X\in\Sigma^*$. If those words exist, the language is bounded. To accommodate that the words $\ell_X$ and $r_X$ can be exponentially long, they are represented by straight-line programs (SLPs) \cite{gawrychowskikrs10}.

For computing boundedness witnesses, our algorithm starts from the words $\ell_X$ and $r_X$. The main idea is that we cluster the dependency graph of nonterminals of a grammar $G$ into strongly connected components (SCC). For each SCC, we know that each cycle induces a derivation $X\Rightarrow^*uXv$ and because of the boundedness of $X$ we know that $u\in\text{left}(X)\subseteq\{\ell_X\}^*$ and $v\in\text{right}(X)\subseteq\{r_X\}^*$. Thus, we know that every word of the language $L(X)$ generated by $X$ has the structure $\ell_X^m w r_X^n$ with a word $w$ that is derived without any cycle in the nonterminals of this SSC (but possibly other SCCs). To determine a boundedness witness for $w$, we therefore only need to consider non-cyclic derivations in this SCC which we compute by unwinding productions (Algorithm \ref{algo:finite-grammar}). As boundedness witnesses may contain blocks that are used by only some words of the language, we can deal with alternative productions by concatenating them (Line \ref{line:concat-alternatives} in Algorithm \ref{algo:finite-grammar}).

Note that the size of the (smallest) boundedness witnesses can be exponential in the size of the (shortest) grammar for the respective language. For example, each boundedness witness for the finite language $\{0,1\}^{m}$ which can be represented by the grammar $A_1 \to A_2 A_2;\;\dots ;\;A_{m-1} \to A_m A_m;\; A_m \to 0 \mid 1$ has size $\bigO(2^m)$. Our algorithm circumvents this by encoding witnesses polynomially via SLPs (see variable $\mathcal{W}$ in Algorithm \ref{algo:compute-witness}).

\begin{algorithm}[t]
    \caption{Computing a boundedness witness for a bounded context-free language.}
    \label{algo:compute-witness}
    \small
    \begin{algorithmic}[1]
        \Function{ComputeBoundednessWitness}{CFG $G$ for a bounded language}
            \State $G_\text{cnf}\gets$ $G$ in Chomsky normal form
            \State $\calG_G\gets$ the dependency graph of $G_\text{cnf}$
            \Statex\Comment{$(X,Y)$ and $(X,Z)$ are edges in $\calG_G$ iff $X\to YZ$ is a rule in $G$}
            \State $\calC_G\gets$ clustering in strongly connected components (SCCs) of $\calG_G$
            \State $\mathcal{W}\gets$ an empty SLP \Comment{SLP storing the witness}
            \ForAll{nonterminals $X$ in $G_\text{cnf}$}
                \State Compute primitive strings $\ell_X$ and $r_X$ such that $\text{left}(X)\subseteq\{\ell_X\}^*$, $\text{right}(X)\subseteq\{r_X\}^*$
                \Statex \Comment{computed as in \cite{gawrychowskikrs10}; if $\text{left}(X)=\emptyset$ or $\text{right}(X)=\emptyset$, choose $\ell_X=\varepsilon$ or $r_X=\varepsilon$, resp.}
                \State $C\gets$ SCC of $X$ in $\calC_G$
                \State Add \Call{ComputeUnwoundRules}{$G_\text{cnf}$, $C$, $X$} to $\mathcal{W}$
                \State Add $X\to\$\ell_X\$X^{|C|-1}\$r_X\$$ to $\mathcal{W}$
            \EndFor
            \State $w\gets$ the (only) string generated by $S$ in $\mathcal{W}$
                \Comment{$S$ is start symbol of $G$}
            \State \Return{list of strings generated from $w$ by splitting at \$}
        \EndFunction
    \end{algorithmic}
\end{algorithm}

\begin{algorithm}[t]
    \caption{Compute unwound rules for a given strongly connected component.}
    \label{algo:finite-grammar}
    \small
    \begin{algorithmic}[1]
        \Function{ComputeUnwoundRules}{CFG $G$, SSC $C$, nonterminal $X$}
            \State $\text{Prods}\gets\emptyset$ \Comment{set of new productions}
            \ForAll{$k\in\{0,\dots,|C|-1\}$}
                \If{$k=0$}
                    \State $\text{Rhs}\gets$ all productions $X\to\alpha$ in $G$ with $\alpha$ containing no nonterminal from $C$
                \Else
                    \State $\text{Rhs}\gets$ all productions $X\to\alpha$ in $G$
                \EndIf
                \ForAll{$\alpha\in \text{Rhs}$}
                    \State In $\alpha$, replace all nonterminals $Y\in C$ by $Y^{k-1}$
                \EndFor
                \If{$\text{Rhs}=\emptyset$}
                    \State Add $X^k\to \$$ to \text{Prods}
                \Else
                    \State Add $X^k\to \$\alpha_1 \$ \dots \$ \alpha_n\$$ to \text{Prods}
                    \Comment{$\text{Rhs}=\{\alpha_1,\dots,\alpha_n\}$}\label{line:concat-alternatives}
                \EndIf
            \EndFor 
            \State \Return{\text{Prods}}
        \EndFunction
    \end{algorithmic}
\end{algorithm}

\subsection{Testing Boundedness with Respect to a Boundedness Witness}\label{section:testing-boundedness}

We can test whether a grammar $H$ is a bounded language with respect to strings $w_1, \ldots, w_\ell$ by testing $L(H)\subseteq L(w_1^*\ldots w_\ell^*)$. This inclusion can be tested by checking whether $L(H)\cap \overline{L(w_1^*\ldots w_\ell^*)}$ is empty, which is possible with elementary formal language methods for a context-free language $L(H)$ and a regular language $L(w_1^*\ldots w_\ell^*)$. More details are provided in Algorithm \ref{algo:bounded-by-witness}.

\begin{algorithm}[t]
    \caption{Testing whether a context-free language is bounded wrt. given strings.}
    \label{algo:bounded-by-witness}
    \small
    \begin{algorithmic}[1]
        \Function{TestingForBoundednessByWitness}{CFG $H$, strings $w_1,\ldots,w_\ell$}
            \State $\mathcal{A}_H\gets$ a PDA for $L(H)$ \Comment{via standard algorithm}
            \State $\mathcal{A}_{\text{witness}}\gets$ a DFA for $L(w_1^*\ldots w_\ell^*)$
            \State $\overline{\mathcal{A}}_{\text{witness}}\gets$ a DFA for the complement of $L(\mathcal{A}_{\text{witness}})$ \Comment{via standard algorithm}
            \State $\mathcal{A}_{\text{subset}}\gets$ a PDA for $\mathcal{A}_H\cap\overline{\mathcal{A}}_{\text{witness}}$ \Comment{via product automaton}
            \State Test $\mathcal{A}_{\text{subset}}$ for emptiness \Comment{via standard algorithm}
            \If{$L(\mathcal{A}_{\text{subset}})$ is empty}
                \State \Return \enquote{$L(H)$ is bounded by $w_1,\ldots,w_\ell$}
            \Else
                \State $u\gets$ a word from $L(\mathcal{A}_{\text{subset}})$
                \State \Return \enquote{$L(H)$ not bounded by $w_1,\ldots,w_\ell$, counterexample $u$}
            \EndIf
        \EndFunction
    \end{algorithmic}
\end{algorithm}

\subsection{Constructing Presburger Formulas}\label{section:constructing-presburger}

In this section we provide an algorithm for constructing a Presburger formula $\varphi_H$ such that $L(H) = \{w_1^{x_1} \ldots w_\ell^{x_\ell} \mid \varphi_H(x_1, \ldots, x_\ell)\}$, for a context-free grammar $H$ describing a bounded language. A very rough idea for the construction of the Presburger formula $\varphi_H$ is to (i) construct a context-free grammar $\widehat{H}$ obtained from $H$ essentially by treating $w_1, \ldots, w_\ell$ as symbols of an alphabet, (ii) constructing Presburger formula $\varphi_{\widehat{H}}$ for the Parikh image of $\widehat{H}$, and (iii) deriving $\varphi_H$ from $\varphi_{\widehat{H}}$. 

We first recall the notion of Parikh images. The \emph{Parikh vector} of a word $w \in \Sigma^*$ is a vector from $\N^\Sigma$ where the component $n_\sigma$ for $\sigma \in \Sigma$ counts the number of occurences of $\sigma$ in $w$. The \emph{Parikh image} of a language $L$ collects all Parikh vectors of words $w \in L$. Parikh's famous theorem states that the Parikh image of a context-free language is a semilinear set, or, equivalently, can be expressed by an (existential) formula of Presburger Arithmetic (i.e., a first-order formula over the structure $(\mathbb{Z},0,1,+,<)$) \cite{Parikh66}. Note that syntactic sugar like subtraction, multiplication by a constant and all the other usual arithmetic comparison relations can be used, but no multiplication of variables.

We first describe how a Presburger formula for the Parikh image of a context-free language specified by a CFG $G=(V,\Sigma,P,S)$ can be computed. Our construction is derived from a proof in \cite{VermaSS05} with slight modifications for efficiency. The idea is to encode components of the Parikh vector by free variables $x_\sigma$, for all $\sigma\in\Sigma$. The formula has to ensure that satisfying valuations for these variables encode  vectors in the Parikh image. To this end, the formula uses two further types of existentially quantified variables. For every production $p\in P$, there is variable $y_p$ encoding how many times the production $p$ is used in a derivation of the word encoded by the~$x_\sigma$. For every nonterminal $X\in V$, there is a variable $z_X$ describing in how many steps $X$ can be reached from the start symbol $S$. The latter is important to ensure that unreachable cycles in the productions do not contribute to the $x_\sigma$. The formula now  encodes dependencies of these variables derived from the productions of $G$. More details of this construction are provided in Algorithm~\ref{algo:bounded-parikh}.

We now describe how, from a context-free grammar $H$ for a bounded language with witness $w_1, \ldots, w_\ell$, one can construct a formula $\varphi_H$ such that $L(H) = \{w_1^{x_1} \ldots w_\ell^{x_\ell} \mid \varphi_H(x_1, \ldots, x_\ell)\}$. Essentially, we want to compute an \emph{adapted Parikh image} with respect to the witness $w_1, \dots, w_\ell$, i.e. an image counting for words how often each $w_i$ is repeated. However, the witness $w_1,\ldots,w_\ell$ are words over $\Sigma$ (as opposed to symbols) and some of the $w_i$ may be identical. The idea is to assign unique symbols $a_1,\ldots,a_\ell$ to $w_1, \ldots, w_\ell$ and compute the Parikh image of a grammar $\widehat{H}$ that adapts $H$ to work with these symbols instead of the original alphabet. See Algorithm~\ref{algo:adapted-parikh-image} for details.

\begin{algorithm}[t]
    \caption{Constructing an existentially quantified Presburger formula $\varphi_H$ for a context-free grammar $H$ for a bounded language such that $L(H) = \{w_1^{x_1} \ldots w_\ell^{x_\ell} \mid \varphi_H(x_1, \ldots, x_\ell)\}$.}
    \label{algo:adapted-parikh-image}
    \small
    \begin{algorithmic}[1]
        \Function{ComputeAdaptedParikhImage}{Bounded CFG $H$; boundedness witness $w_1,\ldots,w_\ell$; symbols $a_1, \ldots, a_\ell$}
\State $h\gets$ homomorphism mapping $w_i\mapsto a_i$
            \State $\mathcal{A}_H\gets$ a PDA for the language $L(H)$ \Comment{via standard algorithm}
            \State $\mathcal{A}_{\text{hom}}\gets$ a PDA for $h^{-1}(L(\mathcal{A}_H))$ \Comment{via standard algorithm}
            \State $\mathcal{A}_{a_i}\gets$ a DFA for the language $L(a_1^*\ldots a_\ell^*)$
            \State $\mathcal{A}_{\text{restr}}\gets$ a PDA for $\mathcal{A}_{\text{hom}}\cap\mathcal{A}_{a_i}$\Comment{via product automaton}
            \State $G_{\text{restr}}\gets$ a CFG for $L(\mathcal{A}_{\text{restr}})$
            \State $\varphi\gets$\Call{ComputeParikhImage}{$G_{\text{restr}}$}
            \State \Return{$\varphi$}
        \EndFunction
    \end{algorithmic}
\end{algorithm}

\begin{algorithm}[t]
    \caption{Constructing a Presburger formula describing the Parikh image of $L(G)$ for a CFG $G$.
}
    \label{algo:bounded-parikh}
    \small
    \begin{algorithmic}[1]
        \Function{ComputeParikhImage}{CFG $G=(V,\Sigma,P,S)$}
            \State $\varphi_S\gets 1 + \sum_{(Y\to u)\in P} \#_S(u)\cdot y_p = \sum_{(S\to v)\in P} y_p$
            \Statex\Comment{$\#_S(u)$ denotes the number of occurrences of $S$ in $u$}
            \ForAll{nonterminals $X \in V$ except $S$}
                \State $\varphi_X\gets\sum_{(Y\to u)\in P} \#_X(u)\cdot y_p = \sum_{(X\to v)\in P} y_p$
            \EndFor
            \ForAll{alphabet symbols $\sigma \in \Sigma$}
                \State $\varphi_\sigma\gets x_\sigma = \sum_{(X\to u)\in P} \#_\sigma(u)\cdot y_p$
            \EndFor
            \State $\psi_S\gets z_S = 1$
            \ForAll{nonterminals $X \in V$ except $S$}
                \State $\psi_X\gets (z_X = 0 \land \bigwedge_{p=(X\to u)\in P} (y_p = 0))$ \\ \hfill $\lor\; \bigvee_{q=(Y\to v)\in P, Y\neq X, \#_X(v)>0} (z_X = z_Y + 1 \land z_Y > 0 \land y_q > 0)$
            \EndFor
            \State\Return $\exists_{p\in P} y_p \exists_{X\in V} z_X \big(\bigwedge_{X\in V} \varphi_X \land \bigwedge_{\sigma\in \Sigma} \land \bigwedge_{X\in V} \psi_X\big)$
        \EndFunction
    \end{algorithmic}
\end{algorithm}

\subsection{Testing Equivalence against a Bounded Language}
\label{section:bounded-equivalence-test}

The equivalence test for a context-free grammar $G$ for a bounded language and a context-free grammar $H$ now proceeds by testing whether $L(H)$ is bounded with respect to the same witness as $G$, computing the Presbruger formulas for the adapted Parikh images, and testing them for equivalence. For the equivalence test of these Presburger formulas a satisfiability solver for Presburger arithmetic such as \emph{Princess}\ \cite{rummer08} can be used. 
More details are provided in Algorithm~\ref{algo:bounded-equivalence}.

\begin{algorithm}[t]
    \caption{Equivalence test for context-free grammars, one of them for a bounded language.}
    \label{algo:bounded-equivalence}
    \small
    \begin{algorithmic}[1]
        \Function{TestForEquivalence}{CFG $G$ for a bounded language with witness $\overline w = w_1,\ldots,w_\ell$; CFG $H$}
\If{$L(H)$ is not bounded by $\overline w$}
							\State\Return \enquote{not equivalent with counterexample $u$ provided by the test}
						\EndIf
            \State Choose unique symbols $\overline a = a_1,\ldots,a_\ell$
\State $\varphi_G\gets$\Call{ComputeAdaptedParikhImage}{$G$, $\overline w$, $\overline a$}
            \State $\varphi_H\gets$\Call{ComputeAdaptedParikhImage}{$H$, $\overline w$, $\overline a$}
            \State $\varphi\gets (\varphi_G\land\lnot \varphi_H) \lor (\varphi_H\land\lnot \varphi_G)$
            \State Test $\varphi$ for satisfiability \Comment{Using a Presburger solver such as Princess}
            \If{$\varphi$ is satisfiable}
                \State Construct a counterexample word $u$ from a satisfiable valuation
                \Statex\Comment{$u\in (L(G)\cap\overline{L(H)})\cup(L(H)\cap\overline{L(G)})$}
                \State\Return \enquote{not equivalent with counterexample $u$}
            \Else
                \State\Return \enquote{equivalent}
            \EndIf
        \EndFunction
    \end{algorithmic}
\end{algorithm}

\subsection{Natural Descriptions of Bounded Languages in Set Notation}
\label{section:bounded-set-notation}

The Presburger formula for the adapted Parikh image computed by Algorithm \ref{algo:adapted-parikh-image} allows to obtain a description of a bounded language given as context-free grammar. However, as this Presburger formula is an existentially quantified formula, this description is not very natural.

Within our framework, we improve readability by (a)  eliminating quantifiers, (b) transforming  the resulting formula into a comprehensible normal form, and (c) introducing syntactic sugar. 

Step (a) leads to a simplified formula only featuring the usual operators of Presburger arithmetic (i.e. addition, subtraction, multiplication by a constant, the usual arithmetic comparison relations) and additional divisibility predicates \cite{presburger1929}. In our implementation, this step is implemented using the Presburger solver \emph{Princess} \cite{rummer08}.

Step (b) transforms this quantifier-free Presburger formula into \emph{comprehensible normal form}, i.e. into a formula of the form $A_{0,1}\land\ldots\land A_{0,m_0}\land\big((A_{1,1}\land\ldots\land A_{1,m_1})\lor\ldots\lor(A_{n,1}\land\ldots\land A_{n,m_n})\big)$ where the $A_{i,j}$ are atomic formulas. This form is obtained by transforming into disjunctive normal form (DNF) and then moving atomic formulas appearing in all clauses to the front.

As final step (c), further simplifications are applied, including the combination of terms for the same variable (e.g. summarizing $i\ge 1\land i\le 2j$ to $1\le i \le 2j$), re-writing equations more naturally (e.g. writing $j-i=-1$ as $j=i-1$), and reducing the number of different exponents in the boundedness witness in set notation by using equality relationships on the exponents (e.g., $\{a^i b^j c^k \mid i,j,k\in\mathbb{N}_0 \land i > 2 \land j = 2i \land k = 1\}$ is re-written into $\{a^i b^{2i} c \mid i,j,k\in\mathbb{N}_0 \land i>2 \}$).

This procedure produces a natural representation in set notation for bounded languages. For instance (see the evaluation in Section \ref{sec:rq3} for more examples), our algorithm produces
\begin{itemize}
 \item $\{a^i b^j \mid i,j \in\mathbb{N}_0 \land j \ge 1 \land i \ge 1\}$ for the grammar $S \to AB; A \to Aa\mid a; B \to Bb\mid b$,
\item $\{a^i b^i \mid i \in \mathbb{N}_0 \land i\equiv_21 \land i \ge 1\}$ for the grammar $S \to A; A \to aBb; B \to aAb\mid\eps$,
\item \(\{a^i b^j \mid i,j \in\mathbb{N}_0 \land ((i \ge 1 \land j = 1+i) \lor (j = i-1 \land i \ge 2))\}\) for the grammar $S \to a X b \mid a Y b; X \to a X b \mid a; Y \to a Y b \mid b$.
\end{itemize}

 \section{Evaluation}
\label{section:evaluation}

In this section, we evaluate our framework with respect to effectiveness, availability of explanations, and efficiency. We address the following research questions:
    \begin{enumerate}[leftmargin=40pt]
        \item[RQ1:] For how many student attempts can our framework decide the equivalence to the solution?
        \item[RQ2:] How many grammars have to be evaluated manually to classify all student attempts?
         \item[RQ3:] For how many incorrect attempts can our framework compute high-level explanations?
        \item[RQ4:] How many student attempts can be classified and explained by caching?
          \item[RQ5:] How efficient is testing student attempts for equivalence to the solution grammar and how efficient is computing high-level explanations?
    \end{enumerate}

The evaluation is performed on authentic student input data for context-free grammar construction exercises from the educational support systems {Iltis} and {AutomataTutor}.

In Section \ref{section:data}, we discuss our datasets and the general experimental set-up. Then, in  Section~\ref{section:pipeline}, we explain the specific transformations used to normalize grammars. Afterwards, in Sections~\ref{sec:rq1}~--~\ref{sec:rq5}, we discuss the research questions RQ1~--~RQ5. 
We illustrate our discussions with selected examples from our data, see Table \ref{tab:clustering} for a summary. The evaluations on the complete dataset can be found in Appendix~\ref{app:alldata}.

\subsection{Data Collection and Preparation}\label{section:data}

Our data consists of authentic student input entered in educational support systems for context-free grammar construction exercises, i.e. exercises in which a (context-free) language description is provided in set notation or in natural language and students are asked to construct a (context-free) grammar that describes the language. Our data comes from three different sources:
\begin{itemize}
    \item[Ia:] Student inputs from the {\Iltis} system collected in an introductory course \emph{Theoretical Computer Science} in winter term 2022/23.
    \item[Ib:] Student inputs from the consecutive iteration of the course above in winter term 2023/24.
    \item[A:] Student inputs from the system {\emph{AutomataTutor}}, collected in multiple courses before 2021. \end{itemize}

As a preprocessing step, we heuristically normalized the datasets to only include representative exercises for the respective courses. To this end, we disregarded single exercises with less than 30 attempts. Then, we formed combined exercises consisting of all exercises with an identical solution grammar within the datasets Ia, Ib, and A (because dataset A comprises multiple iterations of the same courses; dataset Ib contained differently phrased assignments for the same solution grammar). For our evaluation, we only used combined exercises with at least 100 attempts.

After preprocessing, dataset Ia contains 7 different exercises with in total 17\,316 attempts (11\,\% of them equivalent to the respective solution), dataset Ib contains 10 exercises with in total 14\,597 attempts (15\,\% of them equivalent to the solution), and dataset A contains 50 exercises with in total 23\,754 attempts (26\,\% of them equivalent to the solution). In total, we therefore considered 55\,667 attempts with 18\,\% of them being equivalent to the respective solution.\footnote{Merging took only place within one dataset and only if the solution grammars were identical. This is why the datasets still share exercises for the same solution language. Since the course of Ib is an iteration of the course of Ia, 6 out of the 7 exercises of Ia also appear in Ib with the identical solution grammar. There is also an overlap of one language from A with Ia and one other language with Ib. Within dataset A, 5 solution languages appear in two different exercises each (due to different solution grammars).}

The high attempt numbers (especially of incorrect attempts) are likely because both support systems provide feedback after every attempt and students have an unlimited number of attempts per exercise.  For the evaluations in this paper, for each student attempt, we determined the equivalence to the respective solution.
Due to the large number of attempts, we used timeouts for some of the methods (15s or 30s, depending on the method). 

\subsection{Designing a Normalization Pipeline}\label{section:pipeline}
We use grammar normalization via equivalence transformations as one method for proving equivalence of grammars (also see Section~\ref{sec:trafo-overview}). Here, we describe how suitable pipelines of equivalence transformations for simplifying and harmonizing grammars were chosen.\footnote{Note that, unfortunately, standard normalization techniques such as Chomsky normal form are not applicable in our setting because they rather obfuscate grammars instead of highlight construction principles.}

To design the normalization pipeline, a subset of the authors explored dataset Ia (without knowledge of the attempts in the other two datasets) as training data to identify suitable pipelines of equivalence transformations. This normalization pipeline was then used for all evaluations where grammar normalization was used. 

Equivalence transformations used for normalization were selected with the following objectives: The normalized grammars should be small (in terms of number of non-terminals), do not have unnecessary (e.g. redundant) productions, and use the same standard productions for common sub-languages (e.g. $a^*$). All in all, normalized grammars should be easily interpretable and highlight the general construction idea of grammars.
The final normalization pipeline is provided in Appendix~\ref{app:pipeline}.

A recurring issue when designing transformations for normalization is that some transformations simplify some grammars while having the opposite effect on others. For example, removal of $\eps$-productions contributes to reducing the grammar size in some instances, but can also increase the grammar size quite a lot (in general: exponentially). For this reason, our pipelining framework supports branching into different sub-pipelines. Additionally, our pipelining framework supports the iteration of (sequences of) transformations.

\begin{table}
    \caption{A sample of assignments from our dataset used as illustrative examples in the evaluation. In the first part of the table, for selected exercises, the solution language and the number of attempts on these exercises are shown. The columns labeled with \# show the number of attempts of the respective category, the columns labeled with \enquote{\#can.} the number of distinct attempts after canonization, i.e. the number of canons. The second part of the table (column group \enquote{unrecognized}) refers to RQ2: it shows how many attempts cannot be recognized. In the last column, the number of normalization clusters these unrecognized attempts are grouped into is displayed. Data on all assignments can be found in Appendix~\ref{app:alldata:general}.}\label{tab:clustering}
    \Small
    \newcommand{\slimcol}{\hspace{-8pt}}
    \begin{tabular}{l|p{5.3cm}<{\slimcol\raggedright}|r<{\slimcol}r|r<{\slimcol}r|r<{\slimcol}r||r<{\slimcol}r<{\slimcol}r}
        \toprule ex. & solution language & \multicolumn{2}{c|}{all} & \multicolumn{2}{c|}{equivalent} & \multicolumn{2}{c||}{inequivalent} & \multicolumn{3}{c}{unrecognized} \\
        &  & \# & \;\;\#can. & \# & \#can. & \# & \#can. & \# & \;\#can. & \;\#clus.  \\\midrule
        Ia3 & $\{a^n b^m \mid n,m \ge 1 \land (n = m + 1 \lor m = n + 1)\}$ & 2937 & 1474 & 375 & 83 & 2562 & 1391 & 0 & 0 & 0  \\
        Ia6 & $\{w \in \{a,b\}^* \mid \#_a(w) = 2\#_b(w)\}$ & 1089 & 638 & 54 & 26 & 1035 & 612 & 6 & 6 & 5  \\
        Ib5 & $\{w \mid w \text{ is a correctly } \allowbreak \text{parenthesized }$ & 1208 & 488 & 308 & 94 & 900 & 394 & 33 & 23 & 19  \\
        &\multicolumn{1}{r|}{$\text{expression } \allowbreak \text{over } \allowbreak (), \allowbreak [] \allowbreak\text{ and } \allowbreak\{\}\}$}&&&&&&&&&\\
        A5 & $\{ a^n b^m c^p \mid n,m,p \ge 0 \land (n=m \lor n=p) \}$ & 1264 & 858 & 226 & 71 & 1038 & 787 & 0 & 0 & 0  \\
        A6 & $\{a^n b^m c^n \mid m,n \ge 0\}^*$ & 1060 & 627 & 409 & 168 & 651 & 459 & 4 & 4 & 2  \\
        A12 & $\{ w \sigma w^R \mid w \in \{a,b\}^* \land \sigma \in \{a,b,\epsilon\}\}$ & 505 & 270 & 158 & 31 & 347 & 239 & 7 & 6 & 5  \\\bottomrule

    \end{tabular}
\end{table}

\subsection{RQ1: Ratio of Successful Equivalence Tests}\label{sec:rq1}

\paragraph{Setting}
For answering RQ1, we use all data from all three datasets Ia, Ib, and A. For each individual student attempt, we test for (in-)equivalence of the student attempt to the solution grammar specified for the respective exercise. For proving equivalence, we use:
\begin{itemize}
    \item \textbf{Equality test:} We test for syntactical identity between attempt and solution grammar, including the naming of nonterminals. For example, the three grammars $\{S\to SS;\;S\to a;\;S\to b\}$, $\{S\to SS\mid a\mid b\}$, and $\{S\to a\mid SS\mid b\}$ are recognized as identical.\item \textbf{Isomorphy test:} We test for structural identity between student and solution grammar, but with possibly different naming of the variables.
    \item \textbf{Normalization:} We apply the normalization procedure with the pipeline described above for the attempt grammar and the solution grammar, each resulting in a set of equivalent grammars. 
If those sets overlap, the two grammars are equivalent.
\item \textbf{Equivalence test for bounded languages:} For bounded solution languages, we use the algorithm for testing whether the language of the student attempt is bounded by the same boundedness witness, and if so, we test for equivalence (see Sections \ref{section:testing-boundedness} and~\ref{section:bounded-equivalence-test}).
\end{itemize}
For proving non-equivalence, we use the following techniques:
\begin{itemize}
    \item \textbf{Emptiness and finiteness test:} We test whether the attempt grammar and the solution grammar do not agree on whether they describe an empty or finite language.
    \item \textbf{Counterexamples:} We test for all words up to length 15 whether the same words are generated by the attempt and solution grammar.
\item \textbf{Comparison of symbol frequencies:} For all attempt grammars, we compute the Parikh image of the respective language with the algorithm from Section \ref{section:constructing-presburger} and compare it to the Parikh image of the solution. If they do not match, the non-equivalence is proven.
    \item \textbf{Equivalence test for bounded languages:} As for the equivalence test above.
\end{itemize}

\paragraph{Results}

\begin{table}
    \caption{For the six sample exercises, it is shown, how many of the attempts can be proven to be (in-)equivalent by the respective methods. For each exercise, the columns labeled with \# show the number of attempts of the respective category and the columns labeled with \enquote{\#can.} the number of distinct attempts after canonization, i.e. the number of canons. Methods labeled with $^*$ only apply to exercises with bounded solution language. Data on all exercises can be found in Appendix~\ref{app:alldata:methods}.}\label{tab:methods}
    \Small
    \newcommand{\slimcol}{\hspace{-4pt}}
    \begin{tabular}{l|r<{\slimcol}r|r<{\slimcol}r|r<{\slimcol}r|r<{\slimcol}r|r<{\slimcol}r|r<{\slimcol}r}
        \toprule
        & \multicolumn{2}{c|}{Ia3} & \multicolumn{2}{c|}{Ia6} & \multicolumn{2}{c|}{Ib5} & \multicolumn{2}{c|}{A5} & \multicolumn{2}{c|}{A6} & \multicolumn{2}{c}{A12} \\
        & \# & \#can. & \# & \#can. & \# & \#can. & \# & \#can. & \# & \#can. & \# & \#can. \\\midrule
        \textbf{all attempts} & 2937 & 1474 & 1089 & 638 & 1206 & 487 & 1264 & 858 & 1060 & 627 & 505 & 270 \\\midrule
        \multicolumn{13}{c}{\textbf{Proving equivalence}}\\\midrule
        \textbf{all equivalent attempts} & 375 & 83 & 54 & 26 & 306 & 93 & 226 & 71 & 409 & 168 & 158 & 31 \\
        \textbf{equivalence proven} & 375 & 83 & 48 & 20 & 275 & 71 & 222 & 69 & 405 & 164 & 151 & 25 \\
        \textbf{\dots~by test for:} &  &  &  &  &  &  &  &  &  &  &  &  \\
        \; equality & 2 & 1 & 1 & 1 & 48 & 1 & 0 & 0 & 8 & 1 & 91 & 1 \\
        \; isomorphism & 45 & 1 & 1 & 1 & 48 & 1 & 0 & 0 & 48 & 1 & 92 & 1 \\
        \; normalization & 367 & 77 & 48 & 20 & 275 & 71 & 214 & 65 & 405 & 164 & 151 & 25 \\
        \; equivalence$^*$ & 375 & 83 & -- & -- & -- & -- & 208 & 56 & -- & -- & -- & -- \\
        \textbf{equivalence not proven} & 0 & 0 & 6 & 6 & 31 & 22 & 4 & 2 & 4 & 4 & 7 & 6 \\\midrule
        \multicolumn{13}{c}{\textbf{Proving inequivalence}}\\\midrule
        \textbf{all inequivalent attempts} & 2562 & 1391 & 1035 & 612 & 900 & 394 & 1038 & 787 & 651 & 459 & 347 & 239 \\
        \textbf{inequivalence proven} & 2562 & 1391 & 1035 & 612 & 898 & 393 & 1038 & 787 & 651 & 459 & 347 & 239 \\
        \textbf{\dots~by test for:} &  &  &  &  &  &  &  &  &  &  &  &  \\
        \; emptiness & 410 & 222 & 31 & 24 & 54 & 39 & 25 & 20 & 18 & 15 & 19 & 10 \\
        \; finiteness & 886 & 501 & 229 & 146 & 125 & 69 & 57 & 49 & 47 & 37 & 59 & 41 \\
        \; symbol frequencies & 2487 & 1348 & 622 & 381 & 252 & 162 & 987 & 746 & 278 & 231 & 338 & 231 \\
        \; boundedness witness$^*$ & 754 & 491 & -- & -- & -- & -- & 400 & 342 & -- & -- & -- & -- \\
        \; equivalence$^*$ & 1786 & 884 & -- & -- & -- & -- & 565 & 389 & -- & -- & -- & -- \\
        \; minimal counterexample & 2562 & 1391 & 1035 & 612 & 898 & 393 & 1038 & 787 & 651 & 459 & 347 & 239 \\
        \;{} \; \dots{} of size up to 4 & 2204 & 1206 & 640 & 383 & 634 & 306 & 1008 & 762 & 551 & 385 & 337 & 229 \\
        \;{} \; \dots{} of size between 5 and 8 & 344 & 174 & 341 & 199 & 264 & 87 & 25 & 22 & 100 & 74 & 10 & 10 \\
        \;{} \; \dots{} of size between 9 and 15 & 14 & 11 & 54 & 30 & 0 & 0 & 5 & 3 & 0 & 0 & 0 & 0 \\
        \textbf{inequivalence not proven} & 0 & 0 & 0 & 0 & 2 & 1 & 0 & 0 & 0 & 0 & 0 & 0 \\\bottomrule

    \end{tabular}
\end{table}

In 64 out of all 67 exercises in the dataset, our framework can automatically prove the inequivalence for all inequivalent attempts. For the remaining 3 exercises, only few attempts could not be classified (< 5 per exercise). As the data for selected exercises in Table \ref{tab:methods} suggests, most inequivalent attempts are recognized with testing counterexample words up to length 8. However, we also have an attempt (for exercise A1, a parentheses language) for which the shortest counterexample word has length 22, showing that one cannot only rely on counterexamples (of decent size) for catching all incorrect attempts.

Inequivalence for many inequivalent attempts can also be shown by emptiness tests (for 7\,\% of all incorrect attempts), finiteness tests (for 21\,\%), and symbol frequency tests (for 80\,\%). As especially the former two tests are computationally cheap and can already be used to provide non-na\"{\i}ve non-counter example feedback to students, these numbers are interesting in educational contexts.

For 48 out of 67 exercises, our framework could prove the equivalence of more than 95\,\% of all equivalent attempts. In 10 of the remaining exercises, the equivalence could be proven in more than 75\,\% of all equivalent attempts. In RQ2, we will see that by grouping grammars into equivalent clusters of grammars, the number of grammars to be manually evaluated drops significantly.

We would like to point out that in theory, the rows for the equivalence test of bounded languages should always match the number of attempts since we have a complete decision procedure if the solution language is bounded. 
Due to timeouts (30 seconds) for some combinations of solution and student grammar, however, the numbers in the respective rows are often lower than what could be expected (see e.g. for exercise A5 in Table \ref{tab:methods}).
In these cases, we see that the equivalence procedure for bounded languages and the normalization procedure supplement each other nicely.

\subsection{RQ2: Number of Attempts to Check Manually}\label{sec:rq2}

\paragraph{Setting}

To compute the number of grammars that need to be evaluated manually, we compute a grammar clustering which refines the equivalence relation: grammars $G$ and $H$ are in the same cluster if and only if our framework can compute their equivalence by methods listed in Section \ref{sec:rq1}.

We apply this clustering for all attempts and solution grammars of all three data sets. Per exercise, this results in many clusters that can have one out of three possible types: there is exactly one cluster that contains the solution grammar and all the attempts that our framework can prove to be equivalent (by any method) -- called \emph{solution cluster}; there may be clusters containing attempts whose non-equivalence to the solution has been proven by any of the methods mentioned in Section \ref{sec:rq1} -- called \emph{error clusters}; and lastly there may be clusters containing grammars whose equivalence to the solution could neither be proven nor disproven -- called \enquote{unrecognized clusters}.

Since all grammars in the same cluster are equivalent, only one representative per cluster has to be manually evaluated, making the number of unrecognized clusters the key indicator to answer~RQ2.

\paragraph{Results}

Over all datasets, in 31 out of 67 exercises, there is no unrecognized cluster at all; and for 54 exercises, there are less than 5 unrecognized clusters. For our example exercises, the number of unrecognized clusters is shown in the second part of Table \ref{tab:clustering}.
Considering that per unrecognized cluster only one student attempt has to be manually evaluated, for all datasets, on average, less than 5\,\% of all equivalent attempts have to be manually evaluated (3.5\,\% in A, 4.0\,\% in Ia, 2.0\,\% in Ib) and less than 0.1\,\% of all inequivalent attempts. Note that these numbers are considerably lower than the numbers of attempts which could not be classified by our framework (see Table \ref{tab:methods}).

\subsection{RQ3: Explaining Inequivalent Attempts}\label{sec:rq3}
\begin{table}
    \caption{For the six sample exercises, the frequency that high-level feedback can be generated is shown. Methods labeled with $^*$ only apply to exercises with bounded solution language. Data on all exercises can be found in Appendix~\ref{app:alldata:correction-methods}.}\label{tab:correction-methods}
    \Small
    \newcommand{\slimcol}{\hspace{-8pt}}
    \begin{tabular}{l|r<{\slimcol}r|r<{\slimcol}r|r<{\slimcol}r|r<{\slimcol}r|r<{\slimcol}r|r<{\slimcol}r}
        \toprule
        & \multicolumn{2}{c|}{Ia3} & \multicolumn{2}{c|}{Ia6} & \multicolumn{2}{c|}{Ib5} & \multicolumn{2}{c|}{A5} & \multicolumn{2}{c|}{A6} & \multicolumn{2}{c}{A12} \\
        & \# & \#can. & \# & \#can. & \# & \#can. & \# & \#can. & \# & \#can. & \# & \#can. \\\midrule
        \textbf{all attempts} & 2937 & 1474 & 1089 & 638 & 1206 & 487 & 1264 & 858 & 1060 & 627 & 505 & 270 \\
        \textbf{all inequivalent attempts} & 2562 & 1391 & 1035 & 612 & 900 & 394 & 1038 & 787 & 651 & 459 & 347 & 239 \\\midrule
        \multicolumn{13}{c}{\textbf{Explaining inequivalence due to different Parikh images}}\\\midrule
        \textbf{Parikh image different} & 2487 & 1348 & 622 & 381 & 252 & 162 & 987 & 746 & 278 & 231 & 338 & 231 \\
        \; representation concise & 2056 & 1150 & 979 & 569 & 756 & 311 & 553 & 378 & 619 & 428 & 285 & 185 \\
        \textbf{\& bounded by solution witness$^*$} & 1807 & 899 & -- & -- & -- & -- & 636 & 443 & -- & -- & -- & -- \\
        \; set notation computable$^*$ & 1786 & 884 & -- & -- & -- & -- & 563 & 387 & -- & -- & -- & -- \\
        \; set notation concise$^*$ & 1031 & 574 & -- & -- & -- & -- & 230 & 143 & -- & -- & -- & -- \\\midrule
        \multicolumn{13}{c}{\textbf{Explaining inequivalence with bug-fixing transformations}}\\\midrule
        \textbf{error corrected} & 59 & 14 & 0 & 0 & 0 & 0 & 7 & 6 & 0 & 0 & 0 & 0 \\
        \; by C1: add $\eps$ as recursion end & 10 & 7 & 0 & 0 & 0 & 0 & 7 & 6 & 0 & 0 & 0 & 0 \\
        \; by C2: add canonical recursion end & 15 & 5 & 0 & 0 & 0 & 0 & 0 & 0 & 0 & 0 & 0 & 0 \\
        \; by C3: replace $\eps$ by canonical rec. end & 36 & 4 & 0 & 0 & 0 & 0 & 0 & 0 & 0 & 0 & 0 & 0 \\\bottomrule
    \end{tabular}
\end{table}
\paragraph{Setting}

In case of inequivalence, these methods are used to generate high-level explanations:
\begin{itemize}
    \item \textbf{Generating set notation:} If the solution language is bounded and the attempt language is bounded by the same boundedness witness, we use the procedure described in Section \ref{section:bounded-set-notation} to compute a representation of the attempt language in set notation. However, this is only a useful explanation in our educational context if the set notation generated is concise. We call a set notation concise 
    if its right-hand side is in comprehensive normal form, does not contain more than 5 atomic conditions, and satisfies other structural limitations\footnote{A set notation is concise if its right-hand side is a formula $A_{0,1}\land\ldots\land A_{0,m_0}\land\big((A_{1,1}\land\ldots\land A_{1,m_1})\lor\ldots\lor(A_{n,1}\land\ldots\land A_{n,m_n})\big)$ in comprehensive normal form which (a) has at most three clauses in the disjunction (i.e., $n\le 3$) and (b) the clause in front of the disjunction has at most 5 atomic formulas $A_{0,i}$ (i.e., $m_0\le 5$), (c) all clauses in the disjunction have at most 3 atomic formulas $A_{i,j}$ (i.e., $m_i\le 3$ for all $i\ge1$), and (d) in total there are no more than 5 atomic formulas $A_{i,j}$.
    }
    \item \textbf{Representation of differences in symbol frequencies:} When computing the Parikh image of the attempt language and the solution language as explained in Section \ref{section:constructing-presburger}, we get (existential) Presburger formulas over the frequencies of the single alphabet symbols as variables. From here, we can apply a procedure very similar to what is done for generating set notations for bounded languages. This results in a description of the differences of the Parikh sets of the attempt and the solution language as formula in comprehensive normal form. We apply the same criteria for such a formula being concise.
    \item \textbf{Bug-fixing transformations:} We apply proto-typical bug-fixing transformations to the student attempts and then test with our normalization procedure whether the resulting grammar is equivalent to the solution grammar. If the modified grammar is indeed equivalent to the solution, the modification was indeed a fix and can explain what was wrong in the original attempt. Note that these bug-fixing transformations are only used after an attempt has been classified as inequivalent to the solution, since it cannot be guaranteed that the transformations actually change the language of the grammar they are applied to. Definitions of the three proto-typical bug-fixing transformations used in this evaluation can be found in Appendix~\ref{app:correcting-trafos}.
\end{itemize}

\paragraph{Results}

The incorrectness of most attempts -- as already pointed out in the discussion on RQ~1 -- can be explained by a difference between the Parikh sets of the solution and the attempt language, see Table \ref{tab:correction-methods}. In 94\,\% of all attempts that differ from the respective solution in the Parikh image, the representation of this difference is concise and thus potentially helpful as explanation.

If both the solution and the attempt language are bounded, we compute a set notation of the attempt language. Over all exercises, in 73\,\% of the applicable attempts this computation succeeded within 15 seconds, and in 67\,\% of those attempts the set notation was concise.

Over all exercises, the three proto-typical bug-fixing transformations were able to explain 416 of 45458 incorrect attempts. Transformation C1 ($\eps$ is added in recursions) explains 233 attempts, Transformation C2 (a canonical recursion end that is derived from the recursion is added) explains 84 attempts, and Transformation C3 (a generalization of Example \ref{example:replaceEpsByCanonical}) explains 133 attempts. This is promising as it suggests that by adding further transformations one can provide high-level explanations for a significant fraction of wrong grammars.

\subsection{RQ4: Usage of Caching}\label{sec:rq4}
\paragraph{Setting}
To evaluate how effective caching is for classifying and explaining student attempts, we evaluate the predictiveness of previous attempts for future attempts. 

For this evaluation, we focus on the 6 exercises that are shared between dataset Ia and Ib. In particular, we proceed as follows: first, we apply the normalization procedure to all attempts of the respective exercises of dataset Ia. We cache the canonization of the original attempt as well as canonizations of all intermediate grammars and the resulting grammars from the normalization procedure. Then, we test for all attempts of the respective exercises in dataset Ib whether they appear in this cache. For each attempt of Ib, we test containment in the cache with three preprocessing steps: (a) we only canonize the attempt, (b) we apply a basic simplification, i.e. the very first steps of the normalization pipeline featuring e.g. the removal of non-useful non-terminals, before the canonization, and (c) we apply the full normalization procedure (including canonization) for the attempt of Ib and check if any resulting grammar is contained in the cache.

\paragraph{Results}
\begin{table}
    \caption{For the six shared exercises between Ia and Ib, it is shown how many of the attempts of Ib have been cached when applying the normalization procedure for the attempts of Ia before. These results are shown for three preprocessing options for the data of Ib (canonization, basic simplification, normalization).}\label{tab:caching}
    \Small
    \newcommand{\slimcol}{\hspace{-5pt}}
    \begin{tabular}{l|r<{\slimcol}r|r<{\slimcol}r|r<{\slimcol}r|r<{\slimcol}r|r<{\slimcol}r|r<{\slimcol}r}
        \toprule
        & \multicolumn{2}{c|}{Ib1} & \multicolumn{2}{c|}{Ib3} & \multicolumn{2}{c|}{Ib4} & \multicolumn{2}{c|}{Ib5} & \multicolumn{2}{c|}{Ib6} & \multicolumn{2}{c}{Ib7} \\
        & \# & \#can. & \# & \#can. & \# & \#can. & \# & \#can. & \# & \#can. & \# & \#can. \\\midrule

        \textbf{all attempts of Ib} & 965 & 499 & 835 & 268 & 616 & 250 & 1206 & 487 & 4312 & 1947 & 938 & 270 \\\midrule
        \multicolumn{13}{c}{\textbf{Equivalent attempts}}\\\midrule

        \textbf{equivalent attempts of Ib} & 55 & 25 & 218 & 45 & 121 & 14 & 306 & 93 & 551 & 73 & 228 & 26 \\
        \textbf{attempts of Ib known from Ia} &  &  &  &  &  &  &  &  &  &  &  &  \\
        \; after canonization & 31 & 12 & 187 & 27 & 35 & 3 & 216 & 30 & 359 & 14 & 216 & 15 \\
        \; after basic simplification & 40 & 16 & 202 & 32 & 35 & 3 & 224 & 36 & 378 & 22 & 220 & 19 \\
        \; after normalization & 51 & 24 & 217 & 44 & 115 & 13 & 276 & 73 & 466 & 49 & 228 & 26 \\\midrule
        \multicolumn{13}{c}{\textbf{Inequivalent attempts}}\\\midrule

        \textbf{inequivalent attempts of Ib} & 910 & 474 & 617 & 223 & 495 & 236 & 900 & 394 & 3761 & 1874 & 710 & 244 \\
        \textbf{attempts of Ib known from Ia} &  &  &  &  &  &  &  &  &  &  &  &  \\
        \; after canonization & 90 & 46 & 133 & 58 & 2 & 2 & 270 & 72 & 851 & 150 & 415 & 69 \\
        \; after basic simplification & 207 & 116 & 202 & 99 & 41 & 25 & 376 & 136 & 1340 & 425 & 558 & 132 \\
        \; after normalization & 339 & 176 & 227 & 120 & 54 & 34 & 429 & 175 & 1628 & 575 & 597 & 162 \\\bottomrule

    \end{tabular}
\end{table}

For all of the shared exercises, more than 84\,\% of all equivalent attempts of dataset Ib are contained in the Ia cache after applying the normalization procedure as preprocessing on the Ib data; for four of the exercises this percentage even is $\ge95\,\%$. When only performing canonization as preprocessing on the Ib data, the percentage of cache hits still stays at $\ge50\,\%$ of all equivalent attempts for all but one shared exercise.

For inequivalent attempts, caching is considerably less effective as there are more attempts in dataset Ib that have not been seen before when applying normalization for Ia.
The detailed results for all shared exercises are listed in Table \ref{tab:caching}.

\subsection{RQ5: Efficiency}\label{sec:rq5}
\paragraph{Setting}
We present a preliminary evaluation of the performance of our system. For each method, used to answer the previous research questions, we logged the runtime. In this section, we present these times aggregated over the single attempts per exercise. 

We emphasize that our implementation has not been optimized for performance so far, also because we expect that many of the computations will later be done in offline mode due to caching. Also, the framework has been designed to be integrated into an actual educational support system written in Java, so we prioritized maintainability in most cases. 

To make the runtime evaluation of the normalization procedure less dependent on (potentially) unnecessary branching options in the normalization pipeline which have not been optimized for now, we do not give the overall runtime of the pipeline but two other measurements: (a) Per student attempt, we select the fastest branch that computes a grammar that is minimal wrt. the number of non-terminals over all the produced result grammars. Then, per exercise we take the average over all attempts. (b) We also present the runtime for the slowest branch in the pipeline (resulting in an arbitrary grammar), averaged over all attempts of an exercise. The execution of transformation pipelines is highly optimized, e.g. transformations being applied on several branches will be only computed once.

\paragraph{Results}

\begin{table}
    \caption{Runtimes for the six sample exercises for the individual methods. Columns labeled with \enquote{avg} denote the mean of the computation times in milliseconds over all attempts per exercise; columns labeled with \enquote{\#ab.} denotes the number of computations aborted because of a time-out. Methods labeled with $^*$ only apply to exercises with bounded solution language. Data on all exercises can be found in Appendix~\ref{app:alldata:runtime}.}\label{tab:runtime}
    \Small
    \newcommand{\slimcol}{\hspace{-3pt}}
    \begin{tabular}{l|r<{\slimcol}r|r<{\slimcol}r|r<{\slimcol}r|r<{\slimcol}r|r<{\slimcol}r|r<{\slimcol}r}
        \toprule
        & \multicolumn{2}{c|}{Ia3} & \multicolumn{2}{c|}{Ia6} & \multicolumn{2}{c|}{Ib5} & \multicolumn{2}{c|}{A5} & \multicolumn{2}{c|}{A6} & \multicolumn{2}{c}{A12} \\
        & avg & \#ab. & avg & \#ab. & avg & \#ab. & avg & \#ab. & avg & \#ab. & avg & \#ab. \\\midrule
        \textbf{(in-)equivalence tests} &  &  &  &  &  &  &  &  &  &  &  &  \\
        \; isomorphism & 143 & 0 & 158 & 0 & 136 & 0 & 139 & 0 & 154 & 0 & 153 & 0 \\
        \; emptiness & 1 & 0 & 1 & 0 & 33 & 0 & 1 & 0 & 1 & 0 & 1 & 0 \\
        \; finiteness & 33 & 0 & 19 & 0 & 39 & 0 & 23 & 0 & 27 & 0 & 21 & 0 \\
        \; {counterexamples} & 64 & 0 & 92 & 0 & 16744 & 0 & 69 & 0 & 1336 & 0 & 64 & 0 \\
        \; symbol frequencies & 324 & 1 & 106 & 0 & 385 & 14 & 84 & 2 & 43 & 4 & 76 & 0 \\
        \; boundedness witness$^*$ & 531 & 1 & -- & -- & -- & -- & 369 & 2 & -- & -- & -- & -- \\
        \; equivalence$^*$ & 1646 & 507 & -- & -- & -- & -- & 6015 & 413 & -- & -- & -- & -- \\
        \textbf{normalization} &  &  &  &  &  &  &  &  &  &  &  &  \\
        \; fastest useful branch & 369 &  & 712 &  & 3390 &  & 512 &  & 520 &  & 298 &  \\
\; slowest branch & 1633 &  & 18411 &  & 22333 &  & 2834 &  & 2608 &  & 1133 &  \\\bottomrule
    \end{tabular}
\end{table}

The tests for emptiness and finiteness are very fast, they need an average time of less than 150\,ms for all exercises. Canonizing grammars (and therefore testing for isomorphy), for all exercises, costs on average no more than 400\,ms.

The runtime of testing for counterexamples of size up to 15 depends on the specific exercise. For 51 out of the 67 exercises, this takes < 100\,ms on average; for further 11 exercises, it takes less than one second on average.
For all exercises, if there is a counterexample of size up to 15, the counterexample is found on average in less than 150\,ms.

It is also expensive to test whether the symbol frequencies (i.e. the Parikh image) match the ones of the solution language (for all exercises < 1 second on average), to test whether an attempt language is bounded for the witness of the solution language (for all exercises < 5 seconds on average), and -- if the attempt is bounded by the same witness -- to test for equivalence to the solution language (for all exercises < 14 seconds on average). For all these operations, we also registered attempts for which we aborted the computations due to a timeout (15 seconds). We account for them in in Table \ref{tab:runtime}.

The runtime for the normalization procedure strongly depends on the specific exercise, and most specifically on the size of the solution and the expected attempts. For 50 out of 67 exercises, the shortest path leading to a smallest grammar (as explained above) has, on average over all attempts, a computation time of less than 500\,ms; 9 exercises between 500\,ms and 1 second; and the remaining 8 exercises spread up to 5 seconds. The slowest branches are much slower, leading up to 28 seconds.
The average number of branches in the normalization procedure differs from 19 to 1277 per exercise.
The runtimes for the individual methods on our sample exercises can be found in Table \ref{tab:runtime}.

\subsection{Discussion}

We sum up our results, outline some directions for improvements, and discuss some limitations.

For most exercises, our framework performs very well. It can automatically decide equivalence for almost all attempts (RQ1) and by clustering grammars using normalization only few representative grammars have to be manually checked (RQ2). Even though normalization and some of the algorithmic methods have a considerable runtime (RQ5), caching can be used to significantly reduce the computation cost (RQ4). In addition to deciding the equivalence, our framework presents high-level explanations for a considerable fraction of all incorrect attempts (RQ3).

\paragraph{Performance}
For methods which compute and compare Parikh images, several timeouts occurred in the evaluation. This in not surprising, since these methods have high computational complexity. For example, the equivalence test for bounded context-free languages is \textsc{NExpTime}-hard \cite{HuntRS76}. Yet, we expect that these methods can be tweaked for a significantly better performance.

\paragraph{Transformations}
For coming up with suitable equivalence transformations for the normalization pipeline, we only used attempts from the dataset Ia. That the normalization not only works well for Ib (whose exercises overlap to a great extent with Ia), but also for the exercises in dataset A, shows that the rules generalize well. It seems very plausible, that further suitable equivalence transformations can be identified easily. Also using the same equivalence transformations but applying them more often, reduces the number of unclassified normalization clusters in a later experiment. This  highlights the potential of the normalization procedure.

We also tested some bug-fixing transformations which modify incorrect grammar in an attempt to fix typical student mistakes. The choice of transformations and therefore also the respective results should only be seen as a proof of concept, as such bug-fixing transformations have to be validated as part of follow-up CS education research. Other bug-fixing transformations identified in education research could easily be specified and added in our general framework 

\paragraph{Limitations}
Our choice of using authentic student data was motivated to test our framework on real data. We cannot (and do not aim to) make statements about the difficulty of single exercises or typical problems for students. For any responsible statements on this, we would need to look much closer into how the single exercises were phrased and in which setting students worked on them. This is also why we refrain from discussing any assignment-specific indicators.

However, teachers of concrete courses can use the framework introduced in this paper to analyze their students' attempts. Given knowledge about the learners' setting, especially the high-level explanations on symbol frequencies and on fixing typical mistakes by correction transformations, can lead to great insights in the students' understanding.

 \section{Conclusion and Perspectives}
\label{section:conclusion}
We presented a framework that allows for testing context-free grammars for equivalence and for providing high-level explanations for their inequivalence. This framework combines standard methods with more intricate components such as an abstract grammar transformation language, theory-based algorithms for bounded context-free languages, and a graph-theory-inspired grammar canonization. The implementation of the framework, overall, performs well on authentic data.

For us, it is surprising that (in-)equivalence of context-free grammars can be detected and even explained for a large fraction of our data. Even though already testing equivalence is undecidable for general and highly untractable for bounded context-free languages. We hope that our methods can also be applied for testing equivalence of syntax specifications of programming languages.

Our motivation for this work was to support students when learning how to model with context-free representations within educational support systems. Within this setting, our framework is capable of automatically deciding the equivalence for a large fraction of attempts. Our normalization procedure reduces the manual workload for instructors as -- if at all -- only very few attempts have to be manually classified. For a large fraction of inequivalent attempts, our framework can provide more high-level feedback. Compared with more traditional settings, where instructors provide feedback manually for all assignments, this is a significant improvement.

Our work also lays the foundation for addressing follow-up CS education research questions. Formal languages, including context-free languages, have mainly been ignored in CS education research for higher education so far. In particular, misconceptions as well as difficulty-generating factors for modeling with context-free formalisms have not been systematically mapped. Our transformation framework, in particular its ability to easily specify bug-fixing transformations, is an important ingredient for clustering student mistakes according to common types of mistakes. 
Such a clustering is a first step in the process of systematically identifying misconceptions (followed up by interview studies etc.). We plan to use our framework in this context, where it will significantly reduce the manual workload. Another direction is the exploration of the effectiveness of interventions for student errors, for instance, the effectiveness of high-level explanations as provided by our framework.

\section{Data-Availability Statement}

The methods described in this paper have been implemented in scope of the \emph{Iltis} project (\url{https://iltis.rub.de}). The implementation can be found on GitHub (\url{https://github.com/The-Iltis-Project}), see in particular the repository \href{https://github.com/The-Iltis-Project/CFG-Equivalence-Testsuite}{CFG Equivalence Testsuite}. For reproducing our results or running our methods on different data, we provide an artifact \cite{artifact}.
The Datasets Ia and Ib have been published at \cite{attemptData}.

\begin{acks}
    We are grateful to Fabian Vehlken and Nils Vortmeier for insightful discussions and feedback. We also wish to thank the team behind \emph{AutomataTutor} -- in particular Maximilian Weininger -- for generously providing us with their dataset of student attempts.
    This work was supported by the \grantsponsor{DFG}{Deutsche Forschungsgemeinschaft (DFG, German Research Foundation)}{}, grant \grantnum{DFG}{448468041}.
\end{acks}

\printbibliography

\newpage
\appendix

\section{Full Statistics for the Evaluations}\label{app:alldata}

We provide full data for all evaluations where only sample data was provided in Section \ref{section:evaluation}.

\subsection{General Data and RQ2: Number of Attempts to Check Manually}\label{app:alldata:general}

We list all assignments from our dataset used in the evaluation. In the first part of the table, the solution language and the number of attempts on each exercise are shown. The columns labeled with \# show the number of attempts of the respective category, the columns labeled with \enquote{\#can.} the number of distinct attempts after canonization, i.e. the number of canons. The second part of the table (column group \enquote{unrecognized}) refers to RQ2: it shows how many attempts cannot be recognized. In the last column, the number of normalization clusters these unrecognized attempts are grouped into is displayed.
This is an extended version of Table \ref{tab:clustering}.

{
    \Small
    \renewcommand{\arraystretch}{1.5}
    \newcommand{\slimcol}{\hspace{-8pt}}
    \begin{longtable}{l|p{5.3cm}<{\slimcol\raggedright}|r<{\slimcol}r|r<{\slimcol}r|r<{\slimcol}r||r<{\slimcol}r<{\slimcol}r}
            \toprule ex. & solution language & \multicolumn{2}{c|}{all} & \multicolumn{2}{c|}{equivalent} & \multicolumn{2}{c||}{inequivalent} & \multicolumn{3}{c}{unrecognized} \\[-1.5ex]
            & solution grammar & \# & \;\;\#can. & \# & \#can. & \# & \#can. & \# & \;\#can. & \;\#clus.  \\\midrule
        \endfirsthead
            \multicolumn{11}{@{}l}{(continued from last page)}\\[1ex]
            \toprule ex. & solution language & \multicolumn{2}{c|}{all} & \multicolumn{2}{c|}{equivalent} & \multicolumn{2}{c||}{inequivalent} & \multicolumn{3}{c}{unrecognized} \\[-1.5ex]
            & solution grammar & \# & \;\;\#can. & \# & \#can. & \# & \#can. & \# & \;\#can. & \;\#clus.  \\\midrule
        \endhead
            \bottomrule\multicolumn{11}{r@{}}{\raisebox{-1ex}{(to be continued on next page)}}
        \endfoot
            \bottomrule
        \endlastfoot

        Ia1 & $\{a^i b^j a^k b^l \mid i,j,k,l \ge 0 \land i + j = k + l\}$ \newline $S \to  aSb\mid L\mid R\rulesep L \to  bLb\mid B\rulesep R \to  aRa\mid B\rulesep B \to  bBa\mid \varepsilon $ & 5968 & 3607 & 400 & 70 & 5568 & 3537 & 8 & 6 & 4  \\
        Ia2 & $\{ a^n b^{n+m} a^m \mid n,m \ge 0 \}$ \newline $S \to  XY\rulesep X \to  \varepsilon \mid aXb\rulesep Y \to  \varepsilon \mid bYa$ & 3126 & 1685 & 409 & 77 & 2717 & 1608 & 0 & 0 & 0  \\
        Ia3 & $\{a^n b^m \mid n,m \ge 1 \land (n = m + 1 \lor m = n + 1)\}$ \newline $S \to  aTb\rulesep T \to  aTb\mid a\mid b$ & 2937 & 1474 & 375 & 83 & 2562 & 1391 & 0 & 0 & 0  \\
        Ia4 & $\{a^i b^j c^k \mid i,j,k \ge 0 \land (i \neq j \lor j \neq k)\}$ \newline $S \to  AX\mid YC\rulesep X \to  bB\mid cC\mid bXc\rulesep Y \to  aA\mid bB\mid aYb\rulesep A \to  \varepsilon \mid aA\rulesep B \to  \varepsilon \mid bB\rulesep C \to  \varepsilon \mid cC$ & 1848 & 1215 & 104 & 54 & 1744 & 1161 & 71 & 37 & 14  \\
        Ia5 & $\{w \mid w \text{ is a correctly } \allowbreak \text{parenthesized } \allowbreak \text{expression } \allowbreak \text{over } \allowbreak (), \allowbreak [] \allowbreak \text{ and } \allowbreak \{\}\}$ \newline $S \to  \varepsilon \mid SS\mid (S)\mid [S]\mid \{S\}$ & 1605 & 843 & 344 & 97 & 1261 & 746 & 33 & 19 & 15  \\
        Ia6 & $\{w \in \{a,b\}^* \mid \#_a(w) = 2\#_b(w)\}$ \newline $S \to  \varepsilon  \mid  SS \mid  a S a S b \mid  b S a S a \mid  a S b S a$ & 1089 & 638 & 54 & 26 & 1035 & 612 & 6 & 6 & 5  \\
        Ia7 & $\{a^n b^n \mid n \ge 1 \}$ \newline $S \to  ab\mid aSb$ & 743 & 262 & 202 & 28 & 541 & 234 & 0 & 0 & 0  \\
        Ib1 & $\{a^i b^j a^k b^l \mid i,j,k,l \ge 0 \land i + j = k + l\}$ \newline $S \to  aSb\mid L\mid R\rulesep L \to  bLb\mid B\rulesep R \to  aRa\mid B\rulesep B \to  bBa\mid \varepsilon $ & 965 & 499 & 55 & 25 & 910 & 474 & 7 & 4 & 2  \\
        Ib3 & $\{a^n b^m \mid n,m \ge 1 \land (n = m + 1 \lor m = n + 1)\}$ \newline $S \to  aTb\rulesep T \to  aTb\mid a\mid b$ & 835 & 268 & 218 & 45 & 617 & 223 & 0 & 0 & 0  \\
        Ib4 & $\{a^i b^j c^k \mid i,j,k \ge 0 \land (i \neq j \lor j \neq k)\}$ \newline $S \to  AX\mid YC\rulesep X \to  bB\mid cC\mid bXc\rulesep Y \to  aA\mid bB\mid aYb\rulesep A \to  \varepsilon \mid aA\rulesep B \to  \varepsilon \mid bB\rulesep C \to  \varepsilon \mid cC$ & 616 & 250 & 121 & 14 & 495 & 236 & 71 & 9 & 4  \\
        Ib5 & $\{w \mid w \text{ is a correctly } \allowbreak \text{parenthesized } \allowbreak \text{expression } \allowbreak \text{over } \allowbreak (), \allowbreak [] \allowbreak\text{ and } \allowbreak\{\}\}$ \newline $S \to  \varepsilon \mid SS\mid (S)\mid [S]\mid \{S\}$ & 1208 & 488 & 308 & 94 & 900 & 394 & 33 & 23 & 19  \\
        Ib6 & $\{w \in \{a,b\}^* \mid \#_a(w) = 2\#_b(w)\}$ \newline $S \to  \varepsilon  \mid  SS \mid  a S a S b \mid  b S a S a \mid  a S b S a$ & 4312 & 1947 & 551 & 73 & 3761 & 1874 & 87 & 28 & 16  \\
        Ib7 & $\{a^n b^n \mid n \ge 1 \}$ \newline $S \to  ab\mid aSb$ & 938 & 270 & 228 & 26 & 710 & 244 & 0 & 0 & 0  \\
        Ib8 & $\{a^n b^{n+2} c^{m}\mid n,m \in \mathbb{N}_0 \}$ \newline $S\to AbbB\rulesep A\to aAb\mid \varepsilon \rulesep B\to cB\mid \varepsilon $ & 2224 & 1177 & 263 & 71 & 1961 & 1106 & 0 & 0 & 0  \\
        Ib9 & $\{ a^n b^m \mid 0\le n < m\}$ \newline $S\to AB\rulesep A\to aAb\mid \varepsilon \rulesep B\to bB\mid b$ & 1507 & 743 & 246 & 103 & 1261 & 640 & 7 & 2 & 2  \\
        Ib10 & $\{a^n b^{m} c^{n+2}\mid n,m \in \mathbb{N}_0 \}$ \newline $S\to Acc\rulesep A\to aAc\mid B\rulesep B\to bB\mid \varepsilon $ & 1263 & 630 & 101 & 44 & 1162 & 586 & 0 & 0 & 0  \\
        Ib11 & $\{a^{2n} b^{m}\mid n,m\in \mathbb{N}_0, 2n < m \}$ \newline $S\to AB\rulesep A\to aaAbb\mid \varepsilon \rulesep B\to bB\mid b$ & 729 & 393 & 67 & 30 & 662 & 363 & 2 & 2 & 2  \\
        A1 & $\{w \mid w \text{ is a correctly } \allowbreak \text{parenthesized } \allowbreak \text{expression } \allowbreak \text{over } \allowbreak \{\}, \allowbreak [] \allowbreak \text{ and } \allowbreak () \allowbreak \text{ (nesting }\allowbreak \text{is }\allowbreak \text{order-sensitive, }\allowbreak \text{levels }\allowbreak \text{may }\allowbreak \text{be }\allowbreak \text{skipped)} \}$ \newline $S \to  SS \mid  A\rulesep A \to  B \mid  C \mid  D \mid  \varepsilon \rulesep B \to  \{C\}\rulesep C \to  [D] \mid  CC \mid  DC \mid  CD\rulesep D \to  ()D \mid  ()$ & 2927 & 1992 & 281 & 147 & 2646 & 1845 & 205 & 130 & 86  \\
        A2 & $\{ a^m b^n \mid m,n \ge 0, m \neq n \}$ \newline $S \to  a S b \mid  A \mid  B\rulesep A \to  aA \mid  a\rulesep\! B \to  bB \mid  b$ & 1884 & 1149 & 221 & 48 & 1663 & 1101 & 1 & 1 & 1  \\
        A3 & $\{ a^m b^n c^{2m+n} \mid m,n \ge 1\}$ \newline $S \to  a S c c\rulesep\! S \to  a A c c\rulesep A \to  b A c\rulesep A \to  b c$ & 1657 & 782 & 178 & 23 & 1479 & 759 & 0 & 0 & 0  \\
        A4 & $\{ a^m b^n \mid m,n \ge 0, m \neq n \}$ \newline $S \to  aSb \mid  A \mid  B\rulesep A \to  aA \mid  a\rulesep\! B \to  bB \mid  b$ & 1527 & 1094 & 267 & 99 & 1260 & 995 & 2 & 2 & 2  \\
        A5 & $\{ a^n b^m c^p \mid n,m,p \ge 0 \land (n=m \lor n=p) \}$ \newline $S \to  A \mid  DC\rulesep A \to  aAc \mid  B\rulesep B \to  Bb \mid \varepsilon \rulesep C \to  Cc \mid  \varepsilon \rulesep D \to  aDb \mid  \varepsilon $ & 1264 & 858 & 226 & 71 & 1038 & 787 & 0 & 0 & 0  \\
        A6 & $\{a^n b^m c^n \mid m,n \ge 0\}^*$ \newline $S \to  XS \mid  \varepsilon \rulesep X \to  aXc \mid  B\rulesep B \to  bB \mid  \varepsilon $ & 1060 & 627 & 409 & 168 & 651 & 459 & 4 & 4 & 2  \\
        A7 & $\{ w \in \{a,b\}^* \mid \#_a(w) = 0 \bmod 2  \;\land\; \allowbreak \#_b(w) = 1 \bmod 2 \}$ \newline $S \to  a T \mid  b U\rulesep T \to  a S \mid  b V\rulesep U \to  a V \mid  b S \mid  \varepsilon \rulesep V \to  a U \mid  b T$ & 944 & 717 & 55 & 20 & 889 & 697 & 27 & 14 & 11  \\
        A8 & $\{w c^n \mid n \ge 0, w\in\{a,b\}^*, \allowbreak (\#_a(w) = n \lor \allowbreak \#_b(w) = n)\}$ \newline $S \to  A \mid  B\rulesep A \to  D a D A c \mid  D\rulesep\! D \to  b D \mid  \varepsilon \rulesep B \to  C b C B c \mid  C\rulesep C \to  a C \mid  \varepsilon $ & 896 & 671 & 93 & 28 & 803 & 643 & 78 & 23 & 6  \\
        A9 & $\{w \mid w \text{ is a correctly } \allowbreak \text{parenthesized } \allowbreak \text{expression } \allowbreak \text{over } \allowbreak () \allowbreak \text{ and } \allowbreak []\} $ \newline $S \to  SS \mid  (S) \mid [S] \mid  \varepsilon $ & 648 & 301 & 226 & 46 & 422 & 255 & 11 & 11 & 9  \\
        A10 & $\{ a^n b^{n+m} a^m \mid n,m \ge 0 \}$ \newline $S \to  XY\rulesep X \to  aXb \mid  \varepsilon \rulesep Y \to  bYa \mid  \varepsilon $ & 697 & 289 & 410 & 57 & 287 & 232 & 0 & 0 & 0  \\
        A11 & $\{ u a v b \mid u,v \in \{a,b\}^*, |u| = |v| \}$ \newline $S \to  T b\rulesep T \to  a T a\rulesep T \to  a T b\rulesep T \to  b T a\rulesep T \to  b T b\rulesep T \to  a$ & 565 & 266 & 210 & 14 & 355 & 252 & 0 & 0 & 0  \\
        A12 & $\{ w \sigma w^R \mid w \in \{a,b\}^* \land \sigma \in \{a,b,\epsilon\}\}$ \newline $S \to  a S a\rulesep S \to  b S b\rulesep S \to  \varepsilon \rulesep S \to  a\rulesep S \to  b$ & 505 & 270 & 158 & 31 & 347 & 239 & 7 & 6 & 5  \\
        A13 & $\{ w \sigma_1 w^R v \sigma_2 v^R \mid w,v \in\{a,b\}^* \land \allowbreak\sigma_1 ,\sigma_2 \in \{a,b,\epsilon\}\}$ \newline $S \to  TT\rulesep T \to  a T a \mid  b T b \mid  a \mid  b \mid  \varepsilon $ & 440 & 210 & 138 & 16 & 302 & 194 & 1 & 1 & 1  \\
        A14 & $\{ a^n c^m b^n \mid n,m \ge 0\}$ \newline $S \to  a S b \mid  T\rulesep T \to  cT \mid  \varepsilon $ & 422 & 249 & 133 & 25 & 289 & 224 & 0 & 0 & 0  \\
        A15 & $\{ b^n a^m \mid n,m \ge 0 \land m > n\}$ \newline $S \to  b S a \mid  Sa \mid a$ & 427 & 263 & 128 & 33 & 299 & 230 & 2 & 1 & 1  \\
        A16 & $\{ a^n b^m \mid n,m \ge 1 \land n \neq m \bmod 2 \}$ \newline $S \to  aaAb \mid  aAbb\rulesep A \to  aaA \mid  Abb \mid  \varepsilon $ & 414 & 285 & 75 & 34 & 339 & 251 & 2 & 1 & 1  \\
        A17 & $\{ a^n b^n \mid n \ge 0\}^*$ \newline $S \to  T S\mid \varepsilon \rulesep T \to  aTb\mid \varepsilon $ & 365 & 166 & 68 & 13 & 297 & 153 & 0 & 0 & 0  \\
        A18 & $\{ a^n b^m c^p \mid n,m,p \ge 0 \land (n=m \lor n=p) \}$ \newline $S \to  T U \mid  V\rulesep T \to  a T b \mid  \varepsilon \rulesep U \to  c U \mid  \varepsilon \rulesep V \to  a V c \mid  W\rulesep W \to  b W \mid  \varepsilon $ & 393 & 249 & 161 & 42 & 232 & 207 & 0 & 0 & 0  \\
        A19 & $\{a^{2n} b^{2n} \mid n\ge 0\}$ \newline $S \to  aa S bb \mid  \varepsilon $ & 360 & 146 & 133 & 8 & 227 & 138 & 0 & 0 & 0  \\
        A20 & $\{ w \mid w \in \{a,b\}^* \land \#_a(w) = \#_b(w) \}$ \newline $S \to  \varepsilon  \mid  a B \mid  b A\rulesep A \to  a S \mid  b A A\rulesep B \to  b S \mid  a B B$ & 376 & 158 & 173 & 33 & 203 & 125 & 171 & 32 & 10  \\
        A21 & $\{ a^n b^m c^p \mid n,m,p \ge 0 \land (n=m \lor m=p)\}$ \newline $S \to  X \mid  Y\rulesep X \to  T C\rulesep T \to  a T b \mid  \varepsilon \rulesep C \to  c C \mid  \varepsilon \rulesep Y \to  A U\rulesep U \to  b U c \mid  \varepsilon \rulesep A \to  a A \mid  \varepsilon $ & 355 & 262 & 58 & 25 & 297 & 237 & 1 & 1 & 1  \\
        A22 & $\{a^n b^m c^m d^{2n}\mid n \ge 0 , m\ge 1\}$ \newline $S \to  a S d d \mid  A\rulesep A \to  b A c \mid  b c$ & 371 & 168 & 194 & 34 & 177 & 134 & 0 & 0 & 0  \\
        A23 & $\{  a^n b^m c^p \mid m,n,p \ge 0 \land m\neq n\}$ \newline $S \to  TC\rulesep T \to  A \mid  B\rulesep A \to  aAb \mid  aA \mid  a\rulesep B \to  aBb \mid  Bb \mid  b\rulesep C \to  cC \mid  \varepsilon $ & 358 & 244 & 81 & 23 & 277 & 221 & 1 & 1 & 1  \\
        A24 & $\{ a^n c b^n \mid n\ge 0 \}$ \newline $S \to  a S b \mid  c$ & 284 & 166 & 85 & 12 & 199 & 154 & 0 & 0 & 0  \\
        A25 & $\{ a^n b^n \mid n \ge 0\}$ \newline $S \to  a S b \mid  \varepsilon $ & 253 & 83 & 159 & 14 & 94 & 69 & 0 & 0 & 0  \\
        A26 & $\{ a^n b^m a^m b^n \mid n,m \ge 0\}$ \newline $S \to  a S b \mid  I\rulesep I \to  b I a \mid  \varepsilon $ & 260 & 122 & 90 & 13 & 170 & 109 & 0 & 0 & 0  \\
        A27 & $\{ (da)^n aa b^{2m}\mid n,m \ge 0\}$ \newline $S \to  A a a B\rulesep A \to  d a \mid  d a A \mid  \varepsilon \rulesep B \to  b b \mid  b b B \mid  \varepsilon $ & 268 & 175 & 129 & 54 & 139 & 121 & 2 & 2 & 2  \\
        A28 & $\{a w b v c \mid w \in \{a,b\}^* \land v\in \{a,c\}^* \}$ \newline $S \to  A B C D E\rulesep A \to  a\rulesep B \to  aB\mid bB\mid \varepsilon \rulesep C \to  b\rulesep D \to  aD\mid cD\mid \varepsilon \rulesep E \to  c$ & 257 & 99 & 133 & 14 & 124 & 85 & 33 & 7 & 3  \\
        A29 & $\{ w \mid w \in \{a,b\}^* \text{ and at all times } \allowbreak\text{there }\allowbreak\text{are }\allowbreak\text{less }\allowbreak \text{than }\allowbreak \text{twice as many \(b\)s as \(a\)s} \}$ \newline $S \to  a A S \mid  \varepsilon \rulesep A \to  a A A \mid  b \mid  \varepsilon \rulesep $ & 248 & 157 & 94 & 24 & 154 & 133 & 94 & 24 & 7  \\
        A30 & $\{a^n b^m \mid n,m \ge 0 \land n>m\}$ \newline $S \to  aS \mid  aA\rulesep A \to  aAb \mid  \varepsilon $ & 233 & 133 & 85 & 27 & 148 & 106 & 0 & 0 & 0  \\
        A31 & $\{ u a v b \mid u,v \in \{a,b\}^*, |u| = |v| \}$ \newline $S \to  Ab\rulesep A \to  aAa \mid  aAb \mid  bAa \mid  bAb \mid  a$ & 227 & 128 & 82 & 12 & 145 & 116 & 0 & 0 & 0  \\
        A32 & $\{ w\in \{a,b\}^* \mid \#_a(w) = 0 \mod 2 \}$ \newline $S \to  aSaS \mid  bS \mid  \varepsilon $ & 230 & 153 & 92 & 37 & 138 & 116 & 58 & 22 & 13  \\
        A33 & $\{a^n b w a^n \mid n \ge 0 \land w\in \{a,b\}^*\}$ \newline $S \to  a S a \mid  b X\rulesep X \to  a X \mid  b X \mid  \varepsilon $ & 215 & 132 & 66 & 15 & 149 & 117 & 2 & 1 & 1  \\
        A34 & $\{ a^n \mid n\ge 0\}$ \newline $S \to  a S\mid \varepsilon $ & 214 & 37 & 157 & 15 & 57 & 22 & 0 & 0 & 0  \\
        A35 & $\{a^n b^m \mid n,m \ge 0 \land m\le n\le 2m\}$ \newline $S \to  aSb \mid  aaSb \mid  \varepsilon $ & 197 & 117 & 62 & 12 & 135 & 105 & 0 & 0 & 0  \\
        A36 & $\{a^n b^m \mid n,m \ge 0 \land (m=n \lor m=2n)\}$ \newline $S \to  aAb \mid  aBbb \mid  \varepsilon \rulesep A \to  aAb \mid  \varepsilon \rulesep B \to  aBbb \mid  \varepsilon $ & 201 & 101 & 70 & 11 & 131 & 90 & 0 & 0 & 0  \\
        A37 & $\{w \mid w \text{ is a correctly } \allowbreak \text{parenthesized } \allowbreak \text{expression } \allowbreak \text{over } () \}$ \newline $S \to  (S) \mid  SS \mid  \varepsilon $ & 185 & 73 & 75 & 13 & 110 & 60 & 5 & 4 & 2  \\
        A38 & $\{ a^n b^m \mid 0\le n < m\}$ \newline $S \to  a S b\rulesep S \to  S b\rulesep S \to  b$ & 174 & 83 & 75 & 12 & 99 & 71 & 0 & 0 & 0  \\
        A39 & $\{ w c w^R \mid w\in \{a,b\}^*\}$ \newline $S \to  aSa \mid  bSb \mid  c$ & 141 & 49 & 84 & 7 & 57 & 42 & 0 & 0 & 0  \\
        A40 & $\{w \mid w \text{ is a correctly } \allowbreak \text{parenthesized } \allowbreak \text{expression } \allowbreak \text{over } \allowbreak \{\}, \allowbreak [] \allowbreak \text{ and }\allowbreak () \allowbreak \text{ (nesting }\allowbreak \text{is }\allowbreak \text{order-sensitive, }\allowbreak \text{levels }\allowbreak \text{may }\allowbreak \text{be }\allowbreak \text{skipped }\allowbreak\text{and }\allowbreak\text{repeated)} \}$ \newline $S \to  SS \mid  \{S\} \mid  T\rulesep T \to  TT \mid  [T] \mid  U\rulesep U \to  UU \mid  (U) \mid  \varepsilon $ & 142 & 81 & 74 & 20 & 68 & 61 & 20 & 6 & 3  \\
        A41 & $\{a^n b^m \mid 0\le n\le m\}$ \newline $S \to  a S b \mid  S b \mid  \varepsilon $ & 139 & 103 & 53 & 30 & 86 & 73 & 1 & 1 & 1  \\
        A42 & $\{w v \mid w,v \in \{a,b,c\}^* \land |w| = |v| \land\allowbreak \text{ at each position } w \text{ and } v^R \text{ differ}\}$ \newline $S \to  a S b \mid  a S c \mid  b S a\mid b S c\mid c S a\mid c S b\mid \varepsilon $ & 131 & 42 & 82 & 7 & 49 & 35 & 0 & 0 & 0  \\
        A43 & $\{a^n b^m a^{2m+2n} \mid n,m \ge 0\}$ \newline $S \to  aSaa \mid  B\rulesep B \to  bBaa \mid  \varepsilon $ & 136 & 66 & 72 & 12 & 64 & 54 & 2 & 1 & 1  \\
        A44 & $\{ \sigma w \sigma \mid w \in\{a,b\}^* \land \sigma \in \{a,b\}\}\cup\{a,b\}$ \newline $S \to  a W a \mid  b W b \mid  a \mid  b\rulesep\!\! W \to  b W \mid  a W \mid  \varepsilon $ & 122 & 58 & 22 & 5 & 100 & 53 & 2 & 2 & 1  \\
        A45 & $\{s^n w^m \mid 1\le n,m \le 3\}^* \{s^n \mid 1\le n \le 3\}$ \newline $A \to  S \mid  PA\rulesep P \to  SW\rulesep S \to  s\mid ss\mid sss\rulesep W \to  w\mid ww\mid www$ & 125 & 71 & 37 & 9 & 88 & 62 & 26 & 8 & 3  \\
        A46 & $\{ a^n b^n a^m b^m \mid n,m \ge 0\}$ \newline $S \to  AA\rulesep A \to  aAb \mid  \varepsilon $ & 120 & 45 & 73 & 12 & 47 & 33 & 0 & 0 & 0  \\
        A47 & $\{ a^n b^{2m} a^m \mid n,m \ge 0 \}$ \newline $S \to  A B\rulesep A \to  a A \mid  \varepsilon \rulesep B \to  b b B a \mid  \varepsilon $ & 105 & 51 & 67 & 15 & 38 & 36 & 0 & 0 & 0  \\
        A48 & $\{ w\in \{a,b\}^* \mid \#_a(w) = 1 \bmod 2 \}$ \newline $S \to  A E\rulesep B \to  b B \mid  \varepsilon \rulesep A \to  B a B\rulesep E \to  A A E \mid  \varepsilon $ & 103 & 85 & 6 & 5 & 97 & 80 & 6 & 5 & 3  \\
        A49 & $\{ w \sigma w^R \mid w \in \{a,b\}^* \land \sigma \in \{a,b,\epsilon\}\}$ \newline $S \to  a S a \mid  b S b \mid  a \mid  b \mid  \varepsilon $ & 95 & 36 & 42 & 3 & 53 & 33 & 0 & 0 & 0  \\
        A50 & $\{ a^n b^m c^p \mid n,m,p \ge 0 \land (n=m \lor m=p)\}$ \newline $S \to  X C \mid  A Y\rulesep C \to  c C \mid  \varepsilon \rulesep A \to  a A \mid  \varepsilon \rulesep X \to  a X b \mid  \varepsilon \rulesep Y \to  b Y c \mid  \varepsilon $ & 94 & 55 & 21 & 5 & 73 & 50 & 0 & 0 & 0  \\
    \end{longtable}
}

\subsection{RQ1: Ratio of Successful Equivalence Tests}\label{app:alldata:methods}

On the next pages, we show for all exercises how many of their attempts can be proven to be (in-)equivalent by the respective methods. For each exercise, the columns labeled with \# show the number of attempts of the respective category and the columns labeled with \enquote{\#can.} the number of distinct attempts after canonization, i.e. the number of canons. Methods labeled with $^*$ only apply to exercises with bounded solution language.
This is an extended version of Table~\ref{tab:methods}.

Note that for some exercises the number of unrecognized equivalent and inequivalent attempts not necessarily add up to the total number of unrecognized attempts displayed in Appendix~\ref{app:alldata:general}. This is due to detail in our methodology: For the statistics in this section, we consider each attempt individually. In particular, the normalization procedure only checks whether there is an overlap between the normalizations of the solution grammar with the normalizations of each individual attempt. In Appendix~\ref{app:alldata:general}, however, we consier the transitive closure of the normalization procedure over all attempts for the respective exercise. This leads to some attempts being assigned to a solution or error cluster that on their own had been unrecognized.

\begin{landscape}
\begin{raggedright}
\parbox[l][5cm][t]{10cm}{
\scalebox{.95}{
\Small
\newcommand{\slimcol}{\hspace{-3pt}}
\begin{tabular}{l*{10}{|r<{\slimcol}r}}
    \toprule
    & \multicolumn{2}{c|}{Ia1} & \multicolumn{2}{c|}{Ia2} & \multicolumn{2}{c|}{Ia3} & \multicolumn{2}{c|}{Ia4} & \multicolumn{2}{c|}{Ia5} & \multicolumn{2}{c|}{Ia6} & \multicolumn{2}{c|}{Ia7} & \multicolumn{2}{c|}{Ib1} & \multicolumn{2}{c|}{Ib3} & \multicolumn{2}{c}{Ib4} \\
    & \# & \#can. & \# & \#can. & \# & \#can. & \# & \#can. & \# & \#can. & \# & \#can. & \# & \#can. & \# & \#can. & \# & \#can. & \# & \#can. \\\midrule
    \textbf{all attempts} & 5968 & 3607 & 3126 & 1685 & 2937 & 1474 & 1848 & 1215 & 1605 & 843 & 1089 & 638 & 743 & 262 & 965 & 499 & 835 & 268 & 616 & 250 \\\midrule
    \multicolumn{21}{c}{\textbf{Proving equivalence}}\\\midrule
    \textbf{all equivalent attempts} & 400 & 70 & 409 & 77 & 375 & 83 & 104 & 54 & 344 & 97 & 54 & 26 & 202 & 28 & 55 & 25 & 218 & 45 & 121 & 14 \\
    \textbf{equivalence proven} & 392 & 64 & 409 & 77 & 375 & 83 & 33 & 17 & 311 & 78 & 48 & 20 & 202 & 28 & 48 & 21 & 218 & 45 & 50 & 5 \\
    \textbf{\dots~by test for:} &  &  &  &  &  &  &  &  &  &  &  &  &  &  &  &  &  &  &  &  \\
    \; equality & 0 & 0 & 4 & 1 & 2 & 1 & 0 & 0 & 81 & 1 & 1 & 1 & 56 & 1 & 0 & 0 & 0 & 0 & 0 & 0 \\
\end{tabular}
}}
\end{raggedright}
\end{landscape}
\newpage
\begin{landscape}
\begin{raggedright}
\parbox[l][5cm][t]{10cm}{
\scalebox{.95}{
\Small
\newcommand{\slimcol}{\hspace{-3pt}}
\begin{tabular}{l*{10}{|r<{\slimcol}r}}
    \multicolumn{21}{@{}l}{(continued from last page)}\\[1ex]
    \toprule
    & \multicolumn{2}{c|}{Ib5} & \multicolumn{2}{c|}{Ib6} & \multicolumn{2}{c|}{Ib7} & \multicolumn{2}{c|}{Ib8} & \multicolumn{2}{c|}{Ib9} & \multicolumn{2}{c|}{Ib10} & \multicolumn{2}{c|}{Ib11} & \multicolumn{2}{c|}{A1} & \multicolumn{2}{c|}{A2} & \multicolumn{2}{c}{A3} \\
    & \# & \#can. & \# & \#can. & \# & \#can. & \# & \#can. & \# & \#can. & \# & \#can. & \# & \#can. & \# & \#can. & \# & \#can. & \# & \#can. \\\midrule
    \textbf{all attempts} & 1206 & 487 & 4312 & 1947 & 938 & 270 & 2224 & 1177 & 1507 & 743 & 1263 & 630 & 729 & 393 & 2925 & 1990 & 1884 & 1149 & 1657 & 782 \\\midrule
    \multicolumn{21}{c}{\textbf{Proving equivalence}}\\\midrule
    \textbf{all equivalent attempts} & 306 & 93 & 551 & 73 & 228 & 26 & 263 & 71 & 246 & 103 & 101 & 44 & 67 & 30 & 279 & 145 & 221 & 48 & 178 & 23 \\
    \textbf{equivalence proven} & 275 & 71 & 464 & 45 & 228 & 26 & 263 & 71 & 238 & 100 & 101 & 44 & 65 & 28 & 76 & 17 & 219 & 46 & 178 & 23 \\
    \textbf{\dots~by test for:} &  &  &  &  &  &  &  &  &  &  &  &  &  &  &  &  &  &  &  &  \\
\end{tabular}
}}
\end{raggedright}
\end{landscape}
\newpage
\begin{landscape}
\begin{raggedright}
\parbox[l][5cm][t]{10cm}{
\scalebox{.95}{
\Small
\newcommand{\slimcol}{\hspace{-3pt}}
\begin{tabular}{l*{10}{|r<{\slimcol}r}}
    \multicolumn{21}{@{}l}{(continued from last page)}\\[1ex]
    \toprule
    & \multicolumn{2}{c|}{A4} & \multicolumn{2}{c|}{A5} & \multicolumn{2}{c|}{A6} & \multicolumn{2}{c|}{A7} & \multicolumn{2}{c|}{A8} & \multicolumn{2}{c|}{A9} & \multicolumn{2}{c|}{A10} & \multicolumn{2}{c|}{A11} & \multicolumn{2}{c|}{A12} & \multicolumn{2}{c}{A13} \\
    & \# & \#can. & \# & \#can. & \# & \#can. & \# & \#can. & \# & \#can. & \# & \#can. & \# & \#can. & \# & \#can. & \# & \#can. & \# & \#can. \\\midrule
    \textbf{all attempts} & 1527 & 1094 & 1264 & 858 & 1060 & 627 & 944 & 717 & 896 & 671 & 648 & 301 & 697 & 289 & 565 & 266 & 505 & 270 & 440 & 210 \\\midrule
    \multicolumn{21}{c}{\textbf{Proving equivalence}}\\\midrule
    \textbf{all equivalent attempts} & 267 & 99 & 226 & 71 & 409 & 168 & 55 & 20 & 93 & 28 & 227 & 47 & 410 & 57 & 210 & 14 & 158 & 31 & 138 & 16 \\
    \textbf{equivalence proven} & 263 & 95 & 222 & 69 & 405 & 164 & 28 & 6 & 15 & 5 & 216 & 36 & 410 & 57 & 210 & 14 & 151 & 25 & 137 & 15 \\
    \textbf{\dots~by test for:} &  &  &  &  &  &  &  &  &  &  &  &  &  &  &  &  &  &  &  &  \\
\end{tabular}
}}
\end{raggedright}
\end{landscape}
\newpage
\begin{landscape}
\begin{raggedright}
\parbox[l][5cm][t]{10cm}{
\scalebox{.95}{
\Small
\newcommand{\slimcol}{\hspace{-3pt}}
\begin{tabular}{l*{10}{|r<{\slimcol}r}}
    \multicolumn{21}{@{}l}{(continued from last page)}\\[1ex]
    \toprule
    & \multicolumn{2}{c|}{A14} & \multicolumn{2}{c|}{A15} & \multicolumn{2}{c|}{A16} & \multicolumn{2}{c|}{A17} & \multicolumn{2}{c|}{A18} & \multicolumn{2}{c|}{A19} & \multicolumn{2}{c|}{A20} & \multicolumn{2}{c|}{A21} & \multicolumn{2}{c|}{A22} & \multicolumn{2}{c}{A23} \\
    & \# & \#can. & \# & \#can. & \# & \#can. & \# & \#can. & \# & \#can. & \# & \#can. & \# & \#can. & \# & \#can. & \# & \#can. & \# & \#can. \\\midrule
    \textbf{all attempts} & 422 & 249 & 427 & 263 & 414 & 285 & 365 & 166 & 393 & 249 & 360 & 146 & 376 & 158 & 355 & 262 & 371 & 168 & 358 & 244 \\\midrule
    \multicolumn{21}{c}{\textbf{Proving equivalence}}\\\midrule
    \textbf{all equivalent attempts} & 133 & 25 & 128 & 33 & 75 & 34 & 68 & 13 & 161 & 42 & 133 & 8 & 173 & 33 & 58 & 25 & 194 & 34 & 81 & 23 \\
    \textbf{equivalence proven} & 133 & 25 & 126 & 32 & 72 & 32 & 68 & 13 & 161 & 42 & 133 & 8 & 3 & 2 & 57 & 24 & 194 & 34 & 80 & 22 \\
    \textbf{\dots~by test for:} &  &  &  &  &  &  &  &  &  &  &  &  &  &  &  &  &  &  &  &  \\
\end{tabular}
}}
\end{raggedright}
\end{landscape}
\newpage
\begin{landscape}
\begin{raggedright}
\parbox[l][5cm][t]{10cm}{
\scalebox{.95}{
\Small
\newcommand{\slimcol}{\hspace{-3pt}}
\begin{tabular}{l*{10}{|r<{\slimcol}r}}
    \multicolumn{21}{@{}l}{(continued from last page)}\\[1ex]
    \toprule
    & \multicolumn{2}{c|}{A24} & \multicolumn{2}{c|}{A25} & \multicolumn{2}{c|}{A26} & \multicolumn{2}{c|}{A27} & \multicolumn{2}{c|}{A28} & \multicolumn{2}{c|}{A29} & \multicolumn{2}{c|}{A30} & \multicolumn{2}{c|}{A31} & \multicolumn{2}{c|}{A32} & \multicolumn{2}{c}{A33} \\
    & \# & \#can. & \# & \#can. & \# & \#can. & \# & \#can. & \# & \#can. & \# & \#can. & \# & \#can. & \# & \#can. & \# & \#can. & \# & \#can. \\\midrule
    \textbf{all attempts} & 284 & 166 & 253 & 83 & 260 & 122 & 268 & 175 & 257 & 99 & 248 & 157 & 233 & 133 & 227 & 128 & 230 & 153 & 215 & 132 \\\midrule
    \multicolumn{21}{c}{\textbf{Proving equivalence}}\\\midrule
    \textbf{all equivalent attempts} & 85 & 12 & 159 & 14 & 90 & 13 & 129 & 54 & 133 & 14 & 94 & 24 & 85 & 27 & 82 & 12 & 92 & 37 & 66 & 15 \\
    \textbf{equivalence proven} & 85 & 12 & 159 & 14 & 90 & 13 & 126 & 51 & 100 & 7 & 0 & 0 & 85 & 27 & 82 & 12 & 34 & 15 & 64 & 14 \\
    \textbf{\dots~by test for:} &  &  &  &  &  &  &  &  &  &  &  &  &  &  &  &  &  &  &  &  \\
\end{tabular}
}}
\end{raggedright}
\end{landscape}
\newpage
\begin{landscape}
\begin{raggedright}
\parbox[l][5cm][t]{10cm}{
\scalebox{.95}{
\Small
\newcommand{\slimcol}{\hspace{-3pt}}
\begin{tabular}{l*{10}{|r<{\slimcol}r}}
    \multicolumn{21}{@{}l}{(continued from last page)}\\[1ex]
    \toprule
    & \multicolumn{2}{c|}{A34} & \multicolumn{2}{c|}{A35} & \multicolumn{2}{c|}{A36} & \multicolumn{2}{c|}{A37} & \multicolumn{2}{c|}{A38} & \multicolumn{2}{c|}{A39} & \multicolumn{2}{c|}{A40} & \multicolumn{2}{c|}{A41} & \multicolumn{2}{c|}{A42} & \multicolumn{2}{c}{A43} \\
    & \# & \#can. & \# & \#can. & \# & \#can. & \# & \#can. & \# & \#can. & \# & \#can. & \# & \#can. & \# & \#can. & \# & \#can. & \# & \#can. \\\midrule
    \textbf{all attempts} & 214 & 37 & 197 & 117 & 201 & 101 & 185 & 73 & 174 & 83 & 141 & 49 & 142 & 81 & 139 & 103 & 131 & 42 & 136 & 66 \\\midrule
    \multicolumn{21}{c}{\textbf{Proving equivalence}}\\\midrule
    \textbf{all equivalent attempts} & 157 & 15 & 62 & 12 & 70 & 11 & 75 & 13 & 75 & 12 & 84 & 7 & 74 & 20 & 53 & 30 & 82 & 7 & 72 & 12 \\
    \textbf{equivalence proven} & 157 & 15 & 62 & 12 & 70 & 11 & 70 & 9 & 75 & 12 & 84 & 7 & 54 & 14 & 52 & 29 & 82 & 7 & 70 & 11 \\
    \textbf{\dots~by test for:} &  &  &  &  &  &  &  &  &  &  &  &  &  &  &  &  &  &  &  &  \\
\end{tabular}
}}
\end{raggedright}
\end{landscape}
\newpage
\begin{landscape}
\begin{raggedright}
\parbox[l][5cm][t]{10cm}{
\scalebox{.95}{
\Small
\newcommand{\slimcol}{\hspace{-3pt}}
\begin{tabular}{l*{7}{|r<{\slimcol}r}*{6}{@{}l}@{}}
    \multicolumn{21}{@{}l}{(continued from last page)}\\[1ex]
    \toprule
    & \multicolumn{2}{c|}{A44} & \multicolumn{2}{c|}{A45} & \multicolumn{2}{c|}{A46} & \multicolumn{2}{c|}{A47} & \multicolumn{2}{c|}{A48} & \multicolumn{2}{c|}{A49} & \multicolumn{2}{c}{A50}  \\
    & \# & \#can. & \# & \#can. & \# & \#can. & \# & \#can. & \# & \#can. & \# & \#can. & \# & \#can. \\\midrule
    \textbf{all attempts} & 122 & 58 & 125 & 71 & 120 & 45 & 105 & 51 & 103 & 85 & 95 & 36 & 94 & 55 &  &  &  &  &  &  \\\midrule
    \multicolumn{21}{c}{\textbf{Proving equivalence}}\\\midrule
    \textbf{all equivalent attempts} & 22 & 5 & 37 & 9 & 73 & 12 & 67 & 15 & 6 & 5 & 42 & 3 & 21 & 5 &  &  &  &  &  &  \\
    \textbf{equivalence proven} & 20 & 3 & 11 & 1 & 73 & 12 & 65 & 14 & 0 & 0 & 42 & 3 & 21 & 5 &  &  &  &  &  &  \\
    \textbf{\dots~by test for:} &  &  &  &  &  &  &  &  &  &  &  &  &  &  &  &  &  &  &  &  \\
\bottomrule
\end{tabular}
}}
\end{raggedright}
\end{landscape}

\newpage
\begin{landscape}
\subsection{RQ3: Explaining Inequivalent Attempts}\label{app:alldata:correction-methods}

On the next pages, for all exercises the frequency that high-level feedback for the single attempts can be generated is shown. For each exercise, the columns labeled with \# show the number of attempts of the respective category and the columns labeled with \enquote{\#can.} the number of distinct attempts after canonization, i.e. the number of canons. Methods labeled with $^*$ only apply to exercises with bounded solution language. This is an extended version of Table \ref{tab:correction-methods}.

\vspace*{2ex}
\begin{raggedright}
\Small
\newcommand{\slimcol}{\hspace{-8pt}}
\begin{tabular}{l*{10}{|r<{\slimcol}r}}
    \toprule
    & \multicolumn{2}{c|}{Ia1} & \multicolumn{2}{c|}{Ia2} & \multicolumn{2}{c|}{Ia3} & \multicolumn{2}{c|}{Ia4} & \multicolumn{2}{c|}{Ia5} & \multicolumn{2}{c|}{Ia6} & \multicolumn{2}{c|}{Ia7} & \multicolumn{2}{c|}{Ib1} & \multicolumn{2}{c|}{Ib3} & \multicolumn{2}{c}{Ib4} \\
    & \# & \#can. & \# & \#can. & \# & \#can. & \# & \#can. & \# & \#can. & \# & \#can. & \# & \#can. & \# & \#can. & \# & \#can. & \# & \#can. \\\midrule
    \textbf{all attempts} & 5968 & 3607 & 3126 & 1685 & 2937 & 1474 & 1848 & 1215 & 1605 & 843 & 1089 & 638 & 743 & 262 & 965 & 499 & 835 & 268 & 616 & 250 \\
    \textbf{all inequivalent attempts} & 5568 & 3537 & 2717 & 1608 & 2562 & 1391 & 1744 & 1161 & 1261 & 746 & 1035 & 612 & 541 & 234 & 910 & 474 & 617 & 223 & 495 & 236 \\\midrule
    \multicolumn{21}{c}{\textbf{Explaining inequivalence due to different Parikh images}}\\\midrule
    \textbf{Parikh image different} & 4368 & 2891 & 2018 & 1231 & 2487 & 1348 & 1598 & 1079 & 738 & 513 & 622 & 381 & 509 & 220 & 690 & 364 & 598 & 214 & 482 & 227 \\
    \; representation concise & 3091 & 1914 & 2342 & 1381 & 2056 & 1150 & 615 & 392 & 733 & 369 & 979 & 569 & 522 & 219 & 706 & 372 & 512 & 192 & 164 & 91 \\
    \textbf{\& bounded by solution witness$^*$} & 2532 & 1703 & 1575 & 905 & 1807 & 899 & 946 & 656 & -- & -- & -- & -- & 438 & 172 & 339 & 182 & 281 & 152 & 283 & 159 \\
    \; set notation computable$^*$ & 0 & 0 & 1463 & 851 & 1786 & 884 & 0 & 0 & -- & -- & -- & -- & 433 & 169 & 0 & 0 & 279 & 150 & 0 & 0 \\
    \; set notation concise$^*$ & 0 & 0 & 511 & 315 & 1031 & 574 & 0 & 0 & -- & -- & -- & -- & 419 & 157 & 0 & 0 & 129 & 82 & 0 & 0 \\\midrule
    \multicolumn{21}{c}{\textbf{Explaining inequivalence with bug-fixing transformations}}\\\midrule
    \textbf{error corrected} & 0 & 0 & 16 & 14 & 59 & 14 & 0 & 0 & 0 & 0 & 0 & 0 & 3 & 2 & 0 & 0 & 24 & 6 & 0 & 0 \\
    \; by C1: add $\eps$ as recursion end & 0 & 0 & 16 & 14 & 10 & 7 & 0 & 0 & 0 & 0 & 0 & 0 & 3 & 2 & 0 & 0 & 10 & 5 & 0 & 0 \\
    \; by C2: add canonical recursion end & 0 & 0 & 0 & 0 & 15 & 5 & 0 & 0 & 0 & 0 & 0 & 0 & 0 & 0 & 0 & 0 & 0 & 0 & 0 & 0 \\
    \; by C3: replace $\eps$ by canonical rec. end & 0 & 0 & 0 & 0 & 36 & 4 & 0 & 0 & 0 & 0 & 0 & 0 & 0 & 0 & 0 & 0 & 14 & 1 & 0 & 0 \\\bottomrule
    \multicolumn{21}{r@{}}{\raisebox{-1ex}{(to be continued on next page)}}
\end{tabular}
\end{raggedright}
\end{landscape}
\newpage
\begin{landscape}
\begin{raggedright}
\parbox[l][5cm][t]{10cm}{
\scalebox{1}{
\Small
\newcommand{\slimcol}{\hspace{-8pt}}
\begin{tabular}{l*{10}{|r<{\slimcol}r}}
    \multicolumn{21}{@{}l}{(continued from last page)}\\[1ex]
    \toprule
    & \multicolumn{2}{c|}{Ib5} & \multicolumn{2}{c|}{Ib6} & \multicolumn{2}{c|}{Ib7} & \multicolumn{2}{c|}{Ib8} & \multicolumn{2}{c|}{Ib9} & \multicolumn{2}{c|}{Ib10} & \multicolumn{2}{c|}{Ib11} & \multicolumn{2}{c|}{A1} & \multicolumn{2}{c|}{A2} & \multicolumn{2}{c}{A3} \\
    & \# & \#can. & \# & \#can. & \# & \#can. & \# & \#can. & \# & \#can. & \# & \#can. & \# & \#can. & \# & \#can. & \# & \#can. & \# & \#can. \\\midrule
    \textbf{all attempts} & 1206 & 487 & 4312 & 1947 & 938 & 270 & 2224 & 1177 & 1507 & 743 & 1263 & 630 & 729 & 393 & 2925 & 1990 & 1884 & 1149 & 1657 & 782 \\
    \textbf{all inequivalent attempts} & 900 & 394 & 3761 & 1874 & 710 & 244 & 1961 & 1106 & 1261 & 640 & 1162 & 586 & 662 & 363 & 2646 & 1845 & 1663 & 1101 & 1479 & 759 \\\midrule
    \multicolumn{21}{c}{\textbf{Explaining inequivalence due to different Parikh images}}\\\midrule
    \textbf{Parikh image different} & 252 & 162 & 2144 & 1227 & 659 & 218 & 1795 & 1047 & 1220 & 611 & 1048 & 545 & 652 & 355 & 1750 & 1330 & 1588 & 1063 & 1422 & 733 \\
    \; representation concise & 756 & 311 & 3598 & 1774 & 698 & 235 & 1699 & 945 & 1197 & 606 & 983 & 481 & 523 & 317 & 663 & 478 & 1353 & 883 & 1390 & 680 \\
    \textbf{\& bounded by solution witness$^*$} & -- & -- & -- & -- & 539 & 163 & 1086 & 653 & 1039 & 515 & 717 & 382 & 487 & 297 & -- & -- & 1176 & 734 & 1026 & 433 \\
    \; set notation computable$^*$ & -- & -- & -- & -- & 528 & 159 & 1005 & 607 & 975 & 482 & 639 & 339 & 464 & 284 & -- & -- & 1029 & 658 & 1023 & 430 \\
    \; set notation concise$^*$ & -- & -- & -- & -- & 527 & 158 & 739 & 426 & 712 & 359 & 503 & 245 & 338 & 199 & -- & -- & 757 & 459 & 964 & 382 \\\midrule
    \multicolumn{21}{c}{\textbf{Explaining inequivalence with bug-fixing transformations}}\\\midrule
    \textbf{error corrected} & 0 & 0 & 0 & 0 & 5 & 2 & 20 & 11 & 46 & 31 & 10 & 7 & 10 & 5 & 1 & 1 & 10 & 5 & 27 & 8 \\
    \; by C1: add $\eps$ as recursion end & 0 & 0 & 0 & 0 & 5 & 2 & 20 & 11 & 26 & 17 & 10 & 7 & 2 & 2 & 1 & 1 & 9 & 4 & 8 & 3 \\
    \; by C2: add canonical recursion end & 0 & 0 & 0 & 0 & 0 & 0 & 5 & 3 & 15 & 11 & 4 & 2 & 2 & 1 & 0 & 0 & 6 & 3 & 5 & 2 \\
    \; by C3: replace $\eps$ by canonical rec. end & 0 & 0 & 0 & 0 & 0 & 0 & 0 & 0 & 13 & 8 & 0 & 0 & 6 & 2 & 0 & 0 & 0 & 0 & 14 & 3 \\\bottomrule
    \multicolumn{21}{c}{}\\[2mm]
    \toprule
    & \multicolumn{2}{c|}{A4} & \multicolumn{2}{c|}{A5} & \multicolumn{2}{c|}{A6} & \multicolumn{2}{c|}{A7} & \multicolumn{2}{c|}{A8} & \multicolumn{2}{c|}{A9} & \multicolumn{2}{c|}{A10} & \multicolumn{2}{c|}{A11} & \multicolumn{2}{c|}{A12} & \multicolumn{2}{c}{A13} \\
    & \# & \#can. & \# & \#can. & \# & \#can. & \# & \#can. & \# & \#can. & \# & \#can. & \# & \#can. & \# & \#can. & \# & \#can. & \# & \#can. \\\midrule
    \textbf{all attempts} & 1527 & 1094 & 1264 & 858 & 1060 & 627 & 944 & 717 & 896 & 671 & 647 & 300 & 697 & 289 & 565 & 266 & 505 & 270 & 440 & 210 \\
    \textbf{all inequivalent attempts} & 1260 & 995 & 1038 & 787 & 651 & 459 & 889 & 697 & 803 & 643 & 421 & 254 & 287 & 232 & 355 & 252 & 347 & 239 & 302 & 194 \\\midrule
    \multicolumn{21}{c}{\textbf{Explaining inequivalence due to different Parikh images}}\\\midrule
    \textbf{Parikh image different} & 1207 & 952 & 987 & 746 & 278 & 231 & 729 & 589 & 758 & 606 & 242 & 168 & 151 & 128 & 342 & 243 & 338 & 231 & 217 & 136 \\
    \; representation concise & 906 & 694 & 553 & 378 & 619 & 428 & 756 & 583 & 471 & 352 & 381 & 217 & 273 & 218 & 329 & 230 & 285 & 185 & 265 & 157 \\
    \textbf{\& bounded by solution witness$^*$} & 989 & 754 & 636 & 443 & -- & -- & -- & -- & -- & -- & -- & -- & 121 & 95 & -- & -- & -- & -- & -- & -- \\
    \; set notation computable$^*$ & 929 & 703 & 563 & 387 & -- & -- & -- & -- & -- & -- & -- & -- & 112 & 87 & -- & -- & -- & -- & -- & -- \\
    \; set notation concise$^*$ & 603 & 428 & 230 & 143 & -- & -- & -- & -- & -- & -- & -- & -- & 50 & 40 & -- & -- & -- & -- & -- & -- \\\midrule
    \multicolumn{21}{c}{\textbf{Explaining inequivalence with bug-fixing transformations}}\\\midrule
    \textbf{error corrected} & 5 & 5 & 7 & 6 & 0 & 0 & 0 & 0 & 1 & 1 & 0 & 0 & 4 & 3 & 0 & 0 & 0 & 0 & 0 & 0 \\
    \; by C1: add $\eps$ as recursion end & 2 & 2 & 7 & 6 & 0 & 0 & 0 & 0 & 1 & 1 & 0 & 0 & 4 & 3 & 0 & 0 & 0 & 0 & 0 & 0 \\
    \; by C2: add canonical recursion end & 0 & 0 & 0 & 0 & 0 & 0 & 0 & 0 & 0 & 0 & 0 & 0 & 0 & 0 & 0 & 0 & 0 & 0 & 0 & 0 \\
    \; by C3: replace $\eps$ by canonical rec. end & 3 & 3 & 0 & 0 & 0 & 0 & 0 & 0 & 0 & 0 & 0 & 0 & 0 & 0 & 0 & 0 & 0 & 0 & 0 & 0 \\\bottomrule
    \multicolumn{21}{r@{}}{\raisebox{-1ex}{(to be continued on next page)}}
\end{tabular}
}}
\end{raggedright}
\end{landscape}
\newpage
\begin{landscape}
\begin{raggedright}
\parbox[l][5cm][t]{10cm}{
\scalebox{1}{
\Small
\newcommand{\slimcol}{\hspace{-8pt}}
\begin{tabular}{l*{10}{|r<{\slimcol}r}}
    \multicolumn{21}{@{}l}{(continued from last page)}\\[1ex]
    \toprule
    & \multicolumn{2}{c|}{A14} & \multicolumn{2}{c|}{A15} & \multicolumn{2}{c|}{A16} & \multicolumn{2}{c|}{A17} & \multicolumn{2}{c|}{A18} & \multicolumn{2}{c|}{A19} & \multicolumn{2}{c|}{A20} & \multicolumn{2}{c|}{A21} & \multicolumn{2}{c|}{A22} & \multicolumn{2}{c}{A23} \\
    & \# & \#can. & \# & \#can. & \# & \#can. & \# & \#can. & \# & \#can. & \# & \#can. & \# & \#can. & \# & \#can. & \# & \#can. & \# & \#can. \\\midrule
    \textbf{all attempts} & 422 & 249 & 427 & 263 & 414 & 285 & 365 & 166 & 393 & 249 & 360 & 146 & 376 & 158 & 355 & 262 & 371 & 168 & 358 & 244 \\
    \textbf{all inequivalent attempts} & 289 & 224 & 299 & 230 & 339 & 251 & 297 & 153 & 232 & 207 & 227 & 138 & 203 & 125 & 297 & 237 & 177 & 134 & 277 & 221 \\\midrule
    \multicolumn{21}{c}{\textbf{Explaining inequivalence due to different Parikh images}}\\\midrule
    \textbf{Parikh image different} & 246 & 195 & 277 & 214 & 332 & 245 & 132 & 81 & 216 & 193 & 214 & 128 & 61 & 55 & 280 & 224 & 166 & 125 & 260 & 211 \\
    \; representation concise & 261 & 198 & 284 & 217 & 216 & 150 & 295 & 151 & 133 & 118 & 209 & 123 & 202 & 124 & 147 & 114 & 163 & 120 & 225 & 179 \\
    \textbf{\& bounded by solution witness$^*$} & 173 & 127 & 195 & 141 & 289 & 205 & -- & -- & 112 & 100 & 143 & 74 & -- & -- & 165 & 132 & 117 & 81 & 180 & 141 \\
    \; set notation computable$^*$ & 171 & 125 & 191 & 137 & 253 & 185 & -- & -- & 106 & 94 & 140 & 71 & -- & -- & 151 & 124 & 117 & 81 & 162 & 128 \\
    \; set notation concise$^*$ & 150 & 107 & 148 & 101 & 95 & 74 & -- & -- & 22 & 21 & 123 & 56 & -- & -- & 60 & 43 & 108 & 73 & 80 & 57 \\\midrule
    \multicolumn{21}{c}{\textbf{Explaining inequivalence with bug-fixing transformations}}\\\midrule
    \textbf{error corrected} & 9 & 7 & 8 & 7 & 6 & 2 & 0 & 0 & 0 & 0 & 0 & 0 & 2 & 2 & 1 & 1 & 33 & 17 & 2 & 2 \\
    \; by C1: add $\eps$ as recursion end & 9 & 7 & 3 & 3 & 1 & 1 & 0 & 0 & 0 & 0 & 0 & 0 & 2 & 2 & 1 & 1 & 3 & 3 & 2 & 2 \\
    \; by C2: add canonical recursion end & 4 & 3 & 1 & 1 & 0 & 0 & 0 & 0 & 0 & 0 & 0 & 0 & 0 & 0 & 0 & 0 & 9 & 7 & 0 & 0 \\
    \; by C3: replace $\eps$ by canonical rec. end & 0 & 0 & 5 & 4 & 5 & 1 & 0 & 0 & 0 & 0 & 0 & 0 & 0 & 0 & 0 & 0 & 21 & 7 & 0 & 0 \\\bottomrule
    \multicolumn{21}{c}{}\\[2mm]
    \toprule
    & \multicolumn{2}{c|}{A24} & \multicolumn{2}{c|}{A25} & \multicolumn{2}{c|}{A26} & \multicolumn{2}{c|}{A27} & \multicolumn{2}{c|}{A28} & \multicolumn{2}{c|}{A29} & \multicolumn{2}{c|}{A30} & \multicolumn{2}{c|}{A31} & \multicolumn{2}{c|}{A32} & \multicolumn{2}{c}{A33} \\
    & \# & \#can. & \# & \#can. & \# & \#can. & \# & \#can. & \# & \#can. & \# & \#can. & \# & \#can. & \# & \#can. & \# & \#can. & \# & \#can. \\\midrule
    \textbf{all attempts} & 284 & 166 & 253 & 83 & 260 & 122 & 268 & 175 & 257 & 99 & 248 & 157 & 233 & 133 & 227 & 128 & 230 & 153 & 215 & 132 \\
    \textbf{all inequivalent attempts} & 199 & 154 & 94 & 69 & 170 & 109 & 139 & 121 & 124 & 85 & 154 & 133 & 148 & 106 & 145 & 116 & 138 & 116 & 149 & 117 \\\midrule
    \multicolumn{21}{c}{\textbf{Explaining inequivalence due to different Parikh images}}\\\midrule
    \textbf{Parikh image different} & 199 & 154 & 79 & 57 & 79 & 61 & 136 & 118 & 117 & 78 & 77 & 77 & 148 & 106 & 135 & 107 & 64 & 61 & 94 & 85 \\
    \; representation concise & 181 & 138 & 92 & 67 & 170 & 109 & 108 & 93 & 99 & 65 & 146 & 125 & 146 & 104 & 108 & 82 & 135 & 113 & 147 & 115 \\
    \textbf{\& bounded by solution witness$^*$} & 133 & 106 & 62 & 41 & 122 & 81 & 87 & 75 & -- & -- & -- & -- & 136 & 96 & -- & -- & -- & -- & -- & -- \\
    \; set notation computable$^*$ & 133 & 106 & 62 & 41 & 93 & 71 & 80 & 68 & -- & -- & -- & -- & 136 & 96 & -- & -- & -- & -- & -- & -- \\
    \; set notation concise$^*$ & 127 & 100 & 61 & 40 & 40 & 34 & 67 & 58 & -- & -- & -- & -- & 87 & 58 & -- & -- & -- & -- & -- & -- \\\midrule
    \multicolumn{21}{c}{\textbf{Explaining inequivalence with bug-fixing transformations}}\\\midrule
    \textbf{error corrected} & 0 & 0 & 1 & 1 & 7 & 5 & 7 & 7 & 0 & 0 & 7 & 7 & 23 & 8 & 0 & 0 & 1 & 1 & 0 & 0 \\
    \; by C1: add $\eps$ as recursion end & 0 & 0 & 1 & 1 & 7 & 5 & 7 & 7 & 0 & 0 & 7 & 7 & 1 & 1 & 0 & 0 & 1 & 1 & 0 & 0 \\
    \; by C2: add canonical recursion end & 0 & 0 & 0 & 0 & 2 & 1 & 1 & 1 & 0 & 0 & 0 & 0 & 11 & 3 & 0 & 0 & 0 & 0 & 0 & 0 \\
    \; by C3: replace $\eps$ by canonical rec. end & 0 & 0 & 0 & 0 & 0 & 0 & 0 & 0 & 0 & 0 & 0 & 0 & 11 & 4 & 0 & 0 & 0 & 0 & 0 & 0 \\\bottomrule
    \multicolumn{21}{r@{}}{\raisebox{-1ex}{(to be continued on next page)}}
\end{tabular}
}}
\end{raggedright}
\end{landscape}
\newpage
\begin{landscape}
\begin{raggedright}
\parbox[l][5cm][t]{10cm}{
\scalebox{1}{
\Small
\newcommand{\slimcol}{\hspace{-8pt}}
\begin{tabular}{l*{10}{|r<{\slimcol}r}}
    \multicolumn{21}{@{}l}{(continued from last page)}\\[1ex]
    \toprule
    & \multicolumn{2}{c|}{A34} & \multicolumn{2}{c|}{A35} & \multicolumn{2}{c|}{A36} & \multicolumn{2}{c|}{A37} & \multicolumn{2}{c|}{A38} & \multicolumn{2}{c|}{A39} & \multicolumn{2}{c|}{A40} & \multicolumn{2}{c|}{A41} & \multicolumn{2}{c|}{A42} & \multicolumn{2}{c}{A43} \\
    & \# & \#can. & \# & \#can. & \# & \#can. & \# & \#can. & \# & \#can. & \# & \#can. & \# & \#can. & \# & \#can. & \# & \#can. & \# & \#can. \\\midrule
    \textbf{all attempts} & 214 & 37 & 197 & 117 & 201 & 101 & 185 & 73 & 174 & 83 & 141 & 49 & 142 & 81 & 139 & 103 & 131 & 42 & 136 & 66 \\
    \textbf{all inequivalent attempts} & 57 & 22 & 135 & 105 & 131 & 90 & 110 & 60 & 99 & 71 & 57 & 42 & 68 & 61 & 86 & 73 & 49 & 35 & 64 & 54 \\\midrule
    \multicolumn{21}{c}{\textbf{Explaining inequivalence due to different Parikh images}}\\\midrule
    \textbf{Parikh image different} & 57 & 22 & 131 & 101 & 126 & 86 & 58 & 39 & 95 & 68 & 54 & 39 & 20 & 18 & 75 & 65 & 47 & 33 & 47 & 41 \\
    \; representation concise & 57 & 22 & 129 & 100 & 120 & 83 & 110 & 60 & 97 & 69 & 57 & 42 & 59 & 52 & 85 & 72 & 47 & 33 & 57 & 48 \\
    \textbf{\& bounded by solution witness$^*$} & 54 & 19 & 93 & 72 & 108 & 68 & -- & -- & 73 & 47 & -- & -- & -- & -- & 66 & 58 & -- & -- & 33 & 28 \\
    \; set notation computable$^*$ & 54 & 19 & 93 & 72 & 108 & 68 & -- & -- & 73 & 47 & -- & -- & -- & -- & 65 & 57 & -- & -- & 0 & 0 \\
    \; set notation concise$^*$ & 54 & 19 & 47 & 35 & 53 & 40 & -- & -- & 65 & 40 & -- & -- & -- & -- & 51 & 44 & -- & -- & 0 & 0 \\\midrule
    \multicolumn{21}{c}{\textbf{Explaining inequivalence with bug-fixing transformations}}\\\midrule
    \textbf{error corrected} & 26 & 8 & 0 & 0 & 2 & 2 & 0 & 0 & 11 & 9 & 0 & 0 & 2 & 1 & 3 & 3 & 0 & 0 & 0 & 0 \\
    \; by C1: add $\eps$ as recursion end & 26 & 8 & 0 & 0 & 2 & 2 & 0 & 0 & 4 & 4 & 0 & 0 & 2 & 1 & 3 & 3 & 0 & 0 & 0 & 0 \\
    \; by C2: add canonical recursion end & 0 & 0 & 0 & 0 & 1 & 1 & 0 & 0 & 2 & 1 & 0 & 0 & 0 & 0 & 0 & 0 & 0 & 0 & 0 & 0 \\
    \; by C3: replace $\eps$ by canonical rec. end & 0 & 0 & 0 & 0 & 0 & 0 & 0 & 0 & 5 & 4 & 0 & 0 & 0 & 0 & 0 & 0 & 0 & 0 & 0 & 0 \\\bottomrule
    \multicolumn{21}{c}{}\\[2mm]
    \cmidrule[\heavyrulewidth]{1-15}
    & \multicolumn{2}{c|}{A44} & \multicolumn{2}{c|}{A45} & \multicolumn{2}{c|}{A46} & \multicolumn{2}{c|}{A47} & \multicolumn{2}{c|}{A48} & \multicolumn{2}{c|}{A49} & \multicolumn{2}{c}{A50}  \\
    & \# & \#can. & \# & \#can. & \# & \#can. & \# & \#can. & \# & \#can. & \# & \#can. & \# & \multicolumn{1}{r}{\#can.}\\\cmidrule[\lightrulewidth]{1-15}
    \textbf{all attempts} & 122 & 58 & 125 & 71 & 120 & 45 & 105 & 51 & 103 & 85 & 95 & 36 & 94 & \multicolumn{1}{r}{55}  \\
    \textbf{all inequivalent attempts} & 100 & 53 & 88 & 62 & 47 & 33 & 38 & 36 & 97 & 80 & 53 & 33 & 73 & \multicolumn{1}{r}{50}  \\\cmidrule[\lightrulewidth]{1-15}
    \multicolumn{21}{c}{\textbf{Explaining inequivalence due to different Parikh images}}\\\cmidrule[\lightrulewidth]{1-15}
    \textbf{Parikh image different} & 100 & 53 & 87 & 61 & 31 & 18 & 28 & 28 & 85 & 69 & 53 & 33 & 70 & \multicolumn{1}{r}{47}  \\
    \; representation concise & 65 & 35 & 33 & 26 & 46 & 32 & 33 & 31 & 87 & 70 & 51 & 31 & 47 & \multicolumn{1}{r}{28}  \\
    \textbf{\& bounded by solution witness$^*$} & -- & -- & -- & -- & 32 & 19 & 21 & 21 & -- & -- & -- & -- & 47 & \multicolumn{1}{r}{28}  \\
    \; set notation computable$^*$ & -- & -- & -- & -- & 28 & 15 & 19 & 19 & -- & -- & -- & -- & 47 & \multicolumn{1}{r}{28}  \\
    \; set notation concise$^*$ & -- & -- & -- & -- & 16 & 8 & 11 & 11 & -- & -- & -- & -- & 31 & \multicolumn{1}{r}{16}  \\\cmidrule[\lightrulewidth]{1-15}
    \multicolumn{21}{c}{\textbf{Explaining inequivalence with bug-fixing transformations}}\\\cmidrule[\lightrulewidth]{1-15}
    \textbf{error corrected} & 0 & 0 & 6 & 5 & 6 & 3 & 1 & 1 & 4 & 4 & 0 & 0 & 0 & \multicolumn{1}{r}{0} \\
    \; by C1: add $\eps$ as recursion end & 0 & 0 & 6 & 5 & 6 & 3 & 1 & 1 & 4 & 4 & 0 & 0 & 0 & \multicolumn{1}{r}{0}  \\
    \; by C2: add canonical recursion end & 0 & 0 & 0 & 0 & 0 & 0 & 0 & 0 & 1 & 1 & 0 & 0 & 0 & \multicolumn{1}{r}{0} \\
    \; by C3: replace $\eps$ by canonical rec. end & 0 & 0 & 0 & 0 & 0 & 0 & 0 & 0 & 0 & 0 & 0 & 0 & 0 & \multicolumn{1}{r}{0} \\
    \cmidrule[\heavyrulewidth]{1-15}
\end{tabular}
}}
\end{raggedright}
\end{landscape}

\newpage

\begin{landscape}
\subsection{RQ5: Efficiency}\label{app:alldata:runtime}
On the next pages, we provide the runtimes for the individual methods for all exercises considered in our evaluation. Columns labeled with \enquote{avg} denote the mean of the computation times in milliseconds over all attempts per exercise; columns labeled with \enquote{\#ab.} denotes the number of computations aborted because of a time-out. Methods labeled with $^*$ only apply to exercises with bounded solution language. This is an extended version of Table \ref{tab:runtime}.

\vspace*{2ex}
\begin{raggedright}
\Small
\newcommand{\slimcol}{\hspace{-3pt}}
\begin{tabular}{l*{10}{|r<{\slimcol}r}}
    \toprule
    & \multicolumn{2}{c|}{Ia1} & \multicolumn{2}{c|}{Ia2} & \multicolumn{2}{c|}{Ia3} & \multicolumn{2}{c|}{Ia4} & \multicolumn{2}{c|}{Ia5} & \multicolumn{2}{c|}{Ia6} & \multicolumn{2}{c|}{Ia7} & \multicolumn{2}{c|}{Ib1} & \multicolumn{2}{c|}{Ib3} & \multicolumn{2}{c}{Ib4} \\
    & avg & \#ab. & avg & \#ab. & avg & \#ab. & avg & \#ab. & avg & \#ab. & avg & \#ab. & avg & \#ab. & avg & \#ab. & avg & \#ab. & avg & \#ab. \\\midrule
    \textbf{(in)equivalence tests} &  &  &  &  &  &  &  &  &  &  &  &  &  &  &  &  &  &  &  &  \\
    \; isomorphism & 172 & 0 & 157 & 0 & 143 & 0 & 133 & 0 & 137 & 0 & 158 & 0 & 121 & 0 & 172 & 0 & 141 & 0 & 147 & 0 \\
    \; emptiness & 18 & 0 & 9 & 0 & 1 & 0 & 24 & 0 & 16 & 0 & 1 & 0 & 1 & 0 & 1 & 0 & 1 & 0 & 14 & 0 \\
    \; finiteness & 29 & 0 & 50 & 0 & 33 & 0 & 35 & 0 & 33 & 0 & 19 & 0 & 19 & 0 & 21 & 0 & 45 & 0 & 25 & 0 \\
    \; symbol frequencies & 452 & 17 & 62 & 0 & 324 & 1 & 685 & 72 & 272 & 8 & 106 & 0 & 26 & 0 & 91 & 1 & 48 & 0 & 181 & 6 \\
    \; boundedness witness$^*$ & 4041 & 383 & 975 & 6 & 531 & 1 & 1629 & 131 & -- & -- & -- & -- & 117 & 0 & 1187 & 9 & 120 & 0 & 1457 & 8 \\
\end{tabular}
\end{raggedright}
\end{landscape}
\newpage
\begin{landscape}
\begin{raggedright}
\Small
\newcommand{\slimcol}{\hspace{-3pt}}
\begin{tabular}{l*{10}{|r<{\slimcol}r}}
    \multicolumn{21}{@{}l}{(continued from last page)}\\[1ex]
    \toprule
    & \multicolumn{2}{c|}{Ib5} & \multicolumn{2}{c|}{Ib6} & \multicolumn{2}{c|}{Ib7} & \multicolumn{2}{c|}{Ib8} & \multicolumn{2}{c|}{Ib9} & \multicolumn{2}{c|}{Ib10} & \multicolumn{2}{c|}{Ib11} & \multicolumn{2}{c|}{A1} & \multicolumn{2}{c|}{A2} & \multicolumn{2}{c}{A3} \\
    & avg & \#ab. & avg & \#ab. & avg & \#ab. & avg & \#ab. & avg & \#ab. & avg & \#ab. & avg & \#ab. & avg & \#ab. & avg & \#ab. & avg & \#ab. \\\midrule
    \textbf{(in)equivalence tests} &  &  &  &  &  &  &  &  &  &  &  &  &  &  &  &  &  &  &  &  \\
    \; isomorphism & 136 & 0 & 195 & 0 & 379 & 0 & 163 & 0 & 165 & 0 & 156 & 0 & 153 & 0 & 178 & 0 & 168 & 0 & 164 & 0 \\
    \; emptiness & 33 & 0 & 16 & 0 & 1 & 0 & 14 & 0 & 1 & 0 & 1 & 0 & 46 & 0 & 28 & 0 & 15 & 0 & 1 & 0 \\
    \; finiteness & 39 & 0 & 39 & 0 & 51 & 0 & 28 & 0 & 39 & 0 & 30 & 0 & 3 & 0 & 27 & 0 & 45 & 0 & 61 & 0 \\
    \; symbol frequencies & 385 & 14 & 55 & 1 & 27 & 0 & 48 & 0 & 31 & 0 & 34 & 0 & 39 & 0 & 320 & 5 & 120 & 2 & 39 & 0 \\
\end{tabular}
\end{raggedright}
\end{landscape}
\newpage
\begin{landscape}
\begin{raggedright}
\Small
\newcommand{\slimcol}{\hspace{-3pt}}
\begin{tabular}{l*{10}{|r<{\slimcol}r}}
    \multicolumn{21}{@{}l}{(continued from last page)}\\[1ex]
    \toprule
    & \multicolumn{2}{c|}{A4} & \multicolumn{2}{c|}{A5} & \multicolumn{2}{c|}{A6} & \multicolumn{2}{c|}{A7} & \multicolumn{2}{c|}{A8} & \multicolumn{2}{c|}{A9} & \multicolumn{2}{c|}{A10} & \multicolumn{2}{c|}{A11} & \multicolumn{2}{c|}{A12} & \multicolumn{2}{c}{A13} \\
    & avg & \#ab. & avg & \#ab. & avg & \#ab. & avg & \#ab. & avg & \#ab. & avg & \#ab. & avg & \#ab. & avg & \#ab. & avg & \#ab. & avg & \#ab. \\\midrule
    \textbf{(in)equivalence tests} &  &  &  &  &  &  &  &  &  &  &  &  &  &  &  &  &  &  &  &  \\
    \; isomorphism & 158 & 0 & 139 & 0 & 154 & 0 & 170 & 0 & 117 & 0 & 196 & 0 & 138 & 0 & 113 & 0 & 153 & 0 & 105 & 0 \\
    \; emptiness & 1 & 0 & 1 & 0 & 1 & 0 & 39 & 0 & 1 & 0 & 1 & 0 & 1 & 0 & 1 & 0 & 1 & 0 & 1 & 0 \\
    \; finiteness & 39 & 0 & 23 & 0 & 27 & 0 & 23 & 0 & 35 & 0 & 3 & 0 & 46 & 0 & 22 & 0 & 21 & 0 & 30 & 0 \\
    \; symbol frequencies & 60 & 0 & 84 & 2 & 43 & 4 & 114 & 7 & 86 & 1 & 39 & 2 & 26 & 0 & 43 & 0 & 76 & 0 & 53 & 0 \\
\end{tabular}
\end{raggedright}
\end{landscape}
\newpage
\begin{landscape}
\begin{raggedright}
\Small
\newcommand{\slimcol}{\hspace{-3pt}}
\begin{tabular}{l*{10}{|r<{\slimcol}r}}
    \multicolumn{21}{@{}l}{(continued from last page)}\\[1ex]
    \toprule
    & \multicolumn{2}{c|}{A14} & \multicolumn{2}{c|}{A15} & \multicolumn{2}{c|}{A16} & \multicolumn{2}{c|}{A17} & \multicolumn{2}{c|}{A18} & \multicolumn{2}{c|}{A19} & \multicolumn{2}{c|}{A20} & \multicolumn{2}{c|}{A21} & \multicolumn{2}{c|}{A22} & \multicolumn{2}{c}{A23} \\
    & avg & \#ab. & avg & \#ab. & avg & \#ab. & avg & \#ab. & avg & \#ab. & avg & \#ab. & avg & \#ab. & avg & \#ab. & avg & \#ab. & avg & \#ab. \\\midrule
    \textbf{(in)equivalence tests} &  &  &  &  &  &  &  &  &  &  &  &  &  &  &  &  &  &  &  &  \\
    \; isomorphism & 130 & 0 & 112 & 0 & 120 & 0 & 148 & 0 & 107 & 0 & 121 & 0 & 127 & 0 & 146 & 0 & 113 & 0 & 163 & 0 \\
    \; emptiness & 1 & 0 & 1 & 0 & 1 & 0 & 1 & 0 & 1 & 0 & 15 & 0 & 1 & 0 & 1 & 0 & 1 & 0 & 1 & 0 \\
    \; finiteness & 34 & 0 & 22 & 0 & 34 & 0 & 78 & 0 & 28 & 0 & 1 & 0 & 13 & 0 & 25 & 0 & 29 & 0 & 45 & 0 \\
    \; symbol frequencies & 26 & 0 & 30 & 0 & 64 & 0 & 23 & 0 & 44 & 0 & 32 & 0 & 32 & 0 & 51 & 0 & 38 & 0 & 52 & 0 \\
\end{tabular}
\end{raggedright}
\end{landscape}
\newpage
\begin{landscape}
\begin{raggedright}
\Small
\newcommand{\slimcol}{\hspace{-3pt}}
\begin{tabular}{l*{10}{|r<{\slimcol}r}}
    \multicolumn{21}{@{}l}{(continued from last page)}\\[1ex]
    \toprule
    & \multicolumn{2}{c|}{A24} & \multicolumn{2}{c|}{A25} & \multicolumn{2}{c|}{A26} & \multicolumn{2}{c|}{A27} & \multicolumn{2}{c|}{A28} & \multicolumn{2}{c|}{A29} & \multicolumn{2}{c|}{A30} & \multicolumn{2}{c|}{A31} & \multicolumn{2}{c|}{A32} & \multicolumn{2}{c}{A33} \\
    & avg & \#ab. & avg & \#ab. & avg & \#ab. & avg & \#ab. & avg & \#ab. & avg & \#ab. & avg & \#ab. & avg & \#ab. & avg & \#ab. & avg & \#ab. \\\midrule
    \textbf{(in)equivalence tests} &  &  &  &  &  &  &  &  &  &  &  &  &  &  &  &  &  &  &  &  \\
    \; isomorphism & 172 & 0 & 213 & 0 & 146 & 0 & 69 & 0 & 105 & 0 & 105 & 0 & 129 & 0 & 129 & 0 & 110 & 0 & 152 & 0 \\
    \; emptiness & 1 & 0 & 0 & 0 & 1 & 0 & 1 & 0 & 1 & 0 & 1 & 0 & 1 & 0 & 4 & 0 & 1 & 0 & 1 & 0 \\
    \; finiteness & 13 & 0 & 3 & 0 & 1 & 0 & 18 & 0 & 27 & 0 & 5 & 0 & 3 & 0 & 10 & 0 & 1 & 0 & 1 & 0 \\
    \; symbol frequencies & 38 & 0 & 20 & 0 & 19 & 0 & 44 & 0 & 37 & 0 & 32 & 0 & 24 & 0 & 41 & 0 & 30 & 0 & 22 & 0 \\
\end{tabular}
\end{raggedright}
\end{landscape}
\newpage
\begin{landscape}
\begin{raggedright}
\Small
\newcommand{\slimcol}{\hspace{-3pt}}
\begin{tabular}{l*{10}{|r<{\slimcol}r}}
    \multicolumn{21}{@{}l}{(continued from last page)}\\[1ex]
    \toprule
    & \multicolumn{2}{c|}{A34} & \multicolumn{2}{c|}{A35} & \multicolumn{2}{c|}{A36} & \multicolumn{2}{c|}{A37} & \multicolumn{2}{c|}{A38} & \multicolumn{2}{c|}{A39} & \multicolumn{2}{c|}{A40} & \multicolumn{2}{c|}{A41} & \multicolumn{2}{c|}{A42} & \multicolumn{2}{c}{A43} \\
    & avg & \#ab. & avg & \#ab. & avg & \#ab. & avg & \#ab. & avg & \#ab. & avg & \#ab. & avg & \#ab. & avg & \#ab. & avg & \#ab. & avg & \#ab. \\\midrule
    \textbf{(in)equivalence tests} &  &  &  &  &  &  &  &  &  &  &  &  &  &  &  &  &  &  &  &  \\
    \; isomorphism & 146 & 0 & 108 & 0 & 97 & 0 & 112 & 0 & 122 & 0 & 153 & 0 & 131 & 0 & 56 & 0 & 6 & 0 & 118 & 0 \\
    \; emptiness & 1 & 0 & 1 & 0 & 1 & 0 & 1 & 0 & 1 & 0 & 0 & 0 & 1 & 0 & 1 & 0 & 1 & 0 & 0 & 0 \\
    \; finiteness & 1 & 0 & 1 & 0 & 10 & 0 & 1 & 0 & 1 & 0 & 5 & 0 & 1 & 0 & 18 & 0 & 1 & 0 & 38 & 0 \\
    \; symbol frequencies & 14 & 0 & 24 & 0 & 29 & 0 & 27 & 0 & 18 & 0 & 38 & 0 & 49 & 0 & 20 & 0 & 49 & 0 & 30 & 0 \\
\end{tabular}
\end{raggedright}
\end{landscape}
\newpage
\begin{landscape}
\begin{raggedright}
\Small
\newcommand{\slimcol}{\hspace{-3pt}}
\begin{tabular}{l*{7}{|r<{\slimcol}r}*{6}{@{}l}@{}}
    \multicolumn{21}{@{}l}{(continued from last page)}\\[1ex]
    \toprule
    & \multicolumn{2}{c|}{A44} & \multicolumn{2}{c|}{A45} & \multicolumn{2}{c|}{A46} & \multicolumn{2}{c|}{A47} & \multicolumn{2}{c|}{A48} & \multicolumn{2}{c|}{A49} & \multicolumn{2}{c}{A50}  \\
    & avg & \#ab. & avg & \#ab. & avg & \#ab. & avg & \#ab. & avg & \#ab. & avg & \#ab. & avg & \#ab.  \\\midrule
    \textbf{(in)equivalence tests} &  &  &  &  &  &  &  &  &  &  &  &  &  &  &  &  &  &  &  &  \\
    \; isomorphism & 69 & 0 & 99 & 0 & 107 & 0 & 107 & 0 & 121 & 0 & 64 & 0 & 73 & 0 &  &  &  &  &  &  \\
    \; emptiness & 1 & 0 & 1 & 0 & 1 & 0 & 112 & 0 & 1 & 0 & 1 & 0 & 1 & 0 &  &  &  &  &  &  \\
    \; finiteness & 17 & 0 & 22 & 0 & 1 & 0 & 1 & 0 & 11 & 0 & 1 & 0 & 11 & 0 &  &  &  &  &  &  \\
    \; symbol frequencies & 936 & 0 & 66 & 0 & 16 & 0 & 30 & 0 & 34 & 0 & 36 & 0 & 37 & 0 &  &  &  &  &  &  \\
\bottomrule
\end{tabular}
\end{raggedright}
\end{landscape}

\newpage

\section{Full Normalization Pipeline Used in the Evaluation}\label{app:pipeline}

In this appendix, we provide the normalization pipeline that is used for the evaluation in Section~\ref{section:evaluation}. Before presenting the pipeline itself in Section~\ref{app:pipeline-self}, first we present the used transformations in Sections~\ref{app:hard-trafos} and~\ref{app:pattern-trafos}.

\subsection{Pattern-based Transformations}\label{app:pattern-trafos}

In this section, we give definitions for the pattern-based transformations used in the pipeline.

{
\small
\lstset{language=XML,morekeywords={transformations,transformation,targetpattern,sourcepattern,encoding,xs:schema,xs:element,xs:complexType,xs:sequence,xs:attribute}}  

\begin{lstlisting}
<transformations>
    <transformation name="SynchronizeRecursionLevel" type="EQUIVALENCE">
        <sourcepattern>
            X -> phi_i alpha_j Y beta_j psi_i | chi_i
            Y -> alpha_i Y beta_i | gamma_i
            with:
            X is variable
            Y is variable
            X != Y
            phi_i has a value
            alpha_i has a value
            phi_i does not contain Y
            psi_i does not contain Y
            chi_i does not contain Y
            alpha_i does not contain Y
            beta_i does not contain Y
            gamma_i does not contain Y
            alpha_ibeta_i != eps
            Y appears only in matched rules
            Y is not start variable
        </sourcepattern>
        <targetpattern>
            X -> phi_i Y psi_i | chi_i
            Y -> alpha_i Y beta_i | alpha_j gamma_i beta_j
        </targetpattern>
    </transformation>

    <transformation name="SynchronizeRecursionEndFromLeft" 
                    type="EQUIVALENCE">
        <sourcepattern>
            X -> phi_i Y beta_j psi_i | chi_i
            Y -> alpha_i Y beta_i | gamma_i
            with:
            X is variable
            Y is variable
            X != Y
            phi_i has a value
            alpha_i has a value
            phi_i does not contain Y
            psi_i does not contain Y
            chi_i does not contain Y
            alpha_i does not contain Y
            beta_i does not contain Y
            gamma_i does not contain Y
            alpha_ibeta_i != eps
            Y appears only in matched rules
            Y is not start variable
        </sourcepattern>
        <targetpattern>
            X -> phi_i Y psi_i | chi_i
            Y -> alpha_i Y beta_i | gamma_i beta_j
        </targetpattern>
    </transformation>

    <transformation name="SynchronizeRecursionEndFromRight" 
                    type="EQUIVALENCE">
        <sourcepattern>
            X -> phi_i alpha_j Y psi_i | chi_i
            Y -> alpha_i Y beta_i | gamma_i
            with:
            X is variable
            Y is variable
            X != Y
            phi_i has a value
            alpha_i has a value
            phi_i does not contain Y
            psi_i does not contain Y
            chi_i does not contain Y
            alpha_i does not contain Y
            beta_i does not contain Y
            gamma_i does not contain Y
            alpha_ibeta_i != eps
            Y appears only in matched rules
            Y is not start variable
        </sourcepattern>
        <targetpattern>
            X -> phi_i Y psi_i | chi_i
            Y -> alpha_i Y beta_i | alpha_j gamma_i
        </targetpattern>
    </transformation>

    <transformation name="SynchronizeRecursionEndFromLeftAndRight" 
                    type="EQUIVALENCE">
        <sourcepattern>
            X -> phi_i alpha_j Y psi_i | phi_i Y beta_j psi_i | chi_i
            Y -> alpha_i Y beta_i | gamma_i
            with:
            X is variable
            Y is variable
            X != Y
            phi_i has a value
            alpha_i has a value
            phi_i does not contain Y
            psi_i does not contain Y
            chi_i does not contain Y
            alpha_i does not contain Y
            beta_i does not contain Y
            gamma_i does not contain Y
            alpha_ibeta_i != eps
            Y appears only in matched rules
            Y is not start variable
        </sourcepattern>
        <targetpattern>
            X -> phi_i Y psi_i | chi_i
            Y -> alpha_i Y beta_i | alpha_j gamma_i | gamma_i beta_j
        </targetpattern>
    </transformation>

    <transformation name="UnSplit" type="EQUIVALENCE">
        <sourcepattern>
            X -> phi Y psi | phi Z psi | theta_l
            Y -> alpha_k Y beta_k | gamma_j
            Z -> alpha_k Z beta_k | delta_m
            with:
            X is variable
            Y is variable
            Z is variable
            X != Y
            X != Z
            Y != Z
            Y is not start variable
            Z is not start variable
            alpha_k has a value
            phi does not contain Y
            psi does not contain Y
            theta_l does not contain Y
            alpha_k does not contain Y
            beta_k does not contain Y
            gamma_j does not contain Y
            delta_m does not contain Y
            phi does not contain Z
            psi does not contain Z
            theta_l does not contain Z
            alpha_k does not contain Z
            beta_k does not contain Z
            gamma_j does not contain Z
            delta_m does not contain Z
            Y appears only in matched rules
            Z appears only in matched rules
        </sourcepattern>
        <targetpattern>
            X -> phi A psi | theta_l
            A -> alpha_k A beta_k | gamma_j | delta_m
        </targetpattern>
    </transformation>

    <transformation name="EliminateRedundantRecLevel" type="EQUIVALENCE">
        <sourcepattern>
            X -> phi_i Y chi_i Y psi_i | mu_i Y nu_i | alpha_i
            Y -> phi_i Y chi_i Y psi_i | mu_i Y nu_i | alpha_i | beta_i
            with:
            X is variable
            Y is variable
            X != Y
            Y is not start variable
            phi_i or mu_i has a value
            phi_i does not contain Y
            chi_i does not contain Y
            psi_i does not contain Y
            mu_i does not contain Y
            nu_i does not contain Y
            alpha_i does not contain Y
            beta_i does not contain Y
            alpha_i != beta_j
            Y appears only in matched rules
        </sourcepattern>
        <targetpattern>
            X -> phi_i X chi_i X psi_i | mu_i X nu_i | alpha_i | 
                    phi_i beta_j chi_i beta_k psi_i | mu_i beta_j nu_i
        </targetpattern>
    </transformation>

    <transformation name="UnRoll" type="EQUIVALENCE">
        <sourcepattern>
            X -> phi gamma_j psi | alpha_i
            Y -> gamma_j
            with:
            X is variable
            Y is variable
            X != Y
            gamma_j has a value
            gamma_j != Y
            alpha_i != phiYpsi
        </sourcepattern>
        <targetpattern>
            X -> phi Y psi | phi gamma_j psi | alpha_i
            Y -> gamma_j
        </targetpattern>
    </transformation>

    <transformation name="UnRollParts" type="EQUIVALENCE">
        <sourcepattern>
            X -> phi gamma_i psi | phi Z psi | alpha_i
            Y -> gamma_i | delta_i
            Z -> delta_i | beta_i
            with:
            X is variable
            Y is variable
            Z is variable
            X != Y
            X != Z
            Y != Z
            gamma_i has a value
            delta_i has a value
            gamma_j != Y
            alpha_i != phiYpsi
        </sourcepattern>
        <targetpattern>
            X -> phi Y psi | phi Z psi | alpha_i
            Y -> gamma_i | delta_i
            Z -> delta_i | beta_i
        </targetpattern>
    </transformation>

    <transformation name="MoveRecursionInFrontToSeparateRule" 
                    type="EQUIVALENCE">
        <sourcepattern>
            X -> X alpha_i | alpha_i
            with:
            X is variable
            alpha_i != X
            alpha_i != eps
            alpha_i has a value
        </sourcepattern>
        <targetpattern>
            X -> XX | alpha_i
        </targetpattern>
    </transformation>

    <transformation name="MoveRecursionBehindToSeparateRule" 
                    type="EQUIVALENCE">
        <sourcepattern>
            X -> alpha_i X | alpha_i
            with:
            X is variable
            alpha_i != X
            alpha_i != eps
            alpha_i has a value
        </sourcepattern>
        <targetpattern>
            X -> XX | alpha_i
        </targetpattern>
    </transformation>

    <transformation name="AddEpsToRecursion" type="EQUIVALENCE">
        <sourcepattern>
            X -> XX | alpha_i
            with:
            X is variable
            alpha_i does not contain X
            alpha_i != eps
            alpha_i has a value
            X is not start variable
        </sourcepattern>
        <targetpattern>
            Y -> YY | alpha_i | eps
        </targetpattern>
        <replace>
            X -> alpha_i Y
        </replace>
    </transformation>

    <transformation name="MoveRecursionWithEpsInFrontToSeparateRule" 
                    type="EQUIVALENCE">
        <sourcepattern>
            X -> X alpha_i | eps
            with:
            X is variable
            alpha_i != X
            alpha_i has a value
        </sourcepattern>
        <targetpattern>
            X -> XX | alpha_i | eps
        </targetpattern>
    </transformation>

    <transformation name="MoveRecursionWithEpsBehindToSeparateRule" 
                    type="EQUIVALENCE">
        <sourcepattern>
            X -> alpha_i X | eps
            with:
            X is variable
            alpha_i != X
            alpha_i has a value
        </sourcepattern>
        <targetpattern>
            X -> XX | alpha_i | eps
        </targetpattern>
    </transformation>

    <transformation name="EliminateRedundantRecursionInFront" 
                    type="EQUIVALENCE">
        <sourcepattern>
            X -> XX | X alpha_i | beta_i | eps
            with:
            X is variable
            alpha_i has a value
            alpha_i != X
            beta_i does not start with X
            beta_i != eps
        </sourcepattern>
        <targetpattern>
            X -> XX | alpha_i | beta_i | eps
        </targetpattern>
    </transformation>

    <transformation name="EliminateRedundantRecursionBehind" 
                    type="EQUIVALENCE">
        <sourcepattern>
            X -> XX | alpha_i X | beta_i | eps
            with:
            X is variable
            alpha_i has a value
            alpha_i != X
            beta_i does not end with X
            beta_i != eps
        </sourcepattern>
        <targetpattern>
            X -> XX | alpha_i | beta_i | eps
        </targetpattern>
    </transformation>

    <transformation name="EliminateRedundantDoubleRecursion" 
                    type="EQUIVALENCE">
        <sourcepattern>
            X -> XX | alpha_i XX beta_i | gamma_i | eps
            with:
            X is variable
            alpha_i has a value
            alpha_ibeta_i != eps
            gamma_i does not contain XX
            gamma_i != eps;
        </sourcepattern>
        <targetpattern>
            X -> XX | alpha_i X beta_i | gamma_i | eps
        </targetpattern>
    </transformation>

    <transformation name="EliminateRedundantDoubleReferenceInOtherVar" 
                    type="EQUIVALENCE">
        <sourcepattern>
            X -> XX | phi_i | eps
            Y -> alpha_i XX beta_i | gamma_i
            with:
            X is variable
            Y is variable
            X != Y
            phi_i != XX
            phi_i != eps
            alpha_i has a value
            gamma_i does not contain XX
        </sourcepattern>
        <targetpattern>
            X -> XX | phi_i | eps
            Y -> alpha_i X beta_i | gamma_i
        </targetpattern>
    </transformation>

    <transformation name="MoveRecursionStartToFront" type="EQUIVALENCE">
        <sourcepattern>
            X -> phi_i Y alpha_j psi_i | beta_i
            Y -> YY | alpha_i
            with:
            X is variable
            Y is variable
            X != Y
            alpha_i != YY
            alpha_i != eps
            alpha_i has a value
            phi_i has a value
            beta_k != phi_iYalpha_jpsi_i
        </sourcepattern>
        <targetpattern>
            X -> phi_i alpha_j Y psi_i | beta_i
            Y -> YY | alpha_i
        </targetpattern>
    </transformation>

    <transformation name="MoveRecursionStartWithEpsToFront" 
                    type="EQUIVALENCE">
        <sourcepattern>
            X -> phi_i Y alpha_j psi_i | beta_i
            Y -> YY | alpha_i | eps
            with:
            X is variable
            Y is variable
            X != Y
            alpha_i != YY
            alpha_i != eps
            alpha_i has a value
            phi_i has a value
            beta_k != phi_iYalpha_jpsi_i
        </sourcepattern>
        <targetpattern>
            X -> phi_i alpha_j Y psi_i | beta_i
            Y -> YY | alpha_i | eps
        </targetpattern>
    </transformation>

    <transformation name="EliminateRedundantRecursionInFrontInOtherVar" 
                    type="EQUIVALENCE">
        <sourcepattern>
            X -> XX | alpha_i
            Y -> YX | beta_i
            with:
            X is variable
            Y is variable
            X != Y
            alpha_i has a value
            beta_i has a value
            alpha_i != XX
            beta_i does not contain Y
        </sourcepattern>
        <targetpattern>
            X -> XX | alpha_i
            Y -> beta_i X | beta_i
        </targetpattern>
    </transformation>

    <transformation name="EliminateRedundantRecursionBehindInOtherVar" 
                    type="EQUIVALENCE">
        <sourcepattern>
            X -> XX | alpha_i
            Y -> XY | beta_i
            with:
            X is variable
            Y is variable
            X != Y
            alpha_i has a value
            beta_i has a value
            alpha_i != XX
            beta_i does not contain Y
        </sourcepattern>
        <targetpattern>
            X -> XX | alpha_i
            Y -> X beta_i | beta_i
        </targetpattern>
    </transformation>

    <transformation name="EliminateRedundantRecursionInFrontAndBehindInOtherVar" 
                    type="EQUIVALENCE">
        <sourcepattern>
            X -> XX | alpha_i
            Y -> YX | XY | beta_i
            with:
            X is variable
            Y is variable
            X != Y
            alpha_i has a value
            beta_i has a value
            alpha_i != XX
            beta_i does not contain Y
        </sourcepattern>
        <targetpattern>
            X -> XX | alpha_i
            Y -> X beta_i X | X beta_i | beta_i X | beta_i
        </targetpattern>
    </transformation>

    <transformation name="EliminateRedundantReferenceBehindInOtherVar" 
                    type="EQUIVALENCE">
        <sourcepattern>
            X -> alpha_i Y | eps
            Y -> YY | alpha_i | eps
            with:
            X is variable
            Y is variable
            X != Y
            alpha_i != YY
            alpha_i != eps
            alpha_i has a value
        </sourcepattern>
        <targetpattern>
            X -> Y
            Y -> YY | alpha_i | eps
        </targetpattern>
    </transformation>

    <transformation name="EliminateRedundantReferenceInFrontInOtherVar" 
                    type="EQUIVALENCE">
        <sourcepattern>
            X -> Y alpha_i | eps
            Y -> YY | alpha_i | eps
            with:
            X is variable
            Y is variable
            X != Y
            alpha_i != YY
            alpha_i != eps
            alpha_i has a value
        </sourcepattern>
        <targetpattern>
            X -> Y
            Y -> YY | alpha_i | eps
        </targetpattern>
    </transformation>
</transformations>

\end{lstlisting}
}

\subsection{Hard-coded transformations}\label{app:hard-trafos}

In this section, we define the hard-coded transformations used in the normalization pipeline.

\begin{itemize}
    \item EliminateNonGenVars: eliminate non-generating variables with algorithm given in \cite{HopcroftU79}
    \item EliminateUnReachVars: eliminate unreachable variables with algorithm given in \cite{HopcroftU79}
    \item EliminateDelegatingVars: eliminate variables $X$ that have only one production $X \to Y$
    \item EliminateSingleRuleVars: eliminate variables that have only one production (except start variable)
    \item EliminateNonRecVars: eliminate variables $X$ that have a derivation $X \Rightarrow^* uXv$ for some sentential forms $u,v$
    \item EliminateNonSelfRecVars: eliminate variables $X$ that have a derivation $X \Rightarrow uXv$ for some sentential forms $u,v$
    \item EliminateLooselyIsomorphicVar: eliminate isomorphic variables (with a relaxed definition of isomorphy that still preservers equivalence)
    \item EliminateEpsRules: eliminate epsilon rules with algorithm given in \cite{HopcroftU79}
    \item EliminateSelfRecUnitRules: eliminate productions of the type $X \to X$
    \item EliminateUnitRules: eliminate productions of the type $X \to Y$ with algorithm given in \cite{HopcroftU79}
    \item EliminateRedundantRules: eliminate productions that are a \enquote{short-cut}, i.e. that can be generated by a longer derivation
    \item ExplicateEpsRules: for variables $X$ with $X\Rightarrow^*\eps$, add the production $X\to\eps$
\end{itemize}
\newpage
\subsection{The normalization pipeline}\label{app:pipeline-self}

{
\small
\begin{lstlisting}[]
// ######################################
// ### 0. base simplifications - fast ###
// ######################################
EliminateNonGenVars
EliminateUnReachVars
EliminateSelfRecUnitRules
EliminateDelegatingVars
(
    EliminateLooselyIsomorphicVar
    EliminateSelfRecUnitRules
    EliminateDelegatingVars
)*
( GUARD_NUMBER_OF_PRODUCTIONS[>100] eps
| GUARD_NUMBER_OF_PRODUCTIONS[<=100]
    (
        // =====================================
        // ||| A. do not eliminate eps rules |||
        // =====================================
        // ##########################################################
        // ### A1. remove as many non-recursive rules as possible ###
        // ##########################################################
        EliminateSingleRuleVars
        (
            eps
        |
            EliminateNonRecVars
            GUARD_NUMBER_OF_PRODUCTIONS[<=100]
            (
                eps
            |
                {EliminateNonSelfRecVars}
                GUARD_NUMBER_OF_PRODUCTIONS[<=100]
            )
        )
        // ############################################
        // ### A2. add (all possible) unit rules    ###
        // ############################################
        UnRoll*
        UnRollParts*
        {EliminateRedundantRules}
        ( GUARD_NUMBER_OF_NON_CHANGING_TRANSFORMATIONS[>0]
            UnSplit*
        | GUARD_NUMBER_OF_NON_CHANGING_TRANSFORMATIONS[=0]
            UnSplit*
            // ##########################################################
            // ### A3. remove as many non-recursive rules as possible ###
            // ##########################################################
            EliminateSingleRuleVars
            (
                eps
            |
                EliminateNonRecVars
                GUARD_NUMBER_OF_PRODUCTIONS[<=100]
                (
                    eps
                |
                    {EliminateNonSelfRecVars}
                    GUARD_NUMBER_OF_PRODUCTIONS[<=100]
                )
            )
            {EliminateRedundantRules}
        )
        // #############################################################
        // ### A4. optionally remove unit rules (can reduce grammar) ###
        // #############################################################
        (
            eps
        |
            EliminateUnitRules
            EliminateUnReachVars
        )
        (
            EliminateLooselyIsomorphicVar
            EliminateSelfRecUnitRules
            EliminateDelegatingVars
        )*
        // ####################################################
        // ### A5. unifying Kleene recursions (L^+ and L^*) ###
        // ####################################################
        ExplicateEpsRules
        MoveRecursionInFrontToSeparateRule*
        MoveRecursionBehindToSeparateRule*
        (
            eps
        |
            AddEpsToRecursion*
            GUARD_NUMBER_OF_PRODUCTIONS[<=100]
        )
        MoveRecursionWithEpsInFrontToSeparateRule*
        MoveRecursionWithEpsBehindToSeparateRule*
        EliminateRedundantDoubleRecursion*
        EliminateRedundantRecursionInFront*
        EliminateRedundantRecursionBehind*
        EliminateRedundantDoubleReferenceInOtherVar*
        EliminateRedundantReferenceInFrontInOtherVar*
        EliminateRedundantReferenceBehindInOtherVar*
        EliminateDelegatingVars
        MoveRecursionStartWithEpsToFront*
        MoveRecursionStartToFront*
        // #####################################################
        // ### A6. remove redundant recursion in other rules ###
        // #####################################################
        EliminateRedundantRecursionInFrontInOtherVar*
        EliminateRedundantRecursionBehindInOtherVar*
        EliminateRedundantRecursionInFrontAndBehindInOtherVar*
        ( GUARD_NUMBER_OF_NON_CHANGING_TRANSFORMATIONS[>3] eps
        | GUARD_NUMBER_OF_NON_CHANGING_TRANSFORMATIONS[<=3]
            // ##############################################
            // ### A7. reapply rules for Kleene recursion ###
            // ##############################################
            {EliminateRedundantRules}
            // remove as many non-recursive rules as possible
            EliminateSingleRuleVars
            (
                eps
            |
                EliminateNonRecVars
                GUARD_NUMBER_OF_PRODUCTIONS[<=100]
                (
                    eps
                |
                    {EliminateNonSelfRecVars}
                    GUARD_NUMBER_OF_PRODUCTIONS[<=100]
                )
            )
            // applying unifying Kleene recursion trafos a second time
            EliminateRedundantDoubleRecursion*
            EliminateRedundantRecursionInFront*
            EliminateRedundantRecursionBehind*
            EliminateRedundantDoubleReferenceInOtherVar*
            EliminateRedundantReferenceInFrontInOtherVar*
            EliminateRedundantReferenceBehindInOtherVar*
            EliminateDelegatingVars
            MoveRecursionStartWithEpsToFront*
            MoveRecursionStartToFront*
        )
        (
            EliminateLooselyIsomorphicVar
            EliminateSelfRecUnitRules
            EliminateDelegatingVars
        )*
        // #####################################
        // ### A8. unifying other recursions ###
        // #####################################
        UnSplit*
        EliminateRedundantRecLevel*
        SynchronizeRecursionLevel*
        SynchronizeRecursionEndFromLeft*
        SynchronizeRecursionEndFromRight*
        SynchronizeRecursionEndFromLeftAndRight*
        EliminateDelegatingVars
        // =====================================
        // ||| END OF A.                     |||
        // =====================================
    |
        // =====================================
        // ||| B. eliminate eps rules        |||
        // =====================================
        EliminateEpsRules
        EliminateNonGenVars
        EliminateUnReachVars
        GUARD_NUMBER_OF_PRODUCTIONS[<=100]
        (
            EliminateLooselyIsomorphicVar
            EliminateSelfRecUnitRules
            EliminateDelegatingVars
        )*
        // ##########################################################
        // ### B1. remove as many non-recursive rules as possible ###
        // ##########################################################
        EliminateSingleRuleVars
        EliminateNonRecVars
        {EliminateNonSelfRecVars}
        GUARD_NUMBER_OF_PRODUCTIONS[<=100]
        // ############################################
        // ### B2. add (all possible) unit rules    ###
        // ############################################
        UnRoll*
        UnRollParts*
        {EliminateRedundantRules}
        ( GUARD_NUMBER_OF_NON_CHANGING_TRANSFORMATIONS[>0]
            UnSplit*
        | GUARD_NUMBER_OF_NON_CHANGING_TRANSFORMATIONS[=0]
            UnSplit*
            // ##########################################################
            // ### B3. remove as many non-recursive rules as possible ###
            // ##########################################################
            EliminateSingleRuleVars
            EliminateNonRecVars
            {EliminateNonSelfRecVars}
            GUARD_NUMBER_OF_PRODUCTIONS[<=100]
            // #############################################################
            // ### B4. optionally remove unit rules (can reduce grammar) ###
            // #############################################################
            (
                eps
            |
                EliminateUnitRules
                EliminateUnReachVars
            )
        )
        // =====================================
        // ||| END OF B.                     |||
        // =====================================
    )
    // #####################################################
    // ### 9. simplification after recursion unification ###
    // #####################################################
    {EliminateRedundantRules}
    (
        EliminateLooselyIsomorphicVar
        EliminateSelfRecUnitRules
        EliminateDelegatingVars
    )*
)
MinimalAlphabets
CanonicalGrammar
\end{lstlisting}
}

\section{Bug-fixing transformations}\label{app:correcting-trafos}
In this section, we define the three proto-typical bug-fixing transformations used in the evaluation in Section \ref{sec:rq3}.

\subsection[Bug-fixing transformation C1: add epsilon as recursion end]{Bug-fixing transformation C1: add $\varepsilon$ as recursion end}
{
\small
\lstset{language=XML,morekeywords={transformations,transformation,targetpattern,sourcepattern,encoding,xs:schema,xs:element,xs:complexType,xs:sequence,xs:attribute}}  

\begin{lstlisting}
<transformation name="AddEpsilonAsRecursionEnd" type="CORRECTING">
    <sourcepattern>
        X -> alpha_i X beta_i | gamma_i
        with:
        X is variable
        alpha_i has a value
        alpha_ibeta_i != eps
        gamma_i does not contain X
        gamma_i != eps
    </sourcepattern>
    <targetpattern>
        X -> alpha_i X beta_i | gamma_i | eps
    </targetpattern>
</transformation>
\end{lstlisting}
}

\subsection{Bug-fixing transformation C2: add canonical recursion end}

{
\small
\lstset{language=XML,morekeywords={transformations,transformation,targetpattern,sourcepattern,encoding,xs:schema,xs:element,xs:complexType,xs:sequence,xs:attribute}}  

\begin{lstlisting}
<transformation name="AddCanonicalRecursionEnd" type="CORRECTING">
    <sourcepattern>
        X -> alpha_i X beta_i | gamma_i
        with:
        X is variable
        alpha_i has a value
        alpha_ibeta_i != eps
        gamma_i does not contain X
        gamma_i != alpha_jbeta_j
    </sourcepattern>
    <targetpattern>
        X -> alpha_i X beta_i | gamma_i | alpha_i beta_i
    </targetpattern>
</transformation>
\end{lstlisting}
}

\subsection[Bug-fixing transformation C3: replace epsilon by canonical recursion end]{Bug-fixing transformation C3: replace $\varepsilon$ by canonical recursion end}

{
\small
\lstset{language=XML,morekeywords={transformations,transformation,targetpattern,sourcepattern,encoding,xs:schema,xs:element,xs:complexType,xs:sequence,xs:attribute}}  

\begin{lstlisting}
<transformation name="ReplaceEpsilonAsRecursionEndByCanonicalOne" 
                type="CORRECTING">
    <sourcepattern>
        X -> alpha_i X beta_i | gamma_i | eps
        with:
        X is variable
        alpha_i has a value
        alpha_ibeta_i != eps
        gamma_i does not contain X
        gamma_i != alpha_jbeta_j
        gamma_i != eps
    </sourcepattern>
    <targetpattern>
        X -> alpha_i X beta_i | gamma_i | alpha_i beta_i
    </targetpattern>
</transformation>
\end{lstlisting}
}

\section{Proof: ŅP-Hardness of Matching a Pattern Into a Grammar}\label{sec:np-hardness}
We show the \NP-hardness by giving a reduction from \textsc{Clique}.

Let $(\mathfrak{G}=(V=\{1,\dots,n\},E),k)$ be an instance of \textsc{Clique}. From $\mathfrak{G}$, we build a grammar $G_\mathfrak{G}=(N_G=V, \Sigma_G=\{0,1,\#\}, P_G, S_G)$ with productions $P_G=\{v_i\to v_j \mid \{v_i,v_j\}\in E\}$ and arbitrary $S_G\in V$. From $k$, we build a pattern $\pi_k=(V_\pi, R_\pi, C_\pi)$ with $V_\pi = \{X_i,\alpha_i \mid i\in\{1,\ldots,k\}\}$, $R_\pi=\{X_i\to X_j \mid i,j\in\{1,\ldots,k\}\}\cup\{X_i\to \alpha_i \mid i\in\{1,\ldots,k\}\}$ and $C_\pi=\{X_i\neq X_j \mid i,j\in\{1,\ldots,k\}\}$. It is easy to see that the pattern $\pi_k$ matches into $G_\mathfrak{G}$ if and only if the graph $\mathfrak{G}$ contains a $k$-clique.

\end{document}